%% file: main.tex
\documentclass[preprint,12pt,authoryear]{elsarticle}

\usepackage{amssymb}
\usepackage{amssymb}
\usepackage{amsmath}
\usepackage{amsfonts}
\usepackage{algorithm}
\usepackage[noend]{algorithmic}

\newproof{pf}{Proof}

\newlength{\minipagewidth}
\newcommand{\bookbox}[1]{\small
\par\medskip\noindent
\framebox[\columnwidth]{
\begin{minipage}{\minipagewidth} {#1} \end{minipage} } \par\medskip }

\journal{Artificial Intelligence Journal}

\input{symbolsdef}

\allowdisplaybreaks

\begin{document}

\begin{frontmatter}



\title{Truthful Learning Mechanisms for Multi--Slot Sponsored Search Auctions with Externalities}


\author[polimi]{Nicola Gatti\corref{cor1}}
    \ead{nicola.gatti@polimi.it}
    \author[inria]{Alessandro Lazaric}
    \ead{alessandro.lazaric@inria.fr}
    \author[polimi]{Marco Rocco}
    \ead{marco.rocco@polimi.it}
    \author[polimi]{Francesco Trov\`{o}}
    \ead{francesco1.trovo@polimi.it}

\address[polimi]{Politecnico di Milano, piazza Leonardo da Vinci 32, \\20133 Milan, Italy} 
\address[inria]{INRIA Lille - Nord Europe, avenue Halley 40, \\59650 Villeneuve d'Ascq, France }

\begin{abstract}
Sponsored search auctions constitute one of the most successful applications of \emph{microeconomic mechanisms}. In mechanism design, auctions are usually designed to incentivize advertisers to bid their truthful valuations and, at the same time, to assure both the advertisers and the auctioneer a non--negative utility. Nonetheless, in sponsored search auctions, the click--through--rates (CTRs) of the advertisers are often unknown to the auctioneer and thus standard \emph{truthful} mechanisms cannot be directly applied and must be paired with an effective learning algorithm for the estimation of the CTRs. This introduces the critical problem of designing a learning mechanism able to estimate the CTRs at the same time as implementing a truthful mechanism with a revenue loss as small as possible compared to an optimal mechanism designed with the true CTRs. Previous work showed that, when \emph{dominant--strategy} truthfulness is adopted, in single--slot auctions the problem can be solved using suitable exploration--exploitation mechanisms able to achieve a per--step regret (over the auctioneer's revenue) of order $O(T^{-\frac{1}{3}})$  (where $T$ is the number of times the auction is repeated). It is also known that,  when \emph{truthfulness in expectation} is adopted, a per--step regret (over the social welfare) of order $O(T^{-\frac{1}{2}})$ can be obtained. In this paper we extend the results known in the literature to the more complex case of multi--slot auctions. In this case, a model of the user is needed to characterize how the advertisers' valuations change over the slots. We adopt the \emph{cascade model} that is the most famous model in the literature for sponsored search auctions. We prove a number of novel upper bounds and lower bounds both on the auctioneer's revenue loss and social welfare w.r.t. to the VCG auction and we report numerical simulations investigating the accuracy of the bounds in predicting the dependency of the regret on the auction parameters.
\end{abstract}

\begin{keyword}
Economic paradigms, mechanism design, online learning, sponsored search auctions.



\end{keyword}

\end{frontmatter}


\input{sec/01intro}
\input{sec/02preliminaries}

\input{sec/03model}

\input{sec/04posdep}
\input{sec/05posaddep}

\input{sec/06simulation}
\input{sec/07conclusion}
\bibliographystyle{elsarticle-num} 
\bibliography{citations}

\clearpage
\appendix
\input{sec/AppA-monotonicity}
\input{sec/App-proofs}

\input{sec/App-proofsSW}

\end{document}

%% file: symbolsdef.tex
\newcommand{\N}{\mathcal{N}}

\newcommand{\K}{\mathcal{K}}

\newcommand{\F}{\mathcal{F}}
\newcommand{\calG}{\mathcal{G}}

\newcommand{\E}{\mathbb{E}}
\newcommand{\Prob}{\mathbb{P}}
\newcommand{\ind}{\mathbb{I}}

\newcommand{\bv}{\mathbf{v}}
\newcommand{\hbv}{\hat{\mathbf{v}}}
\newcommand{\bs}{\mathbf{s}}
\newcommand{\bx}{\mathbf{x}}
\newcommand{\by}{\mathbf{y}}

\newcommand{\bone}{\boldsymbol{1}}
\newcommand{\tp}{\tilde{p}}

\newcommand{\SW}{\text{SW}}

\newcommand{\tSW}{\widetilde{\text{SW}}}

\newcommand{\hf}{\hat{f}}
\newcommand{\tf}{\tilde{f}}
\newcommand{\tg}{\tilde{g}}

\newcommand{\eps}{\Gamma}

\newcommand{\Gmin}{\Gamma_{\min}}

\newcommand{\hcp}{\tilde{c}^+}

\newcommand{\hq}{\tilde{q}}
\newcommand{\hqp}{\tilde{q}^+}
\newcommand{\qmin}{q_{\min}}

\newcommand{\vmax}{v_{\max}}
\newcommand{\hv}{\hat{v}}
\newcommand{\bR}{\overline{R}}
\newcommand{\tR}{\tilde{R}}

\newcommand{\avcg}{A--VCG}
\newcommand{\padavcg}{PAD--A--VCG}

\newcommand{\cl}{click}

\newcommand{\myremark}[1]{\textit{Remark #1}}

\newtheorem{lemma}{Lemma}

\newtheorem{corollary}{Corollary}
\newtheorem{proposition}{Proposition}

\newtheorem{theorem}{Theorem}

%% file: sec/01intro.tex

\section{Introduction}\label{s:introduction}

Sponsored search auctions (SSAs) constitute one of the most successful applications of \emph{microeconomic mechanisms}, producing a revenue of about \$6 billion dollars in the US alone in the first half of 2010~\cite{IABreport2010}. In a SSA, a number of \emph{advertisers} bid to have their \emph{sponsored links} (from here on \textit{ads}) displayed in some slot alongside the search results of a keyword. Sponsored search auctions currently adopt a \emph{pay--per--click} scheme, requiring positive payments to an advertiser only if its ad has been clicked. Given an allocation of ads over the slots, each ad is associated with a \emph{click--through--rate} (CTR) defined as the probability that such ad will be clicked by the user. CTRs are estimated by the auctioneer and play a crucial role in the auction, since they are used by the auctioneer to find the optimal allocation (in expectation) and to compute the payments for each ad.

There is a large number of works formalizing SSAs as a \emph{mechanism design} problem~\cite{Narahari2009}, where the objective is to design an auction mechanism that incentivizes advertisers to bid their \emph{truthful} valuations (needed for \emph{economic stability}) and that assures both the advertisers and the auctioneer to have a non--negative utility. The most common SSA mechanism is the \emph{generalized second price} (GSP) auction~\cite{Edelman2007,Varian2007}. This mechanism is proved not to be truthful and advertisers may implement bidding strategies that gain more than bidding their truthful valuations as shown in~\cite{Edelman2007}. While in complete information settings the worst Nash equilibrium in the GSP gives a revenue to the auctioneer equal to the revenue given by the Vickrey--Clarke--Groves (VCG) equilibrium~\cite{Edelman2007}, in Bayesian settings the worst Bayes--Nash equilibrium in the GSP can provide a much smaller revenue than the VCG---a lower bound of $\frac{1}{8}$ is provided in~\cite{paes}. The implementation of the VCG mechanism (assuring truthfulness) for SSAs has been investigated in~\cite{Narahari2009}. Although the VCG mechanism is not currently adopted by the search engines (but it is, e.g., by Facebook), a number of scientific theoretical results builds upon it. 

In this paper, we focus on the problem of designing truthful mechanisms when the CTRs are not known and need to be estimated in SSAs with multiple slots. This problem is particularly relevant in practice because the assumption that all the CTRs are known beforehand is rarely realistic. Furthermore, it also poses interesting scientific challenges since it represents one of the first examples where learning theory is paired with mechanism design techniques to obtain effective methods to learn under equilibrium constraints (notably the truthfulness property). Another field where these ideas have been used is crowdsourcing~\cite{crowdsourcing}.
The problem of estimating the CTRs and to identify the best allocation of ads is effectively formalized as a \textit{multi--arm bandit problem}~\cite{robbins1952some} where each ad is an arm and the objective is to minimize the cumulative regret (i.e., the revenue loss w.r.t. an optimal allocation defined according to the exact CTRs). The problem of budgeted advertisers (i.e., auctions where the total amount of money each advertiser is willing to pay is limited) with multiple queries is considered in~\cite{pandey2006handling}. This problem is formalized as a budgeted multi--bandit multi--arm problem, where each bandit corresponds to a query, and an algorithm is proposed with explicit bounds over the regret on the revenue. Nonetheless, the proposed method works in a non--strategic environment, where the advertisers do not try to influence the outcome of the auction and always bid their true values. The strategic dimension of SSAs is partially taken into consideration in~\cite{langford2010maintaining} where the advertisers are assumed to play a bidding strategy at the equilibrium w.r.t. a set of estimated CTRs which are available to both the auctioneer and the advertisers. The authors introduce a learning algorithm which explores different rankings on the ads so as to improve the CTR estimates and, at the same, not to introduce incentives for the advertisers to deviate from the previous equilibrium strategy. A more complete notion of truthfulness for bandit algorithms in multi--slot SSAs is studied in~\cite{gonen2007incentive-compatible}. In particular, they build on the action elimination algorithm in~\cite{even-dar2006action} and they report a probably approximately correct (PAC) analysis of its performance. Unfortunately, as pointed in~\cite{devanur2009price} and \cite{babaioff2008characterizing} the mechanism is not guaranteed to be truthful and thus it only works when the advertisers bid their true values. An extension to the action elimination algorithm is also proposed in~\cite{gonen2007an-adaptive} for the more general setting where budgeted advertisers are allowed to enter and exit the auction at different time instants that they declare along with their bid. The authors derive an algorithm that approximately achieves the best social welfare under the assumption that the gain of untruthful declarations is limited. Finally, single--slot online advertising is studied also in~\cite{nazerzadeh2008dynamic} where the notion of Bayesian incentive compatibility (BIC) is taken into consideration and an asymptotically BIC and \textit{ex ante} efficient mechanism is introduced. The most complete study of truthful bandit mechanisms so far is reported in~\cite{devanur2009price} and \cite{babaioff2008characterizing}. These works first provided a complete analysis on the constraints truthfulness forces on the multi--arm bandit algorithm with single--slot SSAs, showing that no \emph{dominant--strategy} truthful bandit mechanism can achieve a regret (over the social welfare and over the auctioneer's revenue) smaller than $\tilde\Omega(T^\frac{2}{3})$ and that the exploration and exploitation phases must be separate. Furthermore, they also suggest nearly--optimal algorithms. Instead, when the notion of truthfulness is relaxed, adopting truthfulness \emph{in expectation} w.r.t. click (and possibly mechanism) randomness, it is possible to obtain a regret $\tilde O(T^\frac{1}{2})$ (over the social welfare) without  separating the exploration and exploitation phases in the case of single--slot SSAs~\cite{babaioff_impl_pay}.


When multiple slots are present, a user model is needed to describe how the valuations of the advertisers change over the slots. All the models available in the literature assume the separation of the CTR as the product of two terms, the first capturing the probability that an ad will be clicked once observed by the user, while the second capturing the probability that the user will observe such an ad given the displayed allocation. The basic model (commonly referred to as \emph{separability model}) prescribes that the probability of observing an ad depends only on its position~\cite{Narahari2009}. Recently, more accurate models have been proposed and the most famous model is the \emph{cascade model} according to which the user scans the slots from top to bottom and the probability with which the user moves from a slot to the next one depends on the ad and on the slot (this kind of user is commonly called \emph{Markovian user})~\cite{Kempe2008,Aggarwal2008}, while with the remaining probability the user stops to observe ads. As a result, the probability of observing an ad depends on position of the ad and on all the ads allocated above. The validity of the cascade model has been evaluated and supported by a wide range of experimental investigations~\cite{Craswell,Joachims}. The only results on learning mechanisms for SSAs with multiple slots are described in~\cite{sarma2010multi-armed}, where the authors characterize dominant--strategy truthful mechanisms and provide theoretical bounds over the social welfare regret for the separability model. However, these results are partial, e.g., they do not solve the common case in which the slot--dependent parameters are monotonically decreasing in the slots, and they cannot easily be extended to the more challenging case of the cascade model (see discussion in Section~\ref{ss:online.mechanism}). 

In the present paper, we build on the results available in the literature and we provide a number of contributions when the separability model and the cascade model are adopted. More precisely, our results can be summarized as follow.
\begin{itemize}
\item \emph{Separability model with monotone parameters/only position--dependent cascade model}: in this case, there are two groups of parameters, one related to the ads (called \emph{quality}) and one to the slots (called \emph{prominence}). We studied all configurations of information incompleteness. When only qualities are unknown, we provide a regret  analysis in dominant--strategy truthfulness obtaining a regret of $\tilde O(T^\frac{2}{3})$ (while it is open whether it is possible to obtain a better upper bound adopting truthfulness in expectation). When only prominences are unknown, we provide a regret  analysis in truthfulness in expectation obtaining a regret of $0$, whereas we show that any dominant--strategy truthful learning mechanism would have a regret of $\tilde{\Theta}(T)$. When both groups of parameters are unknown, we provide a regret analysis in truthfulness in expectation obtaining a regret of $\tilde {O}(T^\frac{2}{3})$ (while it is open whether it is possible to obtain a better upper bound adopting truthfulness in expectation), whereas any dominant--strategy truthful learning mechanism would have a regret of $\tilde{\Theta}(T)$.
\item \emph{Cascade model}: in the non--factorized cascade model (i.e., when the observation probabilities can be any) we show that it is possible to obtain a regret of $\tilde {O}(T^\frac{2}{3})$ in dominant--strategy truthful learning mechanisms when only the qualities of the ads are unknown. We show also that in the factorized cascade model (i.e., when the observation probabilities are the products of terms depending only on the position or on the ads as used in~\cite{Kempe2008}), in the very special case in which the ad--dependent parameters are unknown we obtain a regret of $\tilde{\Theta}(T)$ in dominant--strategy truthful learning mechanisms (while it is open whether it is possible to obtain a better upper bound adopting truthfulness in expectation).
\item \emph{Learning parameters}: for each setting of uncertainty we study we provide functions, to be used in practice, to set the learning parameters in order to minimize the bound over the regret given the parameters in input.
\item \emph{Numerical simulations}: we investigate the accuracy of all the bounds we provide in the paper in predicting the dependency of the regret on the auction parameters by numerical simulations. We show that the theoretical dependency matches the actual dependency we observed by simulation.
\end{itemize}

The paper is organized as follows. In Section~\ref{s:notation} we briefly review the basics of mechanism design and multi--armed bandit learning. Section~\ref{s:statement} formalizes sponsored search auctions and introduces the corresponding online learning mechanism design problem. In Section~\ref{s:statement} we also provide a more formal overview of existing results in comparison with the findings of this paper. In Sections~\ref{s:constant} and~\ref{s:externalities} we report and discuss the main regret bounds in the case of position--dependent and position-- and ad--dependent externalities. In Section~\ref{s:experiments} we report numerical simulations aiming at testing the accuracy of the theoretical bounds. Section~\ref{s:conclusions} concludes the paper and proposes future directions of investigation. The detailed proofs of the theorems are reported in Appendix.

%% file: sec/02preliminaries.tex

\section{Preliminaries}\label{s:notation}

\subsection{Economic Mechanisms}
\input{sec/021economicMechanisms}

\subsection{Multi--Armed Bandit}
\input{sec/022MultiArmedBandit}

%% file: sec/021economicMechanisms.tex

In this section we provide some background on mechanism design. The aim of mechanism design~\cite{mas-colell1995microeconomic} is to design \emph{allocation} and \emph{payment functions} satisfying some desirable properties when agents are \emph{rational} and have \emph{private} information representing their preferences---also referred to as the \emph{type} of the agent. Without loss of generality, mechanism design focuses on mechanisms, said \emph{direct}, in which the only action available to the agents is to report their (potentially non--truthful) type. On the basis of the agents' reports the mechanism determines the allocation (of resources) to agents and the agents' payments. 

The main desirable property of a mechanism is \emph{truthfulness}---aka \emph{incentive compatibility} (IC)---and requires that reporting  the true types constitutes an \emph{equilibrium strategy profile} for the agents. When a mechanism is not truthful, agents should find their (untruthful) best strategies on the basis of some possible model about the opponents' behavior, but, in absence of common information, no normative model for rational agents exists. This leads the mechanism to be economically unstable, given that the agents continuously change their strategies. As it is customary in game theory, there are different solution concepts and consequently there are different notions of truthfulness. The most common ones are \emph{dominant strategy incentive compatibility} (DSIC)---i.e., reporting the true types is the best action an agent can play independently of the actions of the other agents---, \emph{ex post incentive compatibility} (\emph{ex post} IC)---i.e., reporting the true types is a Nash equilibrium---, and \emph{Bayesian incentive compatibility} (BIC)---i.e., reporting the true types is a Bayes--Nash equilibrium. Interestingly, DSIC and \emph{ex post} IC are equivalent notions of truthfulness in absence of interdependencies, while BIC is weaker than DSIC since it requires that every agent has a Bayesian prior over the types of the other agents and IC is in expectation w.r.t. the prior. When there are sources of randomness in the mechanism design problem (not due to the distribution of probabilities over the types of the agents), e.g., random components of the mechanism or the realization of events, weaker solution concepts, said \emph{in expectation}, are commonly adopted, e.g., DSIC in expectation or \emph{ex post} IC in expectation. Since in the present paper we will only focus on DSIC, whenever some source of randomization is present (e.g., clicks or randomized mechanisms), we will use ``IC'' or ``DSIC'' to refer to DSIC \textit{a posteriori}, and ``IC in expectation'' for ``DSIC in expectation''. Moreover, mechanisms can exploit the realizations of the events adopting different payment functions for each different realization. These mechanisms are said \emph{execution contingent} (EC)~\cite{gerding1,gerding2}. 

In addition to IC, other desirable properties include: \emph{allocative efficiency} (AE)---i.e., the allocation maximizes the social welfare---, \emph{individual rationality} (IR)---i.e., each agent is guaranteed to have no loss when reporting truthfully---, and \emph{weak budget balance} (WBB)---i.e., the mechanism is guaranteed to have no loss. In presence of sources of randomness, IR and WBB can be \emph{in expectation} w.r.t. all the possible realizations, or \emph{a posteriori} if they hold for every possible realization.

The economic literature provides an important characterization of the allocation functions that can be adopted in IC mechanisms when utilities are \emph{quasi linear}~\cite{mas-colell1995microeconomic}. Here, we survey the main results related to DSIC mechanisms. In unrestricted domains (i.e., the agents' types are defined over spaces with arbitrary structure) for the agents' preferences, only \emph{weighted maximal--in--its--range} allocation functions can be adopted in DSIC mechanisms~\cite{nisan2007computationally,Nisan:2007:AGT:1296179}. More precisely, a weighted maximal--in--its--range allocation function chooses, among a subset of allocations that does not depend on the types reported by the agents (i.e., the range), the allocation maximizing the weighted social welfare, where each agent is associated with a positive (type--independent) weight. It trivially follows that, when the range is composed of all the possible allocations and all the agents have the same weights, only AE mechanisms can be DSIC. When weighted maximal--in--its--range allocation functions are adopted, only weighted Groves payments lead to DSIC mechanisms~\cite{mas-colell1995microeconomic}. The most common DSIC mechanism is the Vickrey--Clarke--Groves (VCG), in which the range is composed of all the allocations and all the weights are unitary. VCG satisfies also IR and WBB and, among all the Groves mechanisms, the VCG is the mechanism maximizing the revenue of the auctioneer. We refer to the weighted version of the VCG as WVCG.

When the domain of the agents' preferences is restricted (i.e., the types are defined over spaces with specific structure, e.g., compact sets or discrete values), weighted maximal--in--its--range property is not necessary for DISC. The necessary condition is weakly monotonicity~\cite{mas-colell1995microeconomic}, which is also sufficient for convex domains. In specific restricted domains, weak monotonicity leads to simple and operational tools. For instance, when the preferences of the agents are single--parameter linear---i.e., the agents' value is given as the product between the agent's type and an allocation--dependent coefficient called \emph{load}~\cite{tardos_sp}---, monotonicity requires that the load is monotonically increasing in the type of the agent. In this case, any DSIC mechanism is based on the Myerson's payments defined in~\cite{tardos_sp}.\footnote{See~\ref{ap:monotonicity} for the definition of monotonicity in single--parameter linear environments and Myerson's payments.} Notice that the VCG mechanism is still the mechanism maximizing the auctioneer's revenue among all the DSIC mechanism, including those that are not AE. The payments defined in~\cite{tardos_sp} include an integral that may be not easily computable.  However, by adopting IC in expectation (over the randomness of the mechanism), such integral can be easily estimated by using samples~\cite{archer2003approximate}.  Another drawback of the payments described in~\cite{tardos_sp} is that they require the off--line evaluation of the social welfare of  the allocations for some agents' types different from the reported ones and this may be not possible in many practical situations. A way to overcome this issue is to adopt the result presented in~\cite{babaioff_impl_pay}, in which the authors propose an implicit way to calculate the payments. More precisely, given an allocation function in input, a random component is introduced such that with a small probability the reported types of the agents are modified to obtain the allocations that are needed to compute the payments in~\cite{tardos_sp}. The resulting allocation function is less efficient than the allocation function given in input, but the computation of the payments is possible and it is executed online.

%% file: sec/022MultiArmedBandit.tex

The multi--arm bandit (MAB)~\cite{robbins1952some} is a simple yet powerful framework formalizing the online decision--making problem under uncertainty. Historically, the MAB framework finds its motivation in optimal experimental design in clinical trials, where two new treatments, say $A$ and $B$, need to be tested. In an idealized version of the clinical trial, $T$ patients are sequentially enrolled in the trial, so that whenever a treatment is tested on a patient, the outcome of the test is recorded and it is used to choose which treatment to provide to the next patient. The objective is to provide the best treatment to the largest number of patients. This raises the challenge of balancing the collection of information and the maximization of the performance, a problem usually referred to as the \textit{exploration--exploitation} trade--off. In fact, on the one hand, it is important to gather information about the effectiveness of the two treatments by repeatedly providing them at different patients (\textit{exploration}). On the other hand, in order to meet the objective, as an estimation of effectiveness of the two treatments is available, the (estimated) best treatment should be selected more often (\textit{exploitation}). This scenario matches with a large number of applications, such as online advertisements, adaptive routing, cognitive radio. In general, the MAB framework can be adopted whenever a set of $N$ arms (e.g., treatments, ads) is available and the rewards (e.g., effectiveness of a treatment, click--through--rate of an ad) associated to each of them are random realizations from unknown distributions. Although this problem can be solved by dynamic programming methods and notably by using the Gittins index solution~\cite{gittins1979bandit}, this requires a prior over the distribution of the reward of the arms and it is often computationally heavy (high--degree polynomial in $T$). More recently, a wide range of techniques have been developed to solve the bandit problem. In particular, these algorithms formalize the objective using the notion of \textit{regret}, which corresponds to the difference in performance over $T$ steps between an optimal selection strategy which knows in advance the performance of all the arms and an adaptive strategy which learns over time which arms to select. Although a complete review of the bandit algorithms is beyond the scope of this paper (see~\cite{bubeck2012regret} for a review), we only discuss two results which are relevant to the rest of the paper. The \textit{exploration--separated} algorithms solve the exploration--exploitation trade--off by introducing a strict separation between the exploration and the exploitation phases. While during the exploration phase all the arms are uniformly selected, in the exploitation phase only the best estimated arm is selected until the end of the experiment. The length $\tau$ of the exploration phase is critical to guarantee the success of the experiment and it is possible to show that if properly tuned, the worst--case cumulative regret scales as $O(T^{2/3})$. Another class of algorithms interleave exploration and exploitation and rely on the construction of confidence intervals for the reward of each arm. In particular, the upper--confidence bound (UCB) algorithm~\cite{auer2002finite--time} gives an extra exploration bonus to arms which have been selected only few times in the past and it achieves a worst--case cumulative regret of order $O(T^{1/2})$. Although this represents a clear improvement over the exploration--separated algorithms, as reviewed in the introduction, in some web advertising applications considered in this paper, it is not possible to preserve incentive compatibility when exploration and exploitation are interleaved over time.

%% file: sec/03model.tex

\section{Problem statement}\label{s:statement}

In this section we introduce all the notation used throughout the rest of the paper. In particular, we formalize the sponsored search auction model, we define the mechanism design problem, and we introduce the learning process.

\subsection{Sponsored search auction model}

We resort to the standard model of sponsored search auctions~\cite{Narahari2009}. We denote by $\N=\{1,\ldots,N\}$ the set of ads indexes and by $a_i$ with $i \in \N$ the $i$--th ad (we assume w.l.o.g. each advertiser has only one ad and therefore we can identify by $a_i$ the $i$--th ad and the $i$--th advertiser indifferently).  Each ad~$a_i$ is characterized by a \emph{quality} $q_i$ corresponding to the probability that $a_i$ is clicked once observed by the user, and by a \emph{value} $v_i\in\mathcal V$, with $\mathcal{V}=[0,V]$ and $V\in \mathbb{R}^+$, which $a_i$ receives when clicked ($a_i$ receives a value of zero if not clicked). We denote by $\mathbf{v}$ the profile $(v_1,\ldots,v_N)$ and, as customary in game theory, by $\mathbf{v}_{-i}$ the profile obtained by removing $v_i$ from $\mathbf{v}$. While qualities $\{q_i\}_{i \in \N}$ are commonly known by the auctioneer, values $\{v_i\}_{i \in \N}$ are private information of the advertisers. We denote by $\K=\{1,\ldots,K\}$ with $K < N$,\footnote{Although $K<N$ is the most common case, the results could be smoothly extended to $K>N$.} the set of slot indexes and by $s_m$ with $m\in \K$ the $m$--th slot from top to bottom. For notational convenience, we also define the extended set of slots indexes $\mathcal{K}'=\mathcal{K}\cup\{K+1,\ldots,N\}$.

We denote by the ordered pair $\langle s_m, a_i\rangle$ that ad $a_i$ is allocated into slot $s_m$, by $\theta$ a generic \emph{allocation} and by $\Theta$ the set of all the possible allocations. Although in an auction only $K$ ads can be actually displayed, we define an allocation as $\theta=\{\langle m,i\rangle: m\in \mathcal{K}',i \in \N\}$ where both $m$ and $i$ occur exactly once and any ad assigned to a slot $m>K$ is not displayed.  We define two maps $\pi:\N\times\Theta \rightarrow \K'$ and $\alpha:\K'\times\Theta  \rightarrow \N$ such that $\pi(i;\theta)$ returns the slot in which $a_i$ is displayed in allocation $\theta$ and $\alpha(m;\theta)$ returns the ad allocated in slot~$s_m$ in allocation $\theta$. Given $\theta \in \Theta$, we have that $\pi(i;\theta)=m$ if and only if $\alpha(m;\theta)=i$.

With more than one slot, it is necessary to adopt a model of the user describing how the expected value of an advertiser varies over the slots. We assume that the user behaves according to the popular \emph{cascade model} defined by~\cite{Kempe2008,Aggarwal2008}. In particular, the user's behavior can be modeled as a Markov chain whose states correspond to the slots, which are observed sequentially from the top to the bottom, and the transition probability corresponds to the probability of observing the ad $a_i$ displayed in the next slot; with the remaining probability the user stops observing the ads. This probability may depend on the index of the slot (i.e., $\pi(i;\theta)$), in this case the externalities are said \emph{position--dependent}, and/or on the ad that precedes $a_i$ in the current allocation $\theta$ (i.e., $\alpha(\pi(i;\theta)-1;\theta)$), in this case the externalities are said \emph{ad--dependent}.

In the general case, the cascade model can be described by introducing parameters $\gamma_{m,i}$ defined as the probability that a user observing ad~$a_i$ in slot~$s_{m}$ observes the ad in the next slot $s_{m+1}$.  It can be easily seen that there are $KN$ different parameters $\gamma_{m,i}$. The (cumulative) probability that a user observes the ad displayed at slot $s_m$ in allocation $\theta$ is denoted by $\Gamma_m(\theta)$ and it is defined as:
\begin{align} \label{eq:coeff2}
\eps_m(\theta) = \left\{
  \begin{array}{ll}
    1 & \text{if } m=1 \\
    \prod\limits_{l=1}^{m-1} \gamma_{l,\alpha(l;\theta)} & \text{if } 2 \leq m\leq K\\
    0 & \text{otherwise}
  \end{array} \right.
\end{align}
Given an allocation $\theta$, the \emph{click through rate} (CTR) of ad $a_i$ is the probability to be clicked once allocated according to $\theta$ and it is equal to $\eps_{\pi(i;\theta)}(\theta) q_{i}$. Similarly, the CTR of the ad displayed at slot $m$ can be computed as $\eps_m(\theta) q_{\alpha(m;\theta)}$. We notice that, according to this model, the user might click multiple ads at each impression. Given an allocation $\theta$, the \emph{expected value} (w.r.t. the user's clicks) of advertiser $a_i$ from $\theta$ is $\eps_{\pi(i;\theta)}(\theta) q_{i} v_i$, that is, the product of the CTR $\eps_{\pi(i;\theta)}(\theta) q_{i}$ by the value of the advertiser $v_i$. The advertisers' cumulative expected value from allocation $\theta$, commonly referred to as \emph{social welfare}, is:
\begin{align*}
\SW(\theta,\mathbf{v})= \sum_{i=1}^N \eps_{\pi(i;\theta)}(\theta) q_{i} v_i
\end{align*}
In~\cite{Kempe2008,Aggarwal2008}, the authors factorize the probability $\gamma_{m,i}$ as the product of two independent terms: the \emph{prominence} $\lambda_m$, which only depends on the slot $s_m$, and the \emph{continuation probability} $c_i$, which only depends on the ad $a_i$. This leads to a reduction of the number of the parameters from $KN$ to $K+N$.\footnote{The allocation problem when either all the prominence probabilities $\lambda_m$s or all the continuation probabilities $c_i$s are equal to one can be solved in polynomial time, while, although no formal proof is known, the allocation problem with $\lambda_m$s and $c_i$s different from one is commonly believed to be $\mathcal{NP}$--hard~\cite{Kempe2008}. However, the allocation problem can be solved exactly for concrete settings and for very large settings approximation algorithms can be adopted as shown in~\cite{aamas2013}. In this paper, we just focus on optimal allocation functions.}

Finally, we denote by $\cl^i_{m}\in \{0,1\}$ the click/no--click event for ad $a_i$ allocated in slot $m$.

\subsection{Mechanism design problem} \label{ssec:md}

A direct--revelation economic mechanism for sponsored search auctions is formally defined as a tuple $(\N,\mathcal V,\Theta,f,\{p_i\}_{i\in \N})$ where $\N$ is the set of agents (i.e., the advertisers), $\mathcal V$ is the set of possible actions available to the agents (i.e., the possible reported values), $\Theta$ is the set of the outcomes (i.e., the allocations), $f$ is the allocation function $f:\mathcal V^{N}\rightarrow \Theta$, and $p_i$ is the payment function of advertiser $a_i$ defied as $p_i:\mathcal V^{N}\rightarrow \mathbb R$. We denote by $\hat{v}_i$ the value reported by advertiser $a_i$ to the mechanism, by $\hat{\mathbf{v}}$ the profile of reported values and $\hat{\mathbf{v}}_{-i}$ the profile obtained by removing $\hat{v}_i$ from $\hat{\mathbf{v}}$. 

At the beginning of an auction, each advertiser $a_i$ reports its value $\hat{v}_i$. The mechanism chooses the allocation on the basis of the advertisers' reports as $f(\hat{\mathbf{v}})$ and subsequently computes the payment of each advertiser $a_i$ as $p_i(\hat{\mathbf{v}})$. The expected utility of advertiser $a_i$ is defined as $\eps_{\pi(i;f(\hat{\mathbf{v}}))}f(\hat{\mathbf{v}}) q_{i} v_i - p_i(\hat{\mathbf{v}})$. Since each advertiser is an expected utility maximizer, it will misreport its value (i.e., $\hat v_i \neq v_i$) whenever this may lead its utility to increase. Mechanism design aims at finding an allocation function $f$ and a vector of payments $\{p_i\}_{i \in \N}$ such that some desirable properties---discussed in Section~2.1---are satisfied~\cite{mas-colell1995microeconomic}.

When all the parameters $q_i$ and $\gamma_{m,i}$ are known, the VCG mechanism satisfies IC in expectation (over click realizations), IR in expectation (over click realizations), WBB \emph{a posteriori} (w.r.t. click realizations), and AE. In the VCG mechanism, the allocation function, denoted by $f^*$, maximizes the social welfare given the reported types as:
\begin{align}
\label{eq:efficient-alloc}
\theta^*=f^*(\hat{\mathbf{v}}) \in \arg\max_{\theta \in \Theta}~ \{\SW(\theta,\hat{\mathbf{v}})\}
\end{align}
\noindent and payments are defined as
\begin{align}\label{eq:pay.opt.vcg}
p^*_i(\hat{\mathbf{v}}) =  \SW(\theta^*_{-i},\hat{\mathbf{v}}_{-i}) - \SW_{-i}(\theta^*,\hat{\mathbf{v}}),
\end{align}
where:
\begin{itemize}
\item $\theta^*_{-i}=f^*(\hat{\mathbf{v}}_{-i})$, i.e., the optimal allocation when advertiser $a_i$ is not present,
\item $\SW_{-i}(\theta^*,\hat{\mathbf{v}})=\sum\limits_{j=1,j\neq i}^N \eps_{\pi(j;\theta^*)}(\theta^*) q_{j} \hat{v}_j$, i.e., the cumulative expected value of the optimal allocation $\theta^*$ minus the expected value of advertiser $a_i$.
\end{itemize} 
In words, the payment of advertiser $a_i$ is the difference between the social welfare that could be obtained from allocation $\theta_{-i}^*$ computed removing ad $a_i$ from the auction and the social welfare of the efficient allocation $\theta^*$ without the contribution of advertiser $a_i$. The extension of the VCG mechanism do weighted ads (the WVCG mechanism) is straightforward. The weighted social welfare is $\SW^w(\theta,\mathbf{v})= \sum_{i=1}^N \eps_{\pi(i;\theta)}(\theta) q_{i} v_i w_i$ where $w_i$ is the weight of advertiser $i$. In the WVCG, the allocation maximizing the weighted social welfare is chosen, while the payment is defined as $ p^w_i(\hat{\mathbf{v}}) =  \frac{1}{w_i}(\SW^w(\theta^*_{-i},\hat{\mathbf{v}}_{-i}) - \SW^w_{-i}(\theta^*,\hat{\mathbf{v}}))$.

The previous mechanism is IC and IR in expectation, but it is not DSIC and IR \emph{a posteriori} w.r.t. the clicks (an advertiser may have a positive payment even when its ad has not been clicked). Nonetheless, the mechanism can be easily modified to satisfy DSIC and IR \emph{a posteriori} w.r.t. the clicks by using \emph{pay--per--click} payments $p^{*,c}_i$ as follows:
\begin{align}\label{eq:pay.opt.click.vcg}
p^{*,c}_i(\hat{\mathbf{v}},\cl^i_{\pi(i; \theta^*)}) =  \frac{\SW(\theta^*_{-i},\hat{\mathbf{v}}_{-i}) - \SW_{-i}(\theta^*,\hat{\mathbf{v}})}{\eps_{\pi(i;\theta^*)}(\theta^*) q_i}\ind\{\cl^i_{\pi(i; \theta^*)}\},\end{align}
where $\ind\{\cdot\}$ denotes the indicator function. The contingent formulation of the payments is such that $\mathbb E[ p^c_i(\hat{\mathbf{v}},\cl^i_{\pi(i; \theta^*)})] = p^*_i(\hat{\mathbf{v}})$, where the expectation is w.r.t. the click event, which is distributed as a Bernoulli random variable with parameter coinciding with the CTR of ad $a_i$ in allocation $\theta^*$, i.e., $\eps_{\pi(i; \theta^*)}q_i$. Similar definitions hold for the WVCG.

\subsection{Online learning mechanism design problem}\label{ss:online.mechanism}

In many practical problems, the parameters (i.e., $q_i$ and $\gamma_{m,i}$) are not known in advance by the auctioneer and must be estimated at the same time as the auction is deployed. This introduces a tradeoff between \textit{exploring} different possible allocations so as to collect information about the parameters and \textit{exploiting} the estimated parameters so as to implement a truthful high--revenue auction (i.e., a VCG mechanism). This problem could be easily casted as a multi--arm bandit problem \cite{robbins1952some} and standard techniques could be used to solve it, e.g., \cite{auer2002finite-time}. Nonetheless, such an approach would completely overlook the strategic dimension of the problem: advertisers may choose their reported values at each round $t$ to influence the outcome of the auction at $t$ and/or in future rounds after $t$ in order to increase the cumulative utility over all the rounds of the horizon $T$. Thus, in this context, truthfulness requires that reporting the truthful valuation maximizes the cumulative utility over all the horizon $T$. The truthfulness can be: in dominant strategies if advertisers know everything (including, e.g., the ads that will be clicked at each round $t$ if displayed) or in expectation. As customary, we adopt three forms of truthfulness in expectation: IC in expectation over the click realizations and \emph{a posteriori} w.r.t. the realizations of the random component of the mechanism (if such a component is present), IC in expectation over the realizations of the random component of the mechanism and \emph{a posteriori} w.r.t. the click realizations, and, finally, IC in expectation over both randomizations. We consider IC in expectation over the click realizations weaker than IC in expectation over the realizations of the random mechanism since each advertiser could control the clicks by using software bots.

Thus, here we face the more challenging problem where the exploration--exploitation dilemma must be solved so as to maximize the revenue of the auction under the hard constraint of incentive compatibility. Let $\mathfrak{A}$ be an IC mechanism run over $T$ rounds. We assume, as it is common in practice, that the advertisers' reports can change during these $T$ rounds. At each round~$t$, $\mathfrak{A}$ defines an allocation $\theta_t$ and prescribes an expected payment $p_{i,t}(\hbv)$ for each ad $a_i$. The objective of $\mathfrak{A}$ is to obtain a revenue as close as possible to a VCG mechanism computed on the basis of the actual parameters.\footnote{We refer the reader to~\ref{app:deviation.regret} for a slightly different definition of regret measuring the deviation from the revenue of a VGC mechanism.} More precisely, we measure the performance of $\mathfrak{A}$ as its cumulative regret over $T$ rounds:
\begin{align*}
\mathcal R_T(\mathfrak{A}) = T \sum_{i=1}^n p_i^*(\hbv) - \sum_{t=1}^T \sum_{i=1}^n p_{i,t}(\hbv).
\end{align*}
We remark that the regret is not defined on the basis of the pay--per--click payments asked on a specific sequence of clicks but on the expected payments $p_{i,t}(\hbv)$. Furthermore, since the learning mechanism $\mathfrak{A}$ estimates the parameters from the observed (random) clicks, the expected payments $p_{i,t}(\hbv)$ are random as well. Thus, in the following we will study the expected regret:
\begin{align}\label{eq:regret}
R_T(\mathfrak{A}) = \mathbb E[\mathcal R_T(\mathfrak{A})],
\end{align}
where the expectation is taken w.r.t. random sequences of clicks and possibly the randomness of the mechanism.
The mechanism $\mathfrak{A}$ is a \textit{no--regret} mechanism if its per--round regret $R_T(\mathfrak{A})/T$ decreases to 0 as $T$ increases, i.e., $\lim\limits_{T\rightarrow \infty} R_T(\mathfrak{A}) / T = 0$. Another popular definition of performance \cite{gonen2007incentive-compatible,babaioff2008characterizing} is the social welfare regret, denoted by $R_T^{SW}$ and measured as the difference between the (expected) social welfare of the optimal allocation $\theta^*$ and the (expected) social welfare of the best  allocation $\tilde{\theta}$ found with the estimated parameters  (i.e., $\SW(\theta^*,\hbv) - \SW(\tilde{\theta},\hbv)$). We notice that minimizing the social welfare regret does not coincide with minimizing $R_T$. In fact, once the quality estimates are accurate enough, such that $\theta_t$ is equal to $\theta^*$, the social welfare regret drops to zero. On the other hand, since $p_{i,t}(\hbv)$ is defined according to the estimated qualities, $R_T(\mathfrak{A})$ might still be positive even if $\theta_t = \theta^*$. In addition, we believe that in practical applications providing a theoretical bound over the regret of the auctioneer's revenue is more important rather than a bound on the regret of the social welfare.\footnote{However, we show that our bounds over the regret of auctioneer's revenue can be easily extended also to the regret of the social welfare.}

The study of the problem when $K=1$ is well established in the literature. 
More precisely, the properties required to have a DSIC mechanism are studied in~\cite{devanur2009price} and it is shown that any learning algorithm must split the exploration and the exploitation in two separate phases in order to incentivize the advertisers to report their true values.
This condition has a strong impact on the regret $R_T(\mathfrak{A})$ of the mechanism. In fact, while in a standard bandit problem the distribution--free regret is of order $\Omega(T^{1/2})$, in single--slot auctions, DSIC mechanisms cannot achieve a regret smaller than $\Omega(T^{2/3})$. In~\cite{devanur2009price} a truthful learning mechanism is designed with a nearly optimal regret of order $\tilde O(T^{2/3})$.\footnote{The $\tilde O$ notation hides both constant and logarithmic factors, that is $R_T \leq \tilde O(T^{2/3})$ if there exist $a$ and $b$ such that $R_T \leq a T^{2/3} \log^b T$.} Similar structural properties for DSIC mechanisms are also studied in~\cite{babaioff2008characterizing} and similar lower--bounds are derived for the social welfare regret. 
The authors show in~\cite{babaioff_impl_pay} that, by introducing a random component in the allocation function and resorting to truthfulness in expectation over the realizations of the random component of the mechanism, the separation of exploration  and exploitation phases can be avoided. In this case, the upper bound over the regret over the social welfare is $O(T^{1/2})$ matching the best bound of standard distribution--free bandit problems. However, the payments of this mechanism suffer of potentially high variance. Although it is expected that with this mechanism also the regret over the auctioneer revenue is of the order of $O(T^{1/2})$, no formal proof is known.

On the other hand, the study of the problem when $K>1$ is still mostly open. In this case, a crucial role is played by the CTR model. While with only one slot, the advertisers' CTRs coincide to their qualities $q_i$, with multiple slots the CTRs may also depend on the slots and the allocation of the other ads. The only results on learning mechanisms for sponsored search auction with $K>1$ are described in~\cite{sarma2010multi-armed}, where the authors characterize DSIC mechanisms and provide theoretical bounds over the social welfare regret. More precisely, the authors assume a simple CTR model in which the CTR itself depends on the ad $i$ and the slot $m$. This model differs from the cascade model (see Section~2.1) where the CTR is a more complex function of the quality $q_i$ of an ad and the cumulative probability of observation $\Gamma_m(\theta)$ which, in general, depends on both the slot $m$ and the full allocation $\theta$ (i.e., the ads allocated before slot $s_m$). It can be easily shown that the model studied in~\cite{sarma2010multi-armed} does not include and, at the same time, is not included by the cascade model. However, the two models correspond when the CTRs are separable in two terms in which the first is the agents' quality and the second is a parameter in $[0,1]$ monotonically decreasing in the slots (i.e., only--position--dependent cascade model). Furthermore, while the cascade model is supported by an empirical activity which confirms its validity as a model of the user behavior~\cite{Craswell,Joachims}, the model considered in~\cite{sarma2010multi-armed} has not been empirically studied. In~\cite{sarma2010multi-armed}, the authors show that when the CTRs are unrestricted (e.g., they are not strictly monotonically decreasing in the slots), then the regret over the social welfare is $\Theta(T)$ and therefore at every round (of repetition of the auction) a non--zero regret is accumulated. In addition, the authors provide necessary and, in some situations, sufficient conditions to have DSIC in restricted environments (i.e., higher slot higher click probability, separable CTRs in which only ads qualities need to be estimated), without presenting any bound over the regret (except for reporting an experimental evidence that the regret is $\Omega(T^{2/3})$ when the CTRs are separable). 

We summarize in Tab.~\ref{tab::results} the known results from the literature and, in bold font, the original results provided in this paper.

\begin{table}[h]
\begin{scriptsize}
\begin{center}
\begin{tabular}{|c|c|c|c|c|c|}
\hline
slots			&	CTR model						&		unknown 				&	solution 			& 	regret over 			&	regret over			 	\\ 
			&									&		parameters			&	concept			& 	social welfare			&	auctioneer revenue 			\\ \hline \hline
1			&	--								&		$q_i$				&	DSIC			&	$\Theta(T^{2/3})$		&	$\Theta(T^{2/3})$			\\ \cline{4-6}
			&									&							&	IC in exp.			&	$O(T^{1/2})$			&	$O(T^{2/3})$					\\ \hline
$>1$			&	(unconstrained) $CTR_{i,m}$			&		$CTR_{i,m}$			&	DISC			&	$\Theta(T)$			&	unknown					\\ \cline{2-6}
			&	(unfactorized) 	cascade				&		$q_i$				&	DISC			&	$\mathbf{O(T^{2/3})}$	&	$\mathbf{\Theta(T^{2/3})}$	\\ \cline{3-6}
			&									&		$\gamma_{i,s}$		&	DISC			&	$\mathbf{\Theta(T)}$		&	$\mathbf{\Theta(T)}$			\\ \cline{2-6}
			& 	position--dep. cascade / 				&		$\lambda_m$			&	DSIC			&	$\mathbf{\Theta(T)}$		&	$\mathbf{\Theta(T)}$			\\ \cline{4-6}
			&	separable $CTR_{i,m}$				&							&	IC in exp.	 		&	$\mathbf{0}$ 			&	$\mathbf{0}$				\\ 
			&									&							&	(w.r.t. clicks) 		&			 			&							\\ \cline{4-6}
			&									&							&	IC in exp. 			&	$O(1)$	 			&	$O(1)$				\\ 
			&									&							&	(w.r.t. mechanism)	&			 			&							\\ \cline{3-6}
			&									&		$q_i$, $\lambda_m$		&	DSIC			&	$\mathbf{\Theta(T)}$		&	$\mathbf{\Theta(T)}$			\\ \cline{4-6}
			&									&							&	IC in exp.			&	$\mathbf{O(T^{2/3})}$	&	$\mathbf{O(T^{2/3})}$		\\ \cline{2-6}
			& 	ad--dependent cascade 	 			&		$c_i$				&	DSIC			&	$\mathbf{\Theta(T)}$		&	$\mathbf{\Theta(T)}$			\\ \cline{3-6}
			&									&		$q_i$, $c_i$			&	DSIC			&	$\mathbf{\Theta(T)}$		&	$\mathbf{\Theta(T)}$			\\ \hline
\end{tabular}																																	
\end{center}
\end{scriptsize}	
\caption{Known results on regret bounds for sponsored search auction. We remark with bold font the results provided in this paper.}		
\label{tab::results}
\end{table}

%% file: sec/04posdep.tex
\section{Learning with Position--Dependent Externalities}\label{s:constant}

In this section we study the multi--slot auctions with only position--dependent cascade model. The CTRs depend only on the quality of the ads and on the position of the slots in which the ads are allocated. Formally, parameters $\gamma_{m,i}$ are such that they coincide with the prominence parameter (i.e., $\gamma_{m,i}=\lambda_m$ for every $m$ and $i$). As a result, the cumulative probability of observation, defined in~(\ref{eq:coeff2}), reduces to
\begin{align} \label{eq:coeff}
\Lambda_m = \eps_m(\theta) = \left\{
  \begin{array}{ll}
    1 & \text{if } m=1 \\
    \prod\limits_{l=1}^{m-1} \lambda_{l} & \text{if } 2 \leq m\leq K\\
    0 & \text{otherwise}
  \end{array} \right.,
\end{align}
where we use $\Lambda_m$ instead of $\Gamma_m(\theta)$ for consistency with most of the literature on position--dependent externalities and to stress the difference with respect to the general case.

When all the parameters are known by the auctioneer, the efficient allocation $\theta^*$ prescribes that the ads are allocated to the slots in decreasing order w.r.t. their expected reported value $q_i \hv_i$. More precisely, for any $k\in \mathcal{K}'$, let $\max\limits_{i \in \N} (q_i \hv_i; k)$ be the operator returning the $k$--th largest value in the set $\{q_1 \hv_1, \ldots, q_N \hv_N\}$, then $\theta^*$ is such that, for every $m\in \mathcal{K}'$, the ad displayed at slot $m$ is
\begin{align}\label{eq:pos.dep.efficient.alloc}
\alpha(m; \theta^*) \in \arg\max\limits_{i \in \N} (q_i \hv_i; m).
\end{align}
This condition also simplifies the definition of the efficient allocation $\theta^*_{-i}$ when $a_i$ is removed from $\N$. In fact, for any $i,j\in \N$, if $\pi(j; \theta^*)<\pi(i; \theta^*)$ (i.e., ad $a_j$ is displayed before $a_i$) then $\pi(j; \theta^*_{-i}) = \pi(j;\theta^*)$, while if $\pi(j; \theta^*)>\pi(i; \theta^*)$ then $\pi(j; \theta^*_{-i}) = \pi(j; \theta^*)-1$ (i.e., ad $j$ is moved one slot upward), and $\pi(i; \theta^*_{-i}) = N$. By recalling the definition of VCG payments $p^*_i$ in (\ref{eq:pay.opt.vcg}), in case of position--dependent externalities we obtain the simplified formulation:
\begin{align}\label{eq:pay.opt.click.vcg.posdep.ad}
p^*_i(\hbv) = \begin{cases}\sum\limits_{l=\pi(i; \theta^*)+1}^{K+1} \left[(\Lambda_{l-1} - \Lambda_l) \max\limits_{j\in \mathcal{N}}(q_j \hv_j; l)\right] & \text{if } \pi(i; \theta^*)\leq K\\
0 & \text{otherwise}\end{cases},
\end{align}
which can be easily written as a per--slot payment as:
\begin{align}\label{eq:pay.slot.vgc.posdep.slot}
p^*_{\alpha(m; \theta^*)}(\hat v) = \begin{cases}\sum\limits_{l=m+1}^{K+1} \left[(\Lambda_{l-1} - \Lambda_l) \max\limits_{i\in \mathcal{N}}(q_i \hv_i; l)\right] & \text{if } m\leq K\\
0 & \text{otherwise}\end{cases}.
\end{align}

In the following sections we study the problem of designing incentive compatible mechanisms  under different conditions of lack of information over the parameters $\{q_i\}_{i \in \N}$ and $\{\Lambda_m\}_{m \in \K}$. In particular, in Section~\ref{ssec:uq}, we assume that the actual values of $\{q_i\}_{i \in \N}$ are unknown by the auctioneer, while those of $\{\Lambda_m\}_{m \in \K}$ are known. In Section~\ref{ssec:ul}, we assume that  the actual values of  $\{\Lambda_m\}_{m \in \K}$ are unknown by the auctioneer, while those of $q_i$s are known. Finally, in Section~\ref{ssec:uql}, we assume that the actual values of both $\{q_i\}_{i \in \N}$ and $\{\Lambda_m\}_{m \in \K}$ are unknown.


\subsection{Unknown qualities $\{q_i\}_{i \in \N}$} \label{ssec:uq}

In this section we assume that the qualities of the ads ($\{q_i\}_{i \in \N}$) are unknown, while $\{\Lambda_m\}_{m \in \K}$ are known. We initially focus on DSIC mechanisms and subsequently we discuss about mechanisms IC in expectation.

\begin{figure}[t]
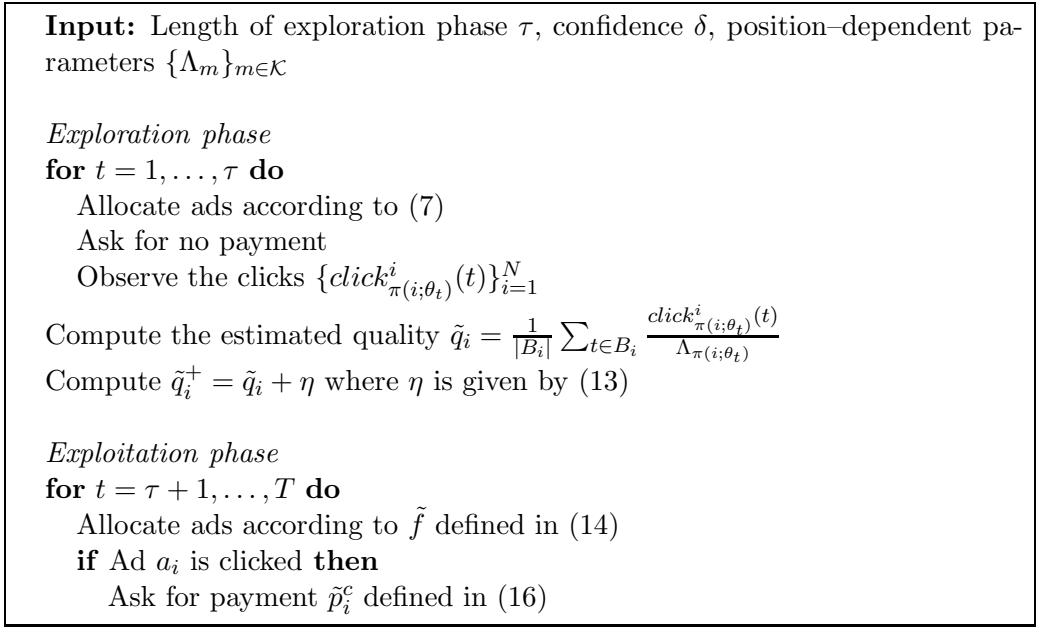

\bookbox{
\begin{algorithmic}
\STATE \textbf{Input:} Length of exploration phase $\tau$, confidence $\delta$, position--dependent parameters $\{\Lambda_m\}_{m \in \K}$
\STATE
\STATE \textit{Exploration phase}
\FOR{$t = 1,\ldots,\tau$}
\STATE Allocate ads according to (\ref{eq:pos.dep.efficient.alloc})
\STATE Ask for no payment
\STATE Observe the clicks $\{\cl_{\pi(i;\theta_t)}^i(t)\}_{i=1}^{N}$
\ENDFOR
\STATE Compute the estimated quality $\tilde{q}_i = \frac{1}{|B_i|}\sum_{t \in B_i} \frac{\cl_{\pi(i; \theta_t)}^i(t)}{\Lambda_{\pi(i; \theta_t)}}$
\STATE Compute $\tilde{q}^+_i = \hq_i + \eta$ where $\eta$ is given by (\ref{eq:eta})
\STATE
\STATE \textit{Exploitation phase}
\FOR{$t = \tau+1,\ldots, T$}
\STATE Allocate ads according to $\tilde{f}$ defined in~(\ref{eq:optimalallocationestimatedq})
\IF{Ad $a_i$ is clicked}
\STATE Ask for payment $\tilde p^c_i$ defined in (\ref{eq:hpay.const.ppc})
\ENDIF
\ENDFOR
\end{algorithmic}}
\caption{Pseudo--code for the A--VCG1 mechanism.}\label{f:alg}
\end{figure}

As in~\cite{devanur2009price,babaioff2008characterizing}, we formalize the problem as a multi--armed bandit problem and we study the properties of a learning mechanism where the exploration and exploitation phases are separated, such that during the exploration phase, we estimate the values of $\{q_i\}_{i \in \N}$ and during the exploitation phase we use the estimated qualities $\{\hq_i\}_{i \in \N}$ to implement an IC mechanism.
The pseudo code of the algorithm A--VCG1 (Adaptive VCG1) is given in Fig.~\ref{f:alg}. The details of the algorithm follow.

\paragraph*{\indent Exploration phase} The exploration phase takes $\tau\geq N/K$ rounds.\footnote{Notice that we need $\tau > N/K$ in order to guarantee that all the ads have at least one sample to initialize the estimates $\hq_i$.} During this phase, the algorithm receives as input the parameters $\{\Lambda_m\}_{m \in \K}$ and collects data to estimate the quality of each ad. Unlike the single--slot case, where we collect only one sample of click or no--click events per round, here we can exploit the fact that each ad $a_i$ has a non--zero CTR whenever it is allocated to a slot $s_m$ with $m\leq K$. As a result, at each round of the exploration phase, we collect $K$ samples (click or no--click events), one from each slot. Let $\theta_t$ (for $t \leq \tau$) be a sequence of (potentially arbitrary) allocations independent from the advertisers' bids. The set $B_i = \{t: \pi(i; \theta_t) \leq K, t\leq\tau\}$ contains all the time instants when ad $a_i$ is allocated to a valid slot, so that $|B_i|$ corresponds to the total number of (click/no--click) samples available for ad $a_i$. We denote by $\cl_{\pi(i; \theta_t)}^i(t)\in\{0,1\}$ the click event at time $t$ for ad $a_i$ when displayed at slot $\pi(i; \theta_t)$. Depending on the slot in which the click event happens, the ad $a_i$ has different CTRs, thus we weigh each click sample by the probability of observation $\Lambda_m$ related to the slot in which the ad was allocated. The estimated quality $\hq_i$ is computed as
\begin{align}\label{eq:est.q}
\tilde{q}_i = \frac{1}{|B_i|}\sum_{t \in B_i} \frac{\cl_{\pi(i; \theta_t)}^i(t)}{\Lambda_{\pi(i; \theta_t)}},
\end{align}
which is an unbiased estimate of $q_i$ (i.e., $\mathbb{E}_{click} [\hq_i] = q_i$, where $\mathbb{E}_{click}$ is the expectation w.r.t. the realization of the clicks). By applying the Hoeffding's inequality~\cite{hoeffding1963probability}, we obtain a bound over the error of the estimated quality $\hq_i$ for each ad $i$.

\begin{proposition}\label{p:hoeffding}
For any ad $i \in \N$
\begin{align}\label{eq:hoeffproposition1}
| q_i - \tilde{q}_i | \leq \sqrt{\Bigg(\sum_{t \in B_i} \frac{1}{\Lambda_{\pi(i; \theta_t)}^2}\Bigg) \frac{1}{2 |B_i|^2} \log \frac{2N}{\delta}},
\end{align}
with probability $1-\delta$ (w.r.t. the click events).
\end{proposition}

During the exploration phase, at each round $t=1,\ldots,\tau$, we adopt the following sequence of allocations
\begin{equation}\label{eq:explorativeallocations}
\theta_t=\{\langle s_1, a_{(t \text{ mod } N) + 1} \rangle, \ldots, \langle s_{N}, a_{(t+N-1 \text{ mod } N) + 1} \rangle\},
\end{equation}
obtaining $|B_i| = \lfloor K \tau / N \rfloor$ for all the ads $a_i$. Thus, given that $\lfloor K \tau / N \rfloor \geq \frac{\tau K}{2N}$,  Equation~(\ref{eq:hoeffproposition1}) becomes
\begin{align}\label{eq:eta}
| q_i - \tilde{q}_i | \leq \sqrt{\Bigg(\sum_{m=1}^{K} \frac{1}{\Lambda_{m}^2}\Bigg) \frac{2 N}{K^2 \tau} \log \frac{2 N}{\delta}} =: \eta.
\end{align}
During this phase, in order to guarantee DSIC, the advertisers cannot be charged with any payment, i.e. all the payments in rounds $t\leq \tau$ are set to 0. In fact, as shown in~\cite{babaioff2008characterizing}, any bid--dependent payment could be easily manipulated by bidders with better estimates of the CTRs, thus obtaining a non--truthful mechanism, whereas non--bid--dependent payments could make the mechanism not to be IR and thus bidders may prefer not to participate to the mechanism.

\paragraph*{\indent Exploitation phase} Once the exploration phase is concluded, an upper--confidence bound over each quality is computed as $\hqp_i = \hq_i + \eta$ and the exploration phase is started and run for the remaining $T-\tau$ rounds. We define the \emph{estimated social welfare} as:
\begin{align*}
\widetilde{\SW}(\theta,\hbv)= \sum_{i=1}^N \Lambda_{\pi(i;\theta)} \hqp_i \hv_i
\end{align*}
and we define $\tilde{f}$ as the allocation function that displays ads in decreasing order of $\hqp_i \hv_i$. $\tf$ returns the efficient allocation $\tilde{\theta}$ on the basis of the estimated qualities as:
\begin{align}\label{eq:optimalallocationestimatedq} 
\tilde{\theta}=\tilde{f}(\hbv) \in \arg~\max\limits_{\theta \in \Theta}~ \{\widetilde{\SW}(\theta,\hbv)\}
\end{align}
Our mechanism adopts $\tilde{f}$ during all the steps of the exploitation phase. Notice that $\tilde{f}$ is an affine maximizer, given that 
\[
\tilde{f}(\hbv) \in \arg\max\limits_{\theta \in \Theta} \sum_{i=1}^N \Lambda_{\pi(i; \theta)} \hqp_i \hv_i = \arg\max\limits_{\theta \in \Theta} \sum_{i=1}^N \frac{\hqp_i }{q_i}  \Lambda_{\pi(i; \theta)} q_i \hv_i = \arg\max\limits_{\theta \in \Theta} \sum_{i=1}^N w_i  \Lambda_{\pi(i; \theta)} q_i \hv_i 
\]
where each weight $w_i= \frac{\hqp_i }{q_i} $ is independent of the advertisers' types $v_i$. Hence, we can apply the WVCG (weighted--VCG) payments (here denoted by $\tilde{p}$ because based on estimated parameters) satisfying the DSIC property. In particular, for any $i$, such that $\pi(i; \tilde\theta) \leq K$, we define the payment
\begin{align}\label{eq:hpay.const}
\tilde p_i (\hbv) &= \frac{1}{w_i} \sum_{l=\pi(i; \tilde{\theta})+1}^{K+1} (\Lambda_{l-1} - \Lambda_l) \max\limits_{j \in \N}(\hqp_j \hv_j; l) \nonumber\\
&=\frac{q_i}{\hqp_i} \sum_{l=\pi(i;\tilde{\theta})+1}^{K+1} (\Lambda_{l-1} - \Lambda_l) \max\limits_{j \in \N}(\hqp_j \hv_j; l).
\end{align}
These payments cannot be computed by the auctioneer, since the actual $\{q_i\}_{i \in \N}$ are unknown. However, we can resort to the \emph{pay--per--click} payments
\begin{align}\label{eq:hpay.const.ppc}
\tilde p^c_i (\hbv,\cl^i_{\pi(i; \tilde{\theta})}) = \frac{1}{\Lambda_{\pi(i; \tilde{\theta})}\hqp_i}\bigg(\sum\limits_{l=\pi(i; \tilde{\theta})+1}^{K+1} (\Lambda_{l-1} - \Lambda_l) \max\limits_{j \in \N}(\hqp_j \hv_j; l)\bigg)\ind\{\cl^i_{\pi(i; \tilde{\theta})}\}.
\end{align}
which in expectation coincide with the WVCG  payments $\tilde{p}_i (\hbv) = \E[\tilde p^c_i(\hbv,\cl^i_{\pi(i; \tilde{\theta})})]$. Unlike the payments $\tilde{p}_i (\hbv)$, these payments can be computed simply relying on the estimates $\hqp_i$ and on the knowledge of the probabilities $\Lambda_m$.

We can state the following.
\begin{proposition}
The A--VCG1 is DSIC, IR \emph{a posteriori}, and WBB \emph{a posteriori}.
\end{proposition}

\begin{pf}
It trivially follows from the fact that the mechanism is a WVCG mechanism and that the payments are pay--per--click.\qed
\end{pf}

We now move to the analysis of the performance of A--VCG1 in terms of  regret the mechanism cumulates through $T$ rounds.

\begin{theorem}\label{thm:constant}
Let us consider a sequential auction with $N$ advertisers, $K$ slots, and $T$ rounds with position--dependent cascade model with parameters $\{\Lambda_m\}_{m=1}^K$ and accuracy $\eta$ as defined in~(\ref{eq:eta}). For any parameter $\tau \in \{0, \ldots, T\}$ and $\delta \in [0,1]$, the A--VCG1  achieves a regret:
\begin{align}\label{eq:regret.const.exact}
R_T &\leq \vmax \left( \sum_{m = 1}^K \Lambda_m \right)\Big( 2(T - \tau) \eta + \tau + \delta T \Big).
\end{align}
\noindent By setting the parameters to
\begin{align*}
\delta &= K^{-\frac{1}{3}} T^{-\frac{1}{3}} N^{\frac{1}{3}}\\
\tau &= 2^{\frac{1}{3}} K^{-\frac{1}{3}} T^{\frac{2}{3}} N^{\frac{1}{3}} \Lambda_{\min}^{-\frac{2}{3}} \left[ \log \left( K^\frac{1}{3} T^\frac{1}{3} N^\frac{2}{3} \right) \right]^{\frac{1}{3}},
\end{align*}
where $\displaystyle \Lambda_{\min} = \min_{m \in \K} \Lambda_m, \ \Lambda_{\min} > 0$, then the regret is
\begin{align}\label{eq:regret.const}
R_T \leq 4 \cdot 2^\frac{1}{3} \vmax \Lambda_{\min}^{-\frac{2}{3}} K^\frac{2}{3} T^\frac{2}{3} N^\frac{1}{3} \left[ \log \big( K^\frac{1}{3} T^\frac{1}{3} N^\frac{2}{3} \big)\right]^{\frac{1}{3}}
\end{align}
\end{theorem}

We initially introduce some remarks about the above results, and subsequently discuss the sketch of the proof of the theorem.

\myremark{1 (The bound).}  Up to numerical constants and logarithmic factors, the previous bound~(\ref{eq:regret.const}) is $R_T \leq \tilde O(T^\frac{2}{3} K^\frac{2}{3} N^\frac{1}{3})$.
We first notice that A--VCG1 is a no--regret algorithm since its per--round regret ($R_T/T$) decreases to 0 as $T^{-\frac{1}{3}}$, thus implying that it asymptotically achieves the same performance as the VCG. Furthermore, we notice that for $K=1$ the bound reduces (up to constants) to the single--slot case analyzed in~\cite{devanur2009price}. Unlike the standard bound for multi--armed bandit algorithms, the regret scales as $\tilde O(T^\frac{2}{3})$ instead of $\tilde O(T^\frac{1}{2})$. As pointed out in~\cite{devanur2009price} and \cite{babaioff2008characterizing} this is the unavoidable price the bandit algorithm has to pay to be DSIC.
Finally, the dependence of the regret on $N$ is sub--linear ($N^\frac{1}{3}$) and therefore an increase of the number of advertisers does not significantly worsen the regret. The dependency on the number of slots $K$ is similar:  according to the bound (\ref{eq:regret.const}) the regret has a sublinear dependency $\tilde O(K^\frac{2}{3})$, meaning that whenever one slot is added to the auction, the performance of the algorithm does not significantly worsen. By analyzing the difference between the payments of the VCG and A--VCG1, we notice that during the exploration phase the regret is $O(\tau K)$ (e.g., if all the ads allocated into the $K$ slots are clicked at each explorative round), while during the exploitation phase the error in estimating the qualities sum over all the $K$ slots, thus suggesting a linear dependency on $K$ for this phase as well. Nonetheless, as $K$ increases, the number of samples available per ad increases as $\tau K/N$, thus improving the accuracy of the quality estimates by $\tilde O(K^{-\frac{1}{2}})$ (see Proposition~\ref{p:hoeffding}). As a result, as $K$ increases, the exploration phase can be shortened (the optimal $\tau$ actually decreases as $K^{-\frac{1}{3}}$), thus reducing the regret during the exploration, and still have accurate enough estimations to control the regret of the exploitation phase.

\myremark{2 (Distribution--free bound).} The bound derived in Theorem~\ref{thm:constant} is a \textit{distribution--free} (or worst--case) bound, since it holds for any set of advertisers (i.e., for any $\{q_i\}_{i\in\N}$ and $\{v_i\}_{i\in\N}$). This generality comes at the price that, as illustrated in other remarks and in the numerical simulations (see Section~\ref{s:experiments}), the bound could be inaccurate for some specific sets of advertisers. On the other hand, distribution--dependent bounds (see e.g., the bounds of UCB~\cite{auer2002finite-time}), where $q$ and $v$ appear explicitly, would be more accurate in predicting the behavior of the algorithm. Nonetheless, they could not be used to optimize the parameters $\delta$ and $\tau$, since they would then depend on unknown quantities.

\myremark{3 (Parameters).} The choice of parameters $\tau$ and $\delta$ reported in Theorem~\ref{thm:constant} is obtained by rough minimizing the upper--bound (\ref{eq:regret.const.exact}). Each parameter can be computed by knowing the characteristics of the auction (number of rounds $T$, number of slots $K$, number of ads $N$, and $\Lambda_m$). Moreover, since the values are obtained optimizing an upper--bound of the regret and not directly the true global regret, these values can provide a good guess for the parametrization, but there could be other values that better optimize the regret. Thus, in practice, the regret could be optimized by searching the space of the parameters around the values suggested in Theorem~\ref{thm:constant}.

\myremark{4 (IC in expectation).}  Two interesting problems we do not solve in this paper once IC in expectation (over the click realizations and/or realizations of the random component of the mechanism) is adopted are whether or not it is possible to avoid the separation of the exploration and exploitation phases and whether it is possible to obtain a regret of $O(T^{1/2})$ as it is possible in the case of $K=1$~\cite{babaioff_impl_pay}. Any attempt we tried to extend the result presented in~\cite{babaioff_impl_pay} to the multi--slot case conducted us to a non--IC mechanism. We briefly provide some examples of adaptation to our framework of the two MAB presented~\cite{babaioff_impl_pay}. None of these attempts provided a monotone allocation function. We have tried to extend the UCB1 in different ways, e.g. introducing $N \cdot K$ estimators, one for each ad for each slot, or maintaining $N$ estimators weighting in different ways click obtained in different slots. The second MAB algorithm, called NewCB, is based on the definition of a set of active ads, the ones that can be displayed. We have considered extensions with a single set for all the slots and with multiple sets, one for each slot, without identifying monotone allocation algorithms.

\textit{(Comments to the proof).} The proof uses relatively standard arguments to bound the regret of the exploitation phase. As discussed in Remark 2, the bound is distribution--free and some steps in the proof are conservative upper--bounds on quantities that might be smaller for specific auctions. For instance, the inverse dependency on the smallest cumulative discount factor $\Lambda_{\min}$ in the final bound could be a quite inaccurate upper--bound on the quantity $\sum_{m=1}^{K} 1/ \Lambda_{m}^2$. In fact, the parameter $\tau$ itself could be optimized as a direct function of $\sum_{m=1}^{K} 1 /\Lambda_{m}^2$, thus obtaining a more accurate tuning of the length of the exploration phase and a slightly tighter bound (in terms of constant terms). Furthermore, we notice that the step $\max\limits_{i \in \N}(\hqp_{i} v_i;h) / \max\limits_{i \in \N}(\hqp_{i} v_i;m) \leq 1$ is likely to become less accurate as the difference between $h$ and $m$ increases (see Eq.~\ref{eq:step.loose} in the proof). For instance, if the qualities $q_i$ are drawn from a uniform distribution in $(0,1)$, as the number of slots increases this quantity reduces as well (on average) thus making the upper--bound by $1$ less and less accurate. The accuracy of the proof and the corresponding bound are further studied in the simulations in Section~\ref{s:experiments}.

In a similar way, adopting the same mechanism as before, it is also possible to derive an upper--bound over the global regret, when the regret, as in~\cite{babaioff_impl_pay} is computed over the social welfare of the allocation. In particular we obtain, that, even in this case, A--VCG1 is a no--regret algorithm and $R^{SW}_T\leq\tilde{O}(T^\frac{2}{3})$.

\begin{theorem} \label{th:pd_q_sw}
Let us consider a sequential auction with $N$ advertisers, $K$ slots, and $T$ rounds with position--dependent cascade model with parameters $\{\Lambda_m\}_{m=1}^K$ and $\eta$ as defined in~(\ref{eq:eta}). For any parameter $\tau \in \{0, \ldots, T\}$ and $\delta \in [0,1]$, the A--VCG1  achieves a regret:
\begin{align}
R^{SW}_T &\leq \vmax K \left( 2 \left(T - \tau \right) \eta + \tau + \delta T \right).
\end{align}
\noindent By setting the parameters to
\begin{align*}
\delta &= \left( \frac{\sqrt{2}}{\Lambda_{\min}} \right)^\frac{2}{3} K^{-\frac{1}{3}} N^\frac{1}{3} T^{-\frac{1}{3}} \\
\tau &=  \left( \frac{\sqrt{2}}{\Lambda_{\min}} \right)^\frac{2}{3} T^\frac{2}{3} N^\frac{1}{3} K^{-\frac{1}{3}} \left( \log 2^\frac{2}{3} \Lambda_{\min}^\frac{2}{3} N^\frac{2}{3} K^\frac{1}{3} T^\frac{1}{3} \right)^\frac{1}{3},
\end{align*}
where $\displaystyle \Lambda_{\min} = \min_{m \in \K} \Lambda_m, \ \Lambda_{\min} > 0$, then the regret is
\begin{align} 
R_T^{SW} \leq 4 \vmax \left( \frac{\sqrt{2}}{\Lambda_{\min}} \right)^\frac{2}{3} K^\frac{2}{3} N^\frac{1}{3} T^\frac{2}{3} \left( \log 2^\frac{2}{3} \Lambda_{\min}^\frac{2}{3} N^\frac{2}{3} K^\frac{1}{3} T^\frac{1}{3}  \right)^\frac{1}{3}
\end{align}
\end{theorem}

Notice that using $\tau$ and $\delta$ defined in Theorem~\ref{thm:constant}, the bound for $R_T^{SW}$ is $\tilde{O}(T^\frac{2}{3})$, even if the parameters are not optimal for this second framework.


\subsection{Unknown $\{\Lambda_m\}_{m \in \K}$} \label{ssec:ul}

We now focus on the situation when the auctioneer knows $\{q_i\}_{i \in \N}$, while $\{\Lambda_m\}_{m \in \K}$ are unknown. By definition of cascade model, $\{\Lambda_m\}_{m \in \K}$ are strictly non--increasing in $m$. This dramatically simplifies the allocation problem since the optimal allocation can be found without knowing the actual values of $\{\Lambda_m\}_{m \in \K}$. Indeed, allocation $\theta^*$ such that $\alpha(m; \theta^*) \in \arg\max\limits_{i \in \N} (q_i \hv_i; m)$ is optimal for all possible $\{\Lambda_m\}_{m \in \K}$. However, the lack of knowledge about $\{\Lambda_m\}_{m \in \K}$ makes the design of a truthful mechanism not straightforward because they appear in the calculation of the payments. Differently from what we presented in the previous section, here we initially focus on IC in expectation mechanisms, providing two mechanisms (the first is IC in expectation over the click realizations and the second is IC in expectation over the realizations of the random component of the mechanism), and subsequently we produce some considerations about DSIC mechanisms.

\subsubsection{IC in expectation over the click realizations mechanism}

\begin{figure}[t]
\bookbox{
\begin{algorithmic}
\STATE \textbf{Input:} Qualities parameters $\{q_i\}_{i \in \N}$
\STATE
\FOR{$t = 1,\ldots, T$}
\STATE Allocate ads according to $f^*$ as prescribed by~(\ref{eq:pos.dep.efficient.alloc})
\IF{Ad $a_i$ is clicked}
\STATE Ask for payment $p^c_i$ defined in (\ref{eq:olppc})
\ENDIF
\ENDFOR
\end{algorithmic}}
\caption{Pseudo--code for the A--VCG2 mechanism.}\label{f:alg2}
\end{figure}

In this case, we do not need any estimation of the parameters $\{\Lambda_m\}_{m \in \K}$ and therefore we do not resort to the multi--armed bandit framework and the mechanism does not present separate phases.  The pseudo code of the algorithm A--VCG2 (Adaptive VCG2) is given in Fig.~\ref{f:alg2}. On the basis of the above considerations, we can adopt the allocatively efficient allocation function $f^*$ as prescribed by~(\ref{eq:pos.dep.efficient.alloc}) even if the mechanism does not know the actual values of the parameters $\{\Lambda_m\}_{m \in \K}$. Nonetheless, the VCG payments defined in (\ref{eq:pay.opt.click.vcg.posdep.ad}) cannot be computed, since $\{\Lambda_m\}_{m \in \K}$ not being known by the mechanism. However, by resorting to execution--contingent payments (generalizing the pay--per--click approach\footnote{In pay--per--click payments, an advertiser pays only once its ad is clicked; in our execution--contingent payments, an advertiser pays also once the ads of other advertisers are clicked.}), we can impose computable payments that, in expectation, are equal to (\ref{eq:pay.opt.click.vcg.posdep.ad}). More precisely, the contingent payments are computed given the bids $\hbv$ and all click events over the slots and take the form:
\begin{align} \label{eq:olppc}
p_{i}^c&(\hat{\mathbf{v}},\{\cl_{\pi(j; \theta^*)}^j\}_{j=1}^K)  \\
&=\sum\limits_{\pi(i;\theta^*) \leq m \leq K } \cl_m^{\alpha(m;\theta^*)} \cdot  \frac{q_{\alpha(m;\theta^*_{-i})} \cdot \hat{v}_{\alpha(m;\theta^*_{-i})}}{q_{\alpha(m;\theta^*)}} \nonumber\\
&\quad\quad\quad- \sum\limits_{\pi(i;\theta^*) < m \leq K } \cl_m^{\alpha(m;\theta^*)} \cdot  \hat{v}_{\alpha(m;\theta^*)}\nonumber
\end{align}
Notice that the payment $p_{i}^c$ depends not only on the click of ad $a_i$, but also on the clicks of all the ads displayed in the slots below. In expectation, the two terms of $p_i^c$ are:
\begin{align*}
\mathbb{E}_{\cl}\left[\sum\limits_{\pi(i;\theta^*) \leq m \leq K } \cl_m^{\alpha(m;\theta^*)} \cdot  \frac{q_{\alpha(m;\theta^*_{-i})} \cdot \hat{v}_{\alpha(m;\theta^*_{-i})}}{q_{\alpha(m;\theta^*)}}\right] 	& = \sum_{\pi(j;\theta^*) \geq  \pi(i;\theta^*)} \Lambda_{\pi(j;\theta^*_{-i})} q_j \hat{v}_j		\\
\mathbb{E}_{\cl}\left[\sum\limits_{\pi(i;\theta^*) < m \leq K } \cl_m^{\alpha(m;\theta^*)} \cdot  \hat{v}_{\alpha(m;\theta^*)}\right]													& = \sum_{\pi(j;\theta^*) >  \pi(i;\theta^*)} \Lambda_{\pi(j;\theta^*)}  q_j \hat{v}_j
\end{align*}
and therefore, in expectation, the payment equals to (\ref{eq:pay.opt.click.vcg.posdep.ad}). Thus, we can state the following.

\begin{proposition}
The A--VCG2 is IC, IR, WBB in expectation (over click realizations) and AE.
\end{proposition}

\begin{pf}
It trivially follows from the fact that the allocation function is AE and the payments in expectation equal the VCG payments. \qed
\end{pf}

We discuss further properties of the mechanism in what follows.
\begin{proposition}
The A--VCG2 is not DSIC \emph{a posteriori} (w.r.t. click realizations).
\end{proposition}

\begin{pf}
The proof is by counterexample. Consider an environment with 3 ads $\N=\{a_1, a_2, a_3\}$ and 2 slots $S=\{s_1,s_2\}$ s.t. $q_1=0.5$, $v_1=4$, $q_2=1$, $v_2=1$, $q_3=1$, $v_3=0.5$, which correspond to expected values of $2$, $1$, and $0.5$.

The optimal allocation $\theta^*$ consists in allocating $a_1$ in $s_1$ and $a_2$ in $s_2$. Consider a time $t$ when both ad $a_1$ and $a_2$ are clicked, from Eq.~\ref{eq:olppc} we have that the payment of $a_2$ is:
\[ 
p_{2}^c = \frac{1}{q_2}q_3v_3 = 0.5
\]
If ad $a_2$ reports a value $\hv_2=3$, the optimal allocation is now $a_2$ in $s_1$ e $a_1$ in $s_2$. In the case both $a_1$ and $a_2$ are clicked, the payment of $a_2$ is: 
\[
p_{2}^c = \frac{1}{q_2} q_1 v_1 + \frac{1}{q_1} q_3 v_3 - v_1 = 2 + 1 - 4 = -1
\]
Given that, in both cases, the utility is $u_2 = v_2 - p_{2}^c$, reporting a non--truthful value is optimal. Thus, we can conclude that the mechanism is not DSIC.
\end{pf}

\begin{proposition}
The A--VCG2 is IR  \emph{a posteriori} (w.r.t. click realizations). \label{prop:AVGC2IRaposteriori}
\end{proposition}

\begin{pf}
Rename the ads $\{a_1, \ldots, a_N\}$ such that $q_1 v_1 \geq q_2 v_2 \geq \ldots \geq q_N v_N$.  We can write payments~(\ref{eq:olppc}) as:
\[
\tilde p_{i}^c = \sum_{j=i}^K \frac{\cl_{j}^j}{q_j} q_{j+1} v_{j+1} - \sum_{j=i+1}^K \cl_{j}^j v_j
\]
Thus, the utility for advertiser $a_i$ is:
\begin{align*}
u_i &= \cl_{j}^j v_i + \sum_{j=i+1}^K \cl_{j}^j v_j - \sum_{j=i}^K \frac{\cl_{j}^j}{q_j} q_{j+1} v_{j+1}\\
&= \sum_{j=i}^K \cl_{j}^j v_j - \sum_{j=i}^K \frac{\cl_{j}^j}{q_j} q_{j+1} v_{j+1}\\
&= \sum_{j=i}^K \left( \cl_{j}^j v_j - \frac{\cl_{j}^j}{q_j} q_{j+1} v_{j+1} \right) \\
&= \sum_{j=i}^K \cl_{j}^j v_j - \frac{\cl_{j}^j}{q_j} q_{j+1} v_{j+1} \\
&= \sum_{j=i}^K \frac{\cl_{j}^j}{q_j} ( q_j v_j - q_{j+1} v_{j+1}).
\end{align*}
Since $\frac{\cl_{j}^j}{q_j} \geq 0$ by definition and $q_j v_j - q_{j+1} v_{j+1} \geq 0$ because of the chosen ordering of the ads, then the utility is always positive and we can conclude the mechanism is IR \emph{a posteriori}. \qed
\end{pf}

\begin{proposition}
The A--VCG2 is not WBB \emph{a posteriori} (w.r.t. click realizations).
\end{proposition}

\begin{pf}
The proof is by counterexample. Consider an environment with 3 ads $\N=\{a_1, a_2, a_3\}$ and 2 slots $S=\{s_1,s_2\}$ s.t. $q_1=1$, $v_1=2$, $q_2=0.5$, $v_2=1$, $q_3=1$, $v_3=\epsilon$, where $\epsilon > 0$ is a small number.

The optimal allocation $\theta^*$ consists in allocating $a_1$ in $s_1$ e $a_2$ in $s_2$. Consider a time instant $t$ when both ad $a_1$ and $a_2$ are clicked, their payments are:

\[ 
p_{1}^c = \frac{1}{q_1}q_2v_2 + \frac{1}{q_2} q_3 v_3 -  v_2 = 0.5 + 2 \epsilon - 1 = 2 \epsilon - 0.5 < 0
\]
\[
p_{2}^c = \frac{1}{q_2} q_3 v_3 = 2 \epsilon
\]

Thus, $\sum_{i=1}^3 p_{i}^c = 4 \epsilon - 0.5 < 0$, and we can conclude that the mechanism is not WBB \emph{a posteriori}.\qed
\end{pf}

Now we state the following theorem, whose proof is straightforward. 

\begin{theorem}\label{thm:constant.l}
Let us consider an auction with $N$ advertisers, $K$ slots, and $T$ rounds, with position--dependent cascade model with parameters $\{\Lambda_m\}_{m=1}^K$.  The A--VCG2 achieves an expected regret $R_T=0$.
\end{theorem}
An important property of this mechanism is that the expected payments are exactly the VCG payments for the optimal allocation when all the parameters are known. Moreover, the absence of an exploration phase allows us to obtain an instantaneous expected regret of zero and, thus, the cumulative regret over the $T$ rounds of auction $R_T=0$. Similar considerations can be applied to the study of the regret over the social welfare, obtaining the following.

\begin{corollary}
The A--VCG2 has an expected regret over the social welfare  of zero.
\end{corollary}

\subsubsection{IC in expectation over random component realizations mechanism} \label{sssec:l.uc.m}

\begin{figure}[t]
\bookbox{
\begin{algorithmic}
\STATE \textbf{Input:} Length of exploration phase $\tau$, confidence $\delta$
\STATE
\STATE \textit{Exploitation phase}
\FOR{$t = \tau+1,\ldots, T$}
\STATE Allocate ads according to $f^{*'}$ as prescribed by Algorithm~1
\IF{Ad $a_i$ is clicked}
\STATE Ask for payment $p^{B,*,x}_i$ defined in (\ref{eq:pay.opt.babaioff.click})
\ENDIF
\ENDFOR
\end{algorithmic}}
\caption{Pseudo--code for the A--VCG2$^\prime$ mechanism.}\label{f:alg22}
\end{figure}

As for the previous mechanism, here we have only the exploitation phase. Differently from the previous mechanism, the mechanism has a random component as proposed in~\cite{babaioff_impl_pay}. The mechanism, called A--VCG2$^\prime$ is reported in Fig.~\ref{f:alg22}. It is obtained applying the approach described in~\cite{babaioff_impl_pay} to allocation function $f^*$.

Since $f^*$ is monotonic~(see~\ref{ap:monotonicity}) and the problem is with single parameter and linear utilities, payments assuring DSIC can be written as~\cite{tardos_sp}:
\begin{equation} \label{eq:pay.vcg.emp.tardos}
p_i^*(\hat{\bv}) = \Lambda_{\pi(i;f^*(\hbv))} q_i \hv_i - \int_{0}^{\hv_i} \Lambda_{\pi(i;f^*(\hbv_{-i},u))} q_i du,
\end{equation}
which coincide with the VCG payments defined in~\ref{eq:pay.opt.vcg} (hence the use of the same notation $p_i^*$). This is justified by the fact that when a mechanism is AE, IR and WBB the only payments that lead to a DSIC mechanism are the VCG payments with Clacke's pivot~\cite{greenLaffont}, thus (\ref{eq:pay.vcg.emp.tardos}) must coincide.
However, these payments are not directly computable, because parameters $\{\Lambda_m\}_{m \in \K}$ in the integral are unknown (and, as in the case discussed in Section~\ref{ssec:ul}.1, we cannot replace them by empirical estimates). We could obtain these payments in expectation by using execution--contingent payments associated with non--optimal allocations where the report $\hat{v}_i$ is modified between 0 and the actual value. This can be obtained by resorting to the approach proposed in~\cite{babaioff_impl_pay}. More precisely, the approach  proposed in~\cite{babaioff_impl_pay} takes in input a generic allocation function $f$ and introduces a randomized component into it, producing a new allocation function that we denote by $f'$. This technique, at the cost of reducing the efficiency of $f$, allows the computation of the allocation and the payments at the same time even when payments described in~\cite{tardos_sp} cannot be computed directly.  

We apply the approach proposed in~\cite{babaioff_impl_pay} to our $f^*$ obtaining a new allocation function $f^{*'}$. With $f^{*'}$, the advertisers' reported values $\{\hv_i\}_{i \in \N}$ are modified, each with a (small) probability $\mu$. The (potentially) modified values are then used to compute the allocation (using $f^*$) and the payments. More precisely, with a probability of $(1-\mu)^N$, $f^{*'}$ returns the same allocation  $f^*$ would return, while it does not with a probability of $1 - (1-\mu)^N$. The reported values $\{\hv_i\}_{i \in \N}$ are modified through the \emph{canonical self--resampling procedure} (cSRP) described in~\cite{babaioff_impl_pay} that generates two samples: $x_i(\hv_i,\omega_i)$ and $y_i(\hv_i,\omega_i)$, where $\omega_i$ is the random seed. We sketch the result of cSRP where the function `rec' is defined in~\cite{babaioff_impl_pay}:
\begin{align*}
(x_i,y_i) = cSRP(\hv_i)=\begin{cases} (\hv_i,\hv_i) & \mbox{w.p. } 1-\mu \\ (\hv''_i,\hv'_i) & \mbox{otherwise }\end{cases},
\end{align*} 
where $\hv_i'\sim\mathcal{U}([0,\hv_i])$ and $\hv_i''=\text{rec}(\hv_i')$.

\begin{algorithm}
\begin{algorithmic}[1]
\begin{scriptsize}
\FORALL {$a_i \in N$}
	\STATE $(x_i, y_i)=cSRP(\hv_i)$ \label{s:csrp}
	\STATE $\mathbf{x}=(x_1,\ldots,x_N)$
\ENDFOR
\STATE $\theta = f^*(\mathbf{x})$ \label{s:alloc}
\end{scriptsize}
\end{algorithmic}
\caption{$f^{*'}(\hbv)$}
\label{alg:babalg}
\end{algorithm}

Algorithm~\ref{alg:babalg} shows how $f^{*'}$ works when the original allocation function is $f^*$.  The reported values $\{\hv_i\}_{i \in \N}$ are perturbed through the canonical self--resampling procedure~(Step~\ref{s:csrp}) and then it returns the allocation found by applying the original allocation function $f^*$ to the new values $\mathbf{x}$ (Step~\ref{s:alloc}).
Finally, the payments are computed as
%
%
\begin{multline} \label{eq:pay.opt.babaioff.click}
p^{B,*,c}_i(\bx, \cl_{\pi(i; f^*(\bx))}^{i}) = \begin{cases} \frac{p_i^{B,*}(\bx,\by;\hbv)}{\Lambda_{\pi(i; f^*(\bx))} q_i} & \mbox{if } \cl_{\pi(i; f^*(\bx))}^{i}=1 \\ 0 & \mbox{otherwise} \end{cases} \\
=  \begin{cases} \hv_i -
\begin{cases}  \frac{1}{\mu} \hv_i &\mbox{if $y_i<\hv_i$}  \\
0 & \mbox{otherwise}, \end{cases} 
 & \mbox{if } \cl_{\pi(i; f^*(\bx))}^{i}=1 \\ 0 & \mbox{otherwise} \end{cases}
\end{multline}
where
\begin{align}
p_i^{B,*}(\bx,\by;\hbv) = \Lambda_{\pi(i; f^*(\bx))}q_i \hv_i - 
\begin{cases}  \frac{1}{\mu}\Lambda_{\pi(i; f^*(\bx))}q_i \hv_i &\mbox{if $y_i<\hv_i$}  \\
0 & \mbox{otherwise}, \end{cases}
\end{align}
$\mathbf{y}=(y_1,\ldots,y_N)$ and the expected value of payments~(\ref{eq:pay.opt.babaioff.click}) w.r.t. the randomization of the mechanism are the payments~\cite{tardos_sp} for the randomized allocation function $f^{*'}$. The result presented in~\cite{babaioff_impl_pay} assures that the resulting mechanism is IC in expectation over the realizations of the random component and \emph{a posteriori} w.r.t. the click realizations.

We state the following results on the properties of the above mechanism.

\begin{theorem}\label{thm:constant.l.baba}
Let us consider an auction with $N$ advertisers, $K$ slots, and $T$ rounds, with position--dependent cascade model with parameters $\{\Lambda_m\}_{m=1}^K$.  The A--VCG2 $^\prime$ achieves an expected regret $R_T \leq 2 K^2 \mu \vmax T$.
\end{theorem}

Adopting $\mu = \frac{1}{T^\alpha}$ with $\alpha>1$ then $R_T \rightarrow 0$, but, as we will show in Section~\ref{s:experiments}, the smaller $\mu$ the larger the variance of the payments. We provide a similar result for the regret over the social welfare.

\begin{theorem}\label{thm:constant.l.sw.baba}
Let us consider an auction with $N$ advertisers, $K$ slots, and $T$ rounds, with position--dependent cascade model with parameters $\{\Lambda_m\}_{m=1}^K$.  The A--VCG2 $^\prime$ achieves an expected regret $R^{SW}_T \leq K^2 \mu \vmax T$.
\end{theorem}

\subsubsection{Considerations about DSIC mechanisms}

At the cost of worsening the regret, one may wonder whether there exists some no--regret DSIC mechanism. In what follows, resorting to the same arguments used in~\cite{sarma2010multi-armed}, we show that the answer to such question is negative.

\begin{theorem}\label{thm:constant.l}
Let us consider an auction with $N$ advertisers, $K$ slots, and $T$ rounds, with position--dependent cascade model with parameters $\{\Lambda_m\}_{m=1}^K$  whose value are unknown. Any online learning DSIC \textit{a posteriori} (w.r.t. click realizations) mechanism achieves an expected regret $R_T=\Theta(T)$.
\end{theorem}

\begin{pf}\textbf{(sketch)} Basically, the A--VCG2 mechanism is only IC in expectation (and not DSIC) because it adopts execution--contingent payments in which the payment of advertiser $a_i$ depends also on the clicks over ads different from $a_i$. The above payment technique---i.e., payments reported in~(\ref{eq:olppc})---is necessary to obtain in expectation the values  $\SW(\theta^*_{-i},\hat{\mathbf{v}}_{-i})$ and $\SW_{-i}(\theta^*,\hat{\mathbf{v}})$,  since parameters $\{\Lambda_m\}_{m \in \K}$ are not known. In order to have DSIC \textit{a posteriori} (i.e., truthful for any realization of the clicks), we need payments $p_i$ that are deterministic w.r.t. the clicks over other ads different from $a_i$ (i.e., pay--per--click payments are needed).

We notice that even if $\Lambda_m$ have been estimated (e.g., in an exploitation phase), we cannot have payments leading to DSIC. Indeed, with estimates $\tilde{\Lambda}_m$, the allocation function maximizing $\widetilde{SW}$ (computed with $\tilde{\Lambda}_m$) is not an affine maximizer and therefore the adoption of WVCG mechanism would not guarantee DSIC. As a result, only mechanisms with payments defined as in~\cite{tardos_sp} can be used. However, these payments, if computed exactly (and not estimated in expectation), require the knowledge about the actual $\Lambda_m$ related to each slot $s_m$ in which an ad can be allocated for each report $\hat{v}\leq v$. 

To prove the theorem, we provide a characterization of DSIC mechanisms. Exactly, we need a monotonic allocation function and the payments defined in~\cite{tardos_sp}. These payments, as said above, require the knowledge about the actual $\Lambda_m$ related to the slot $s_m$ in which an ad can be allocated for each report $\hat{v}\leq v$. Thus we have two possibilities:
\begin{itemize}
\item In the first case, an ad can be allocated only in one slot and its report determines only whether it is displayed or not. That is, the ads are partitioned and each partition is associated with a slot and the ad with the largest expected valuation is chosen at each slot independently. This case is equivalent to multiple separate--single slot auctions and therefore each auction is DSIC as shown in~\cite{devanur2009price}. However, as shown in~\cite{sarma2010multi-armed}, this mechanism would have a regret $\Theta(T)$.
\item In the second case, an ad can be allocated in more than one slot on the basis of its report. In this case, to compute the payments, it would be necessary to know the exact CTRs of the ad for each possible slot, but this is possible only in expectation either by using the above execution--contingent as we do in Section~4.2.1 or by generating non--optimal allocation as we do in Section~4.2.2.
\end{itemize}
Thus, in order to have DSIC, we need to adopt the class of mechanisms described in the first case, obtaining $R_T=\Theta(T)$.\qed
\end{pf}


\subsection{Unknown $\{\Lambda_{m}\}_{m \in \K}$ and $\{q_i\}_{i \in \N}$} \label{ssec:uql}

In this section we study the situation in which both $\{q_i\}_{i \in \N}$ and $\{\Lambda_{m}\}_{m \in \K}$ are unknown. From the results discussed in the previous section, we know that adopting DSIC as solution concept we would obtain $R_T=\Theta(T)$. Thus, we focus only on IC in expectation. 

\begin{figure}[t]
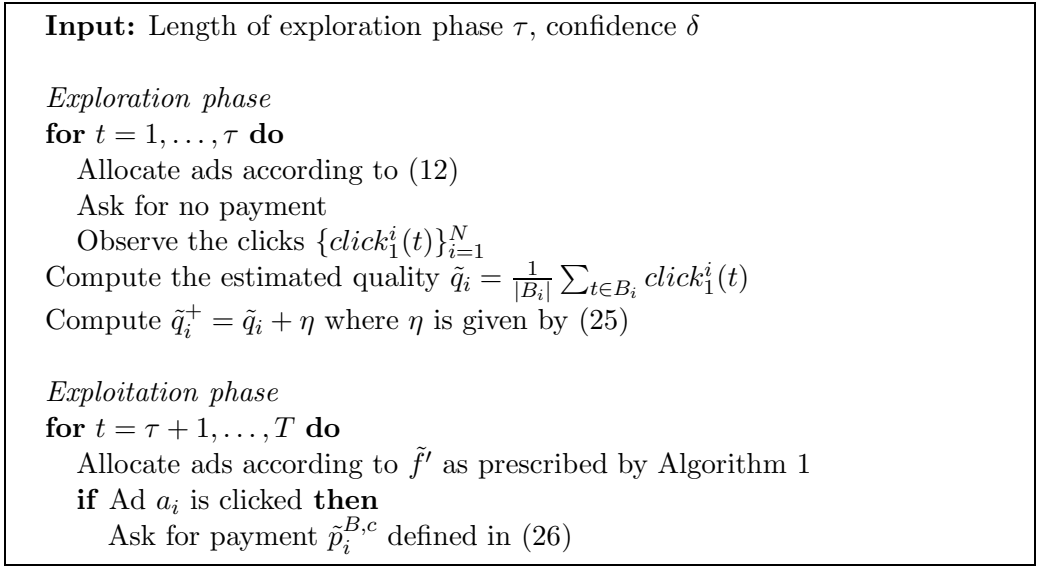

\bookbox{
\begin{algorithmic}
\STATE \textbf{Input:} Length of exploration phase $\tau$, confidence $\delta$
\STATE
\STATE \textit{Exploration phase}
\FOR{$t = 1,\ldots,\tau$}
\STATE Allocate ads according to (\ref{eq:explorativeallocations})
\STATE Ask for no payment
\STATE Observe the clicks $\{\cl_{1}^i(t)\}_{i=1}^{N}$
\ENDFOR
\STATE Compute the estimated quality $\hq_i = \frac{1}{|B_i|}\sum_{t \in B_i} \cl_{1}^i(t)$
\STATE Compute $\hqp_i = \hq_i + \eta$ where $\eta$ is given by (\ref{eq:hoeff})
\STATE
\STATE \textit{Exploitation phase}
\FOR{$t = \tau+1,\ldots, T$}
\STATE Allocate ads according to $\tilde{f}'$ as prescribed by Algorithm~1
\IF{Ad $a_i$ is clicked}
\STATE Ask for payment $\tilde{p}^{B,c}_i$ defined in (\ref{eq:pay.babaioff.ppc})
\ENDIF
\ENDFOR
\end{algorithmic}}
\caption{Pseudo--code for the A--VCG3 mechanism.}\label{f:alg3}
\end{figure}

First of all, we remark that the mechanisms presented in Sections~\ref{ssec:uq} and~\ref{ssec:ul} cannot be adopted here, but the study of a new mechanism is required. The mechanism we design is given by the combination of A--VCG1 and A--VCG2$^\prime$. The pseudo code of the algorithm A--VCG3 (Adaptive VCG3) is given in Fig.~\ref{f:alg3}. As in the case in which only $\{q_i\}_{i \in \N}$ are unknown, we formalize the problem as a multi--armed bandit where the exploration and exploitation phases are separate and where, during the exploration phase,  we estimate the values of $\{q_i\}_{i \in \N}$. Details of the algorithm follow.


\paragraph*{\indent Exploration phase} During the first $\tau$ rounds of the auction, estimates of $\{q_i\}_{i \in \N}$ are computed. We use the same exploration policy of Section~\ref{ssec:uq}, but the estimations are computed just using samples from the first slot, since $\Lambda_m$ with $m>1$ are unknown.\footnote{In the following, we report some considerations about the case in which also the samples from the slots below the first are considered.} Define $B_i = \{t: \pi(i; \theta_t) = 1, t\leq\tau\}$ the set of rounds $t\leq \tau$ where $a_i$ is displayed in the first slot, the number of samples collected for $a_i$ is $|B_i| = \lfloor \frac{\tau}{N} \rfloor \geq \frac{\tau}{2N}$. The estimated value of $q_i$ is computed as:

\begin{align*}
\hq_i = \frac{1}{|B_i|}\sum_{t \in B_i} \cl_1^i(t).
\end{align*}
such that $\hq_i$ is an unbiased estimate of $q_i$ (i.e., $\mathbb{E}_{click} [\hq_i] = q_i$, where $\mathbb{E}_{click}$ is in expectation w.r.t. the realization of the clicks). By applying the Hoeffding's inequality we obtain an upper bound over the error of the estimated quality $\hq_i$ for each ad $a_i$.

\begin{proposition}\label{p:hoeffding.ql}
For any ad $\{a_i\}_{i \in \N}$
\begin{align}\label{eq:hoeff}
| q_i - \hq_i | \leq \sqrt{\frac{1}{2 |B_i|} \log \frac{2N}{\delta}} \leq \sqrt{\frac{N}{\tau} \log \frac{2N}{\delta}} =: \eta,
\end{align}
with probability $1-\delta$ (w.r.t. the click events).
\end{proposition}
After the exploration phase, an upper--confidence bound over each quality is computed as $\hqp_i = \hq_i + \eta$. 

\paragraph*{\indent Exploitation phase} We first focus on the allocation function. During the exploitation phase we want to use an allocation $\tilde{\theta}=\tilde{f}(\hbv)$ maximizing the estimated social welfare with estimated $\{\hqp_i\}_{i \in \N}$ and the parameters $\{\Lambda_m\}_{m \in \K}$. Since the actual parameters $\{\Lambda_m\}_{m \in \K}$ are monotonically non--increasing we can use an allocation $\{\langle s_m, a_{\alpha(m; \tilde{\theta})} \rangle\}_{m \in \K'}$, where
\begin{align*}
\alpha(m; \tilde{\theta}) \in \arg\max_{i \in \N} (\hqp_i \hv_i; m) = \arg\max_{i \in \N} (\hqp_i \Lambda_m \hv_i; m).
\end{align*}

We now focus on payments. Allocation function $\tilde{f}$ is an affine maximizer (due to weights depending on $\tilde{q}_i$ as in Section~\ref{ssec:uq}), but WVCG payments cannot be computed given that parameters $\{\Lambda_m\}_{m \in \K}$ are unknown. Neither the adoption of execution--contingent payments, like in~(\ref{eq:olppc}), is allowed, given that $q_i$ is unknown and only estimates $\tilde{q}_i$ are available.

Thus, we resort to implicit payments as in Section~4.2.2. More precisely, we use the same exploitation phase we used in Section~4.2.2 except that we adopt $\tilde{f}$ in place of $f^*$. In this case, we have that the per--click payments are:

\begin{multline} \label{eq:pay.babaioff.ppc}
\tilde p^{B,c}_i(\bx, \cl_{\pi(i; \tilde{f}(\bx))}^{i}) = \begin{cases} \frac{\tilde p_i^B(\bx,\by;\hbv)}{\Lambda_{\pi(i; \tilde{f}(\bx))} q_i} & \mbox{if } \cl_{\pi(i; \tilde{f}(\bx))}^{i}=1 \\ 0 & \mbox{otherwise} \end{cases}
= \\ \begin{cases} \hv_i -
\begin{cases}  \frac{1}{\mu} \hv_i &\mbox{if $y_i<\hv_i$}  \\
0 & \mbox{otherwise}, \end{cases} 
 & \mbox{if } \cl_{\pi(i; \tilde{f}(\bx))}^{i}=1 \\ 0 & \mbox{otherwise} \end{cases}
\end{multline}
where
\begin{align}
\tilde p_i^B(\bx,\by;\hbv) = \Lambda_{\pi(i; \tf(\bx))}q_i \hv_i - 
\begin{cases}  \frac{1}{\mu}\Lambda_{\pi(i; \tf(\bx))}q_i \hv_i &\mbox{if $y_i<\hv_i$}  \\
0 & \mbox{otherwise}, \end{cases}
\end{align}

We can state the following.
\begin{theorem}
The A--VCG3 is IC and WBB in expectation (over the realizations of the random component of the mechanism) and IR \emph{a posteriori} (w.r.t. the random component of the mechanism). These properties hold \emph{a posteriori} w.r.t. the click realizations.
\end{theorem}
\begin{pf}
The proof of IC in expectation and WBB in expectation easily follows from the definition of the adopted mechanism as discussed in~\cite{babaioff_impl_pay}. The proof of IR \emph{a posteriori} is similar to the proof of Proposition~\ref{prop:AVGC2IRaposteriori}. The fact that the properties hold \emph{a posteriori} w.r.t. the click realizations follows from~\cite{babaioff_impl_pay}.\qed
\end{pf}

Now we want to analyze the performance of the mechanism in terms of regret cumulated through $T$ rounds. Notice that in this case we have to focus on two different potential sources of regret: the adoption of a sub--optimal (randomized) allocation function and the estimation of the unknown parameters.

\begin{theorem}\label{thm:constant.ql}
Let us consider an auction with $N$ advertisers, $K$ slots, and $T$ rounds, with position--dependent cascade model with parameters $\{\Lambda_m\}_{m=1}^K$. For any parameter $\tau$ and $\delta$, the A--VCG3 achieves a regret

\begin{align*}
R_T & \leq \vmax K \left[\left( T-\tau \right) \left(2 \eta + 2 \mu N \right) + \tau + \delta T \right]
\end{align*}

\noindent By setting the parameters to

\begin{itemize}
	\item $\mu = N^{-\frac{2}{3}} T^{-\frac{1}{3}}$. $\mu$ is always $\leq 1$
	\item $\delta = N^\frac{1}{3} T^{-\frac{1}{3}}$. $\delta \leq 1$, thus $T \geq N$
	\item $\tau = T^\frac{2}{3} N^\frac{1}{3} \left( \log{\frac{2N}{\delta}} \right)^\frac{1}{3}$
\end{itemize}
then the regret is
\begin{align}
			R_T & \leq 6\vmax K T^\frac{2}{3} N^\frac{1}{3} \Big( \log \big(2 N^\frac{2}{3} T^\frac{1}{3} \big)\Big)^\frac{1}{3} \label{eq:regret.posdep.qlu}
\end{align}
\end{theorem}

\myremark{1 (The bound).}  Up to numerical constants and logarithmic factors, the previous bound is $R_T \leq \tilde O(T^\frac{2}{3} K N^\frac{1}{3})$.
We first notice we match the lowest possible complexity for the parameter $T$ when exploration and exploitation phases are separate. Moreover observe that the proposed mechanism is a no--regret algorithm, thus asymptotically it achieves the same performances of VGC (when all the parameter are known), since its per--round regret ($R_T/T$) decreases to 0 as $T^{-\frac{1}{3}}$.
We can observe that, with respect to the case of Section~\ref{ssec:uq}, the dependence of the cumulative regret in the parameter $K$ is augmented by a factor $K^\frac{1}{3}$. The reason resides in the exploration phase, indeed, in this last case, we cannot take advantage of all data we can collect, given that we estimate the qualities only on the basis of their visualization in the first slot. Instead, the dependency on $N$ is the same of the one in the case studied in Section~\ref{ssec:uq}.

\myremark{2 (Non--separate phases and $O(T^{1/2})$).} The questions whether or not it is possible to avoid the separation of the exploration and exploitation phases preserving IC in expectation (in some form) and whether or not it is possible to obtain a regret of $O(T^{1/2})$ are open. We conjecture that, if it is possible to have $R_T=O(T^{1/2})$ when only $\{q_i\}_{i \in \N}$ are unknown, then it is possible to have $R_T=O(T^{1/2})$ also when $\{q_i\}_{i \in \N}$ and $\{\Lambda_m\}_{m \in \K}$ are unknown.  However, such a problem is still open.
%
%
%

\myremark{3 (Using samples from multiple slots).} The question whether it is possible to exploit the samples from the slots below the first one to improve the accuracy of the estimates and to reduce the length of the exploration phase is open. The critical issue here is that the samples from those slots are about the product of two random variables, i.e., $\Lambda_s$ and $q_i$, and it is not trivial to find a method to use these samples to improve the esteems. However, in the case it is possible to exploit these samples, we would obtain a reduction of the regret bound of at most $K^{1/3}$, given that the dependency from $K$ cannot be better than in the case discussed in Section~\ref{ssec:uq} (i.e., $O(K^{\frac{2}{3}})$).

A--VCG3 allows also the identification of an upper--bound over the regret on the social welfare. The derivation is not straightforward with respect to the bound over the regret on the payments, but, using the value of the parameters identified in Theorem~\ref{thm:constant.ql}, the bound is $\tilde{O}(T^\frac{2}{3})$. Optimising the parameters w.r.t. to the regret over the social welfare, we obtain the following.

\begin{theorem} \label{th:pd_lq_sw}
Let us consider an auction with $N$ advertisers, $K$ slots, and $T$ rounds, with position--dependent cascade model with parameters $\{\Lambda_m\}_{m=1}^K$. For any parameter $\tau$ and $\delta$, the A--VCG3 achieves a regret

\begin{align*}
R_T^{SW} & \leq \vmax K \left[ (T - \tau) ( 2 \eta + N \mu) + \tau + \delta T \right]\\
& \leq \vmax K \left[ (T - \tau) \left( 2 \sqrt{\frac{N}{\tau} \log{\frac{2N}{\delta}}} + N \mu \right) + \tau + \delta T \right]
\end{align*}
\noindent By setting the parameters to
\begin{align*}
	\mu &= K^{-1} N^\frac{1}{3} T^{-\frac{1}{3}}.\ \mu \leq 1 \textit{ when } T > \frac{N}{K^3}\\
	\delta &= N^\frac{1}{3} T^{-\frac{1}{3}}\\
	\tau &= T^\frac{2}{3} N^\frac{1}{3} \left( \log{\frac{2N}{\delta}} \right)^\frac{1}{3}
\end{align*}
then the regret is
\begin{align*}
			R_T^{SW} &\leq 5 \cdot \vmax  K N^\frac{1}{3} T^\frac{2}{3} \left( \log{N^\frac{2}{3}  T^\frac{1}{3}} \right)^\frac{1}{3}.
			\end{align*}%
\end{theorem}

%% file: sec/05posaddep.tex

\section{Learning with Position-- and Ad--Dependent Externalities}\label{s:externalities}

In this section we deal with the general model where both position-- and ad--dependent externalities are present, as formalized in~(\ref{eq:coeff2}), and we provide several partial results. In Section~\ref{sse:uqpad}, we analyze the problem of designing a DSIC mechanism when only the qualities of the ads are unknown. In Section~\ref{sse:pepad} we highlight some problems that rise when also other parameters are uncertain.

\subsection{Unknown quality} \label{sse:uqpad}

\begin{figure}[t]
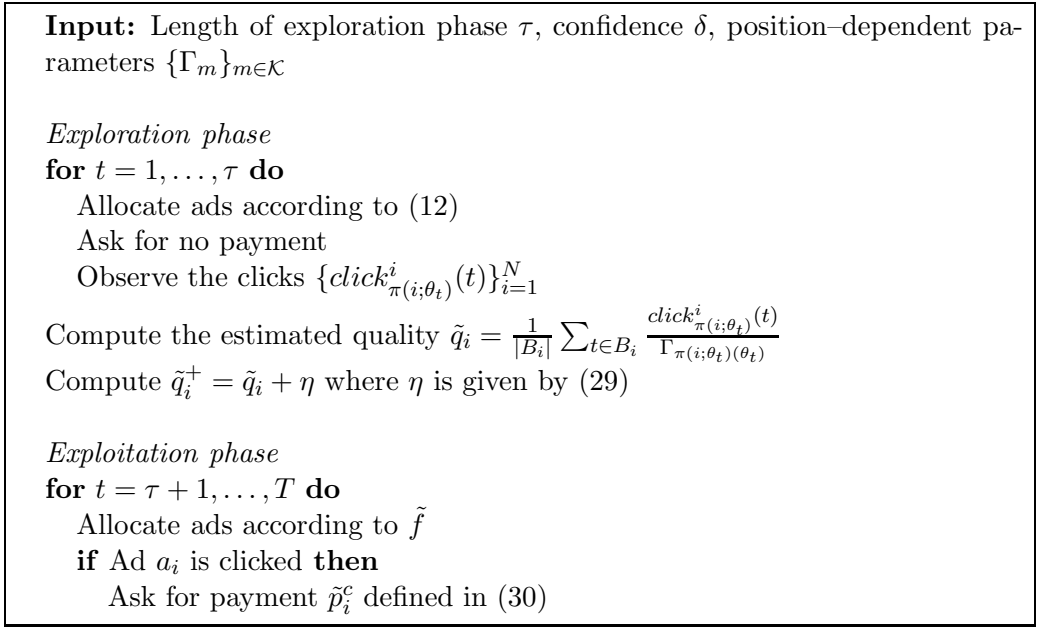

\bookbox{
\begin{algorithmic}
\STATE \textbf{Input:} Length of exploration phase $\tau$, confidence $\delta$, position--dependent parameters $\{\Gamma_m\}_{m \in \K}$
\STATE
\STATE \textit{Exploration phase}
\FOR{$t = 1,\ldots,\tau$}
\STATE Allocate ads according to (\ref{eq:explorativeallocations})
\STATE Ask for no payment
\STATE Observe the clicks $\{\cl_{\pi(i;\theta_t)}^i(t)\}_{i=1}^{N}$
\ENDFOR
\STATE Compute the estimated quality $\tilde{q}_i = \frac{1}{|B_i|}\sum_{t \in B_i} \frac{\cl_{\pi(i; \theta_t)}^i(t)}{\Gamma_{\pi(i; \theta_t)(\theta_t)}}$
\STATE Compute $\hqp_i = \hq_i + \eta$ where $\eta$ is given by (\ref{eq:eta2})
\STATE
\STATE \textit{Exploitation phase}
\FOR{$t = \tau+1,\ldots, T$}
\STATE Allocate ads according to $\tilde{f}$
\IF{Ad $a_i$ is clicked}
\STATE Ask for payment $\tilde p^c_i$ defined in (\ref{eq:hpay.extern})
\ENDIF
\ENDFOR
\end{algorithmic}}
\caption{Pseudo--code for the \padavcg\ mechanism.}\label{f:algpad}
\end{figure}

In this section we analyze the problem where the only unknown parameters are the qualities $\{q_i\}_{i \in \N}$ of the ads and the externality model includes position-- and ad--dependent externalities. As we do in Section~\ref{ssec:uq}, we focus on DSIC mechanisms and we leave open the question whether better bounds over the regret can be found by employing IC in expectation. Therefore we study MAB algorithms that separate the exploration and exploitation phases. The structure of the mechanism we propose, called \padavcg, is similar to the \avcg1 and is reported in Fig.~\ref{f:algpad}.
\paragraph{\indent Exploration phase.} During the exploration phase with length $\tau \leq T$ steps we collect $K$ samples of click or no--click events. Given a generic exploration policy $\{\theta_t\}_{0 \leq t \leq \tau}$, the estimate quality $\hq_i$ is computed as:
\begin{align*}
\tilde{q}_i = \frac{1}{|B_i|}\sum_{t \in B_i} \frac{\cl_{\pi(i; \theta_t)}^i(t)}{\eps_{\pi(i; \theta_t)}(\theta_t)},
\end{align*}
\noindent where we identify the set $B_i  = \{t: \pi(i; \theta_t) \leq K, t \leq \tau\}$.

The explorative allocations $\theta_t$ have an impact on the discount $\eps_m(\theta_t)$ and thus a variation of Proposition~\ref{p:hoeffding} holds in which (\ref{eq:hoeffproposition1}) is substituted by:
\begin{align*}
| q_i - \hq_i | \leq \sqrt{\Bigg(\sum_{t \in B_i} \frac{1}{\eps_{\pi(i; \theta_t)}(\theta_t)^2}\Bigg) \frac{1}{2 |B_i|^2} \log \frac{2N}{\delta}}.
\end{align*}
 For each exploration policy  such that $|B_i| = \lfloor K\tau / N \rfloor$ $\forall i \in \N$, e.g. policy (\ref{eq:explorativeallocations}),  we redefine $\eta$ as
\begin{align}\label{eq:eta2}
| q_i - \hq_i | \leq \frac{1}{\eps_{\min}}\sqrt{\frac{N}{2 K \tau} \log \frac{N}{\delta}} := \eta,
\end{align}
where $\eps_{\min} = \min\limits_{\theta \in \Theta, m \in \K} \{\eps_m(\theta)\}$. We define the upper--confidence bound $\hqp_i = \hq_i + \eta$. 
During the exploration phase, in order to preserve the DSIC property, the allocations $\{\theta_t\}_{0 \leq t \leq \tau}$ do not depend on the reported values of the advertisers and no payments are imposed to the advertisers.
\paragraph{\indent Exploitation phase} We define the estimated social welfare as
\begin{align*}
\tSW(\theta, \hbv) = \sum_{i=1}^N \eps_{\pi(i; \theta)}(\theta) \hqp_i \hv_i = \sum_{m=1}^K \eps_m(\theta) \hqp_{\alpha(m;\theta)} \hv_{\alpha(m;\theta)} .
\end{align*}
We denote by $\tilde{\theta}$ the allocation maximizing $\tSW(f(\hbv), \hbv)$ and by $\tf$ the allocation function returning $\tilde{\theta}$:
\[
\tilde{\theta} = \tf(\hbv) \in \arg\max_{\theta\in\Theta}\tSW(\theta, \hbv).
\]

Once the exploration phase is over, the ads are allocated on the basis of $\tf$. Since $\tf$ is an affine maximizer, the mechanism can impose WVCG payments to the advertisers satisfying the DSIC property. In a \emph{pay--per--click} fashion, if ad $a_i$ is clicked, the advertiser is charged
\begin{align}\label{eq:hpay.extern}
\tilde p_i^c(\hbv, click_{\pi(i;\tilde{\theta})}^i) = \frac{\tSW(\tilde{\theta}_{-i}) - \tSW_{-i}(\tilde{\theta})}{\eps_{\pi(i; \tilde{\theta})}(\tilde{\theta}) \hqp_i}
\end{align}
which corresponds, in expectation, to the WVCG payment $\tilde p_i = \tilde p_i^c \eps_{\pi(i; \tilde{\theta})}(\tilde{\theta}) q_i$.

We are interested in bounding the regret of the auctioneer's revenue due to \padavcg\ compared to the auctioneer's revenue of the VCG mechanism when all the parameters are known.

\begin{theorem}\label{thm:extern}
Let us consider an auction with $N$ advs, $K$ slots, and $T$ rounds. The auction has position/ad--dependent externalities and cumulative discount factors $\{\eps_m(\theta)\}_{m=1}^K$ and $\eta$ defined as in (\ref{eq:eta2}). For any parameter $\tau \in \{0, \ldots, T\}$ and $\delta \in [0,1]$, the \padavcg\  achieves a regret:

\begin{align}\label{eq:regret.extern.exact}
R_T \leq \vmax K \left[ (T - \tau) \left( \frac{3\sqrt{2}n}{\eps_{\min}q_{\min}} \sqrt{\frac{N}{K\tau} \log \frac{N}{\delta}} \right) + \tau + \delta T \right],
\end{align}

\noindent where $q_{\min} = \min_{i \in \N} q_i$. By setting the parameters to
%
\begin{align*}
\delta &=K^\frac{1}{3} N^\frac{1}{3} \left( \frac{5}{\sqrt{2} \Gmin} \right)^\frac{2}{3} T^{-\frac{1}{3}},\\
\tau &= \left( \frac{5}{\sqrt{2} \Gmin} \right)^{\frac{2}{3}} K^{\frac{1}{3}} T^{\frac{2}{3}} N^{\frac{1}{3}} \left( \log{\frac{N}{\delta}} \right)^{\frac{1}{3}},
\end{align*} 
the regret is
%
%
%
\begin{align}\label{eq:regret.extern}
R_T \leq 4 \vmax K^\frac{4}{3} T^\frac{2}{3} N^\frac{1}{3} \frac{5^\frac{2}{3}}{2^\frac{1}{3} \Gmin^\frac{2}{3} \qmin} \left(\log{\frac{2^\frac{1}{3} \Gmin^\frac{2}{3} N^\frac{2}{3} T^\frac{1}{3}}{K^\frac{1}{3} 5^\frac{2}{3}}}\right)^\frac{1}{3}.
\end{align}
\end{theorem}

\myremark{1 (Differences w.r.t. position--dependent externalities.)} Up to constants and logarithmic factors, the previous distribution--free bound is $R_T\leq \tilde O(T^\frac{2}{3} N^\frac{1}{3} K^\frac{4}{3})$.\footnote{We notice that in~\cite{glt} the authors provide a bound $O(T^\frac{2}{3} N K^\frac{2}{3})$ that does not match with their numerical simulations and thus they conjecture that the actual bound is  $O(T^\frac{2}{3} N^\frac{1}{3} K^\frac{4}{3})$. Here we show that the conjecture is correct.} We first notice that moving from position-- to position/ad--dependent externalities does not change the dependency of the regret on both the number of rounds $T$ and the number of ads $N$. Moreover, the per--round regret still decreases to 0 as $T$ increases.
The main difference w.r.t. the bound in Theorem~\ref{thm:constant} is in the dependency on $K$ and on the smallest quality $q_{\min}$. We believe that the augmented dependence in $K$ is mostly due to an intrinsic difficulty of the position/ad--dependent externalities. The intuition is that now, in the computation of the payment for each ad $a_i$, the errors in the quality estimates cumulate through the slots (unlike the position--dependent case where they are scaled by $\eps_{k}-\eps_{k+1}$). This cumulated error should impact only on a portion of the ads (i.e., those which are actually impressed according to the optimal and the estimated optimal allocations) whose cardinality can be upper--bounded by $2K$. Thus we observe that the bound shows a super--linear dependency in the number of slots. 
The other main difference is that now the regret has an inverse dependency on the smallest quality $q_{\min}$. Inspecting the proof, this dependency appears because the error of a quality estimation for an ad $a_i$ might be amplified by the inverse of the quality itself $\frac{1}{q_i}$. As discussed in Remark 2 of Theorem~\ref{thm:constant}, this dependency might follow from that fact the we have a distribution--free bound.
We investigate whether this dependency is an artifact of the proof or it is intrinsic in the algorithm in the numerical simulations reported in Section~\ref{s:experiments}. 

\myremark{2 (Optimization of the parameter $\tau$).} We are considering an environment where $\{q_i\}_{i \in \N}$ are unknown, but if, at least, a guess about the value of $q_{\min}$ is available, it could be used to better tune  $\tau$ by multiplying it by $(q_{\min})^{-\frac{2}{3}}$, thus reducing the regret from $\tilde O((q_{\min})^{-1})$ to $\tilde O((q_{\min})^{-\frac{2}{3}})$.

\myremark{3 (Externalities--dependent bound).} We notice that the above bound does not reduce to the bound (\ref{eq:regret.const}) in which only position--dependent externalities are present even disregarding the constant terms. Indeed, the dependency on $K$ is different in the two bounds: in (\ref{eq:regret.const}) we have $K^{\frac{2}{3}}$ while in (\ref{eq:regret.extern}) we have $K^{\frac{4}{3}}$. This means that bound (\ref{eq:regret.extern}) over--estimates the dependency on $K$ whenever the auction has position--dependent externalities. It is an interesting open question whether it is possible to derive an \textit{auction--dependent} bound where the specific values of the discount factors $\gamma_k(f)$ explicitly appear in the bound and that it reduces to (\ref{eq:regret.const}) for position--dependent externalities.

\textit{(Comment to the proof).} 
While the proof of Thm.~\ref{thm:constant} could exploit the specific definition of the payments for position--dependent slots and it is a fairly standard extension of~\cite{devanur2009price}, in this case the proof is more complicated because of the dependency of the discount factors on the actual allocations and decomposes the regret of the exploitation phase in components due to the different allocations ($\tf$ instead of $f^*$) and the different qualities as well ($\hqp$ instead of $q$).

Using the mechanism described before, it is possible to derive an upper--bound over the global regret, when the regret, as in~\cite{babaioff_impl_pay}, is computed over the social welfare of the allocation. We obtain the same dependence over $T$, as for the regret on the payment. Thus $R^{SW}_T\leq\tilde{O}(T^\frac{2}{3})$. In particular notice that \padavcg\ is a zero--regret algorithm.

\begin{theorem} \label{th:pad_q_sw}
Let us consider an auction with $N$ advs, $K$ slots, and $T$ rounds. The auction has position/ad--dependent externalities and cumulative discount factors $\{\eps_m(\theta)\}_{m=1}^K$ and $\eta$ defined as in (\ref{eq:eta2}). For any parameter $\tau \in \{0, \ldots, T\}$ and $\delta \in [0,1]$, the \padavcg\  achieves a regret:

\begin{align}
R^{SW}_T \leq \vmax K \left[ (T - \tau) \frac{2}{\eps_{\min}} \sqrt{\frac{N}{2K\tau} \log \frac{N}{\delta}} + \tau + \delta T \right],
\end{align}

By setting the parameters to
\begin{align*}
\delta & = \left( \frac{\sqrt{2}}{\eps_{\min}} \right)^\frac{2}{3} K^{-\frac{1}{3}} N^\frac{1}{3} T^{-\frac{1}{3}}\\
\tau & = \left( \frac{\sqrt{2}}{\eps_{\min}} \right)^\frac{2}{3} T^\frac{2}{3} N^\frac{1}{3} K^{-\frac{1}{3}} \left( \log \frac{2N}{\delta} \right)^\frac{1}{3},
\end{align*}
the regret is
\begin{align}
R^{SW}_T \leq 4 \vmax \left( \frac{\sqrt{2}}{\eps_{\min}} \right)^\frac{2}{3} K^\frac{2}{3} N^\frac{1}{3} T^\frac{2}{3} \left( \log 2^\frac{2}{3} \eps_{\min}^{-\frac{2}{3}} N^\frac{2}{3} K^\frac{1}{3} T^\frac{1}{3} \right)^\frac{1}{3}.
\end{align}

\end{theorem}

Notice that using $\tau$ and $\delta$ defined in Theorem~\ref{thm:extern}, the bound for $R_T^{SW}$ is $\tilde{O}(T^\frac{2}{3})$, even if the parameters are not optimal for this second framework.


\subsection{Further extensions} \label{sse:pepad}

In this section we provide a negative, in terms of regret, result under DSIC truthfulness when  the parameter $\gamma_{i,m}$ depends only on the ad $i$ (as in~\cite{Kempe2008}, we denote it by $c_i$) and this parameter is the only uncertain parameter. 

We focus on the exploitation phase, supposing the exploration phase has produced the estimates $\{\hcp_i\}_{i \in \N}$ for the continuation probabilities $\{c_i\}_{i \in \N}$. The allocation function $f$ presented in~\cite{Kempe2008} is able to compute the optimal allocation when $\{c_i\}_{i \in \N}$ values are known, but it is not an affine maximizer when applied to the estimated values $\{\hcp_i\}_{i \in \N}$. Indeed, we call this allocation function $\tf$:

\begin{equation} \label{eq:tf}
\tf(\hbv) \in \arg\max_{\theta \in \Theta} \sum_{m=1}^K q_{\alpha(m;\theta)} \hv_{\alpha(m;\theta)} \prod_{h=1}^{m-1} \hcp_{\alpha(h; \theta)}.
\end{equation}

In this case, a weight depending only on a single ad cannot be isolated. Furthermore, we show also that this allocation function is not monotonic.

\begin{proposition}
The allocation function $\tf$ is not monotonic.
\end{proposition}

\begin{pf}
The proof is by counterexample.
Consider an environment with 3 ads and 2 slots such that:

\begin{table}[!h]
	\begin{center}
		\begin{tabular}{|c|c|c|c|}
			\hline
			ad		&	$v_i$		&	$\hcp_i$	&	$c_i$\\
			\hline
			$a_1$	&	$0.85$		&	$1$			&	$0.89$\\
			\hline
			$a_2$	&	$1$			&	$0.9$		&	$0.9$\\
			\hline
			$a_3$	&	$1.4$		&	$0$			&	$0$\\
			\hline
		\end{tabular}	
	\end{center}
\end{table}

\noindent and $q_i = 1$ $\forall i \in \N$. The optimal allocation $\tilde{\theta}$ found by $\tf$ when agents declare their true values $\bv$ is: ad $a_2$ is allocated in the first slot and $a_3$ in the second one. We have $CTR_{a_3}(\tilde{\theta}) = 0.9$.

If advertiser $a_3$ reports a larger value: $\hv_3 = 1.6$, in the allocation $\hat{\theta}$ found by $\tf(\hv_3, \bv_{-3})$, ad $a_1$ is displayed into the first slot and $a_3$ into the second one. In this case $CTR_{a_3}(\hat{\theta}) = 0.89 < CTR_{a_3}(\tilde{\theta})$, thus the allocation function $\tf$ is not monotonic.  \qed
\end{pf}

On the basis of the above result, we can state the following theorem.

\begin{theorem}\label{thm:regretsocialwelfarec}
Let us consider an auction with $N$ advertisers, $K$ slots, and $T$ rounds, with ad--dependent cascade model with parameters $\{c_i\}_{i=1}^N$ whose value are unknown. Any online learning DSIC mechanism achieves an expected regret $R^{SW}_T=\Theta(T)$ over the social welfare.
\end{theorem}

\begin{pf}
Call $f(\hat{\mathbf{v}}| \mathbf{c})$ the allocation function maximizing the social welfare given parameters $\mathbf{c}$. As shown above, $f(\hat{\mathbf{v}}| \tilde{\mathbf{c}})$ cannot be adopted in the exploitation phase, the mechanism would not be DSIC otherwise. However, it can be easily observed that a necessary condition to have a no--regret algorithm is that the allocation function used in the exploitation phase, say $g(\hat{\mathbf{v}}| \tilde{\mathbf{c}})$, is such that $g(\hat{\mathbf{v}}| \mathbf{c}) = f(\hat{\mathbf{v}}| \mathbf{c})$ for every $\hat{\mathbf{v}}$ and $\mathbf{c}$ (that is, they always return the same allocation) given that $\tilde{\mathbf{c}}$ are consistent estimates and  $\tilde{\mathbf{c}}\rightarrow \mathbf{c}$ as $T\rightarrow +\infty$. Otherwise, since allocations are finite and the difference between the values of the allocations is generically strictly positive, the algorithm would suffer from a strictly positive regret when $T\rightarrow +\infty$ and therefore it would not be a no--regret mechanism. However, any such a $g$ would not be monotonic and therefore it cannot be adopted in a DSIC mechanism. As a result, any online learning DSIC mechanism is not a no--regret mechanism.

To complete the proof, we need to provide a mechanism with regret $\Theta(T)$. Such a mechanism can be easily obtained by partitioning ads in groups such that in each group the ads compete only for a single slot. Therefore, each ad can appear in only one slot. \qed
\end{pf}

The above result shows that no approach similar to the approach described in~\cite{babaioff_impl_pay} can be adopted even for IC in expectation. Indeed, the approach described in~\cite{babaioff_impl_pay} requires in input a monotonic allocation function. This would suggest a negative result in terms of regret also when  IC in expectation. However, in this paper we leave the study of this case open.

Finally, we provide a result on the regret over the auctioneer's revenue, whose proof is straightforward given that the (W)VCG cannot be adopted due to the above result and therefore the regret over the payments cannot go to zero as $T$ goes to $\infty$.

\begin{theorem}\label{thm:constant.l}
Let us consider an auction with $N$ advertisers, $K$ slots, and $T$ rounds, with ad--dependent cascade model with parameters $\{c_i\}_{i=1}^N$  whose value are unknown. Any online learning DSIC mechanism achieves an expected regret over the auctioneer's revenue $R_T=\Theta(T)$.
\end{theorem}

%% file: sec/06simulation.tex
\section{Numerical Simulations}\label{s:experiments}

In this section we report numerical simulations to validate the theoretical bounds over the regret of the auctioneer's revenue presented in the previous sections.\footnote{The bounds over the regret of the social welfare present a structure similar to those over the auctioneer's revenue and their empirical analysis is omitted, providing similar results.} In particular, we analyze the accuracy with which our bounds predict the dependency of the regret on the main parameters of the auctions such as $T$, $N$,  $K$, and $q_{\min}$. All the simulations share the way the ads are generated. The qualities $\{q_i\}_{\N}$ are drawn from a uniform distribution in $[0.01, 0.1]$, while the values $\{v_i\}_{\N}$ are randomly drawn from a uniform distribution on $[0, 1]$ ($\vmax = 1$). Since the main objective is to test the accuracy of the bounds, we report the \textit{relative regret}
$$\overline{R}_T = \frac{R_T}{B(T,K,N, q_{\min}, \eps_{\min})},$$
where $B(T,K,N, q_{\min}, \eps_{\min})$ is the value of the bound for the specific setting (i.e., (\ref{eq:regret.const}) and (\ref{eq:regret.posdep.qlu}) for position--dependent, and (\ref{eq:regret.extern}) for position/ad--dependent externalities). We analyze the accuracy of the bound w.r.t. each specific parameter, changing only its value and keeping the values of all the others fixed.  We expect the relative regret to be always smaller than $1$, indeed we expect $B$ to be an actual upper--bound on the real regret $R_T$.  All the results presented in the following sections have been obtained by setting $\tau$ and $\delta$ as suggested by the bounds derived in the previous sections and, where it is not differently specified, by averaging over 100 independent runs.


\subsection{Position--Dependent Externalities}\label{s:exp.constant}

\subsubsection{Unknown $\{q_i\}_{i \in \N}$}

\begin{figure*}[t]
\begin{center}
\includegraphics[width=0.45\textwidth]{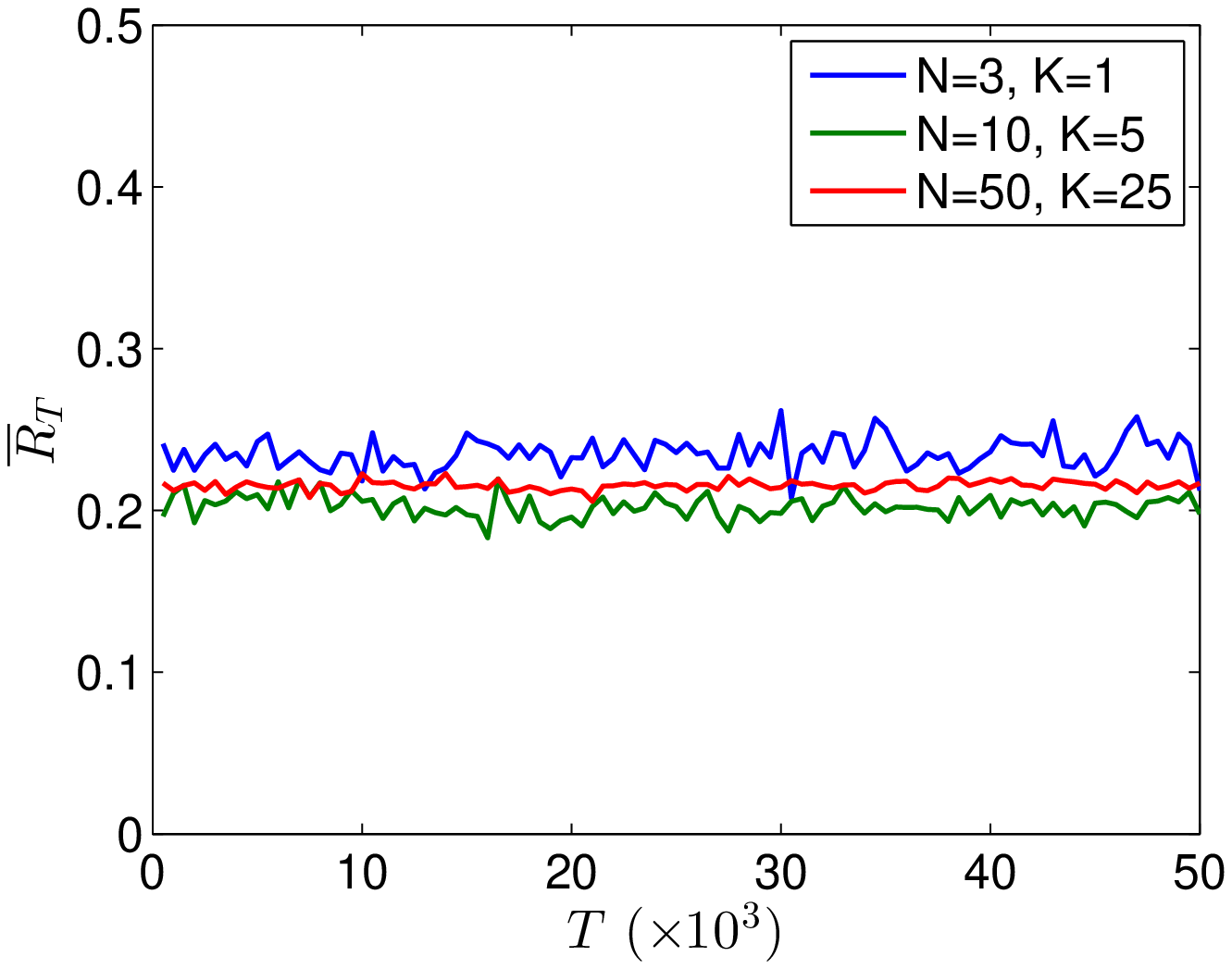}
\includegraphics[width=0.45\textwidth]{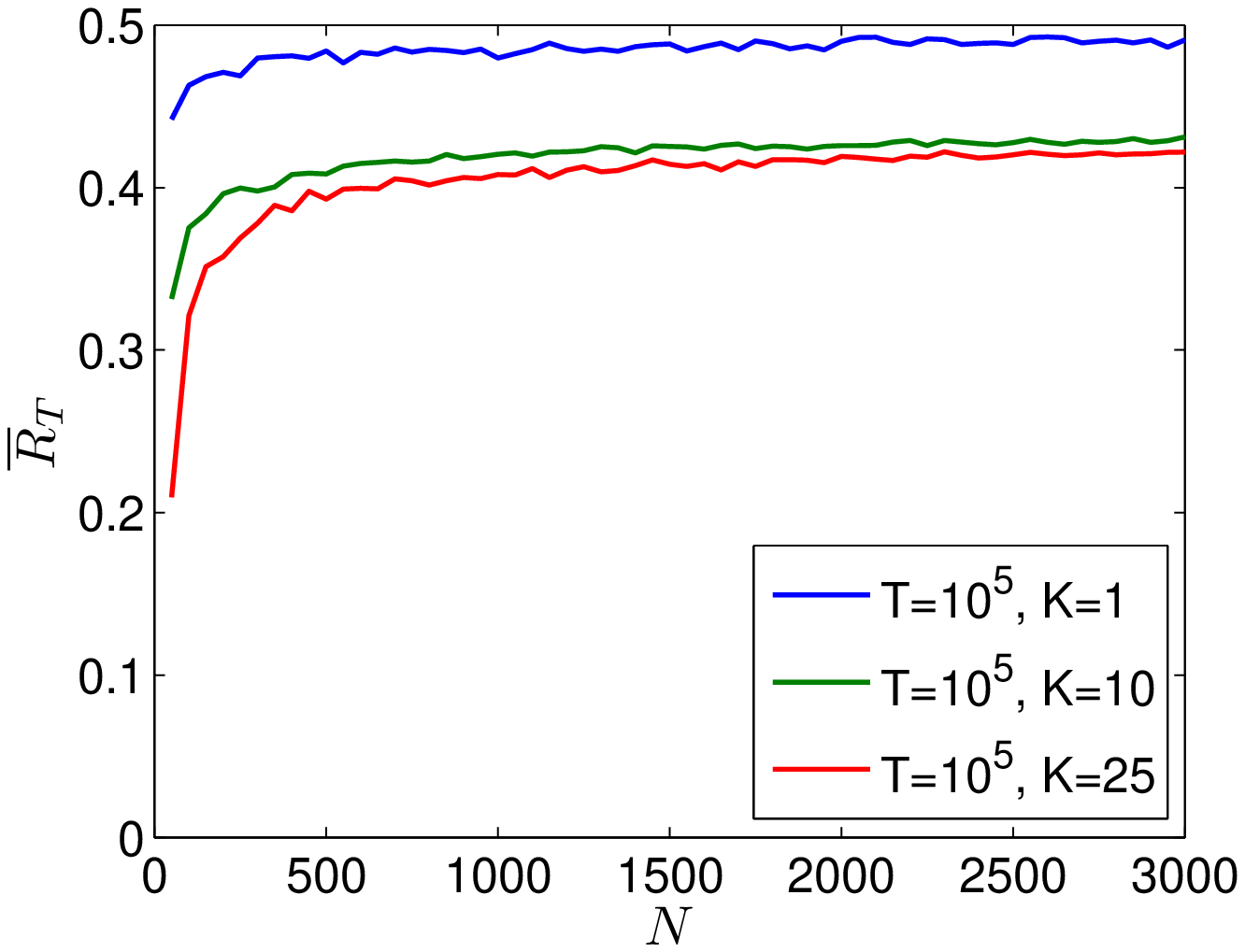}
\end{center}
\vspace{-0.4cm}
\caption{Position--dependent externalities with unknown $\{q_i\}_{i \in \N}$. Dependency of the relative regret on $T$, $N$.}\label{f:const}
\vspace{-0.4cm}
\end{figure*}

\begin{figure*}[t]
\begin{center}
\includegraphics[width=0.45\textwidth]{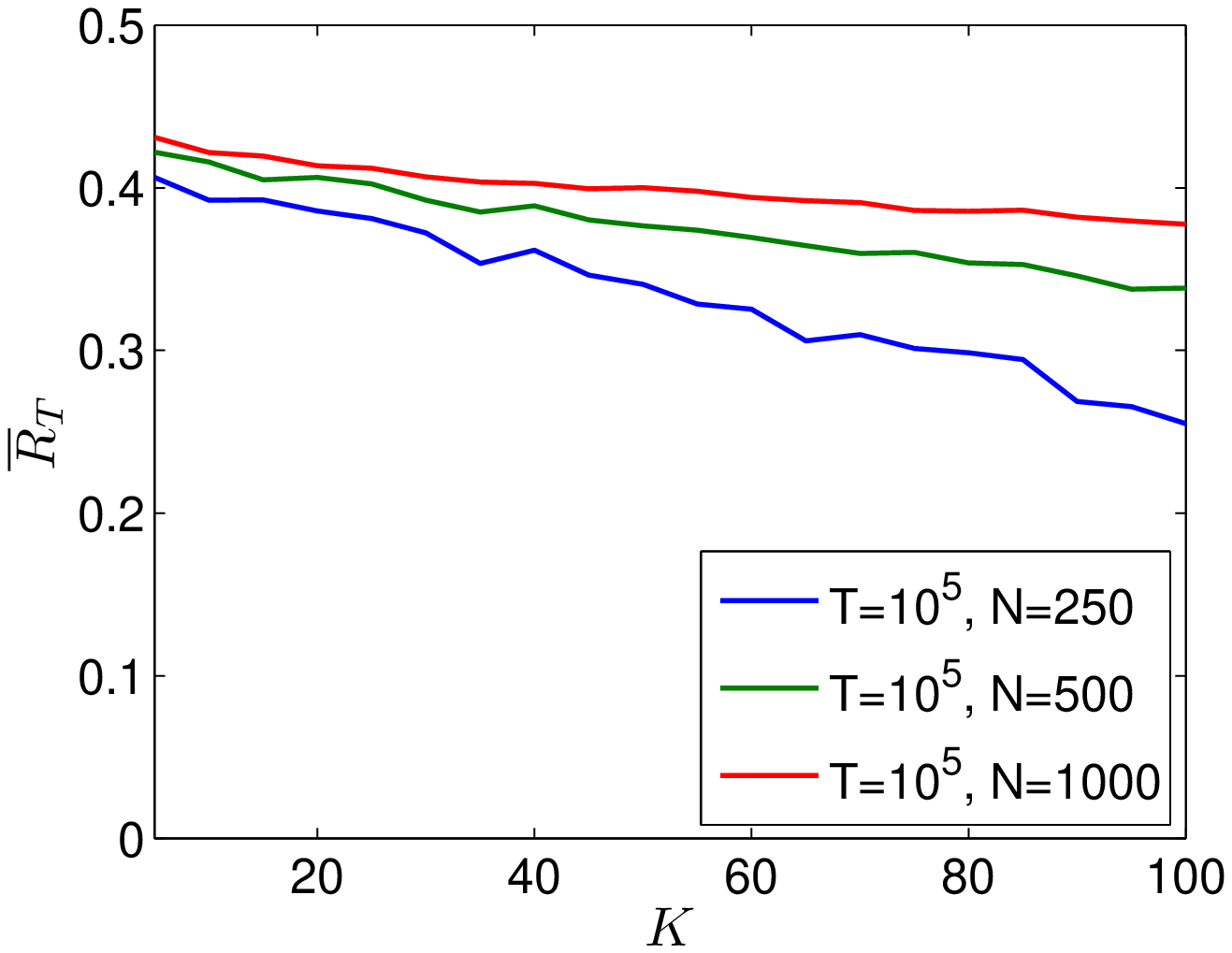}
\includegraphics[width=0.45\textwidth]{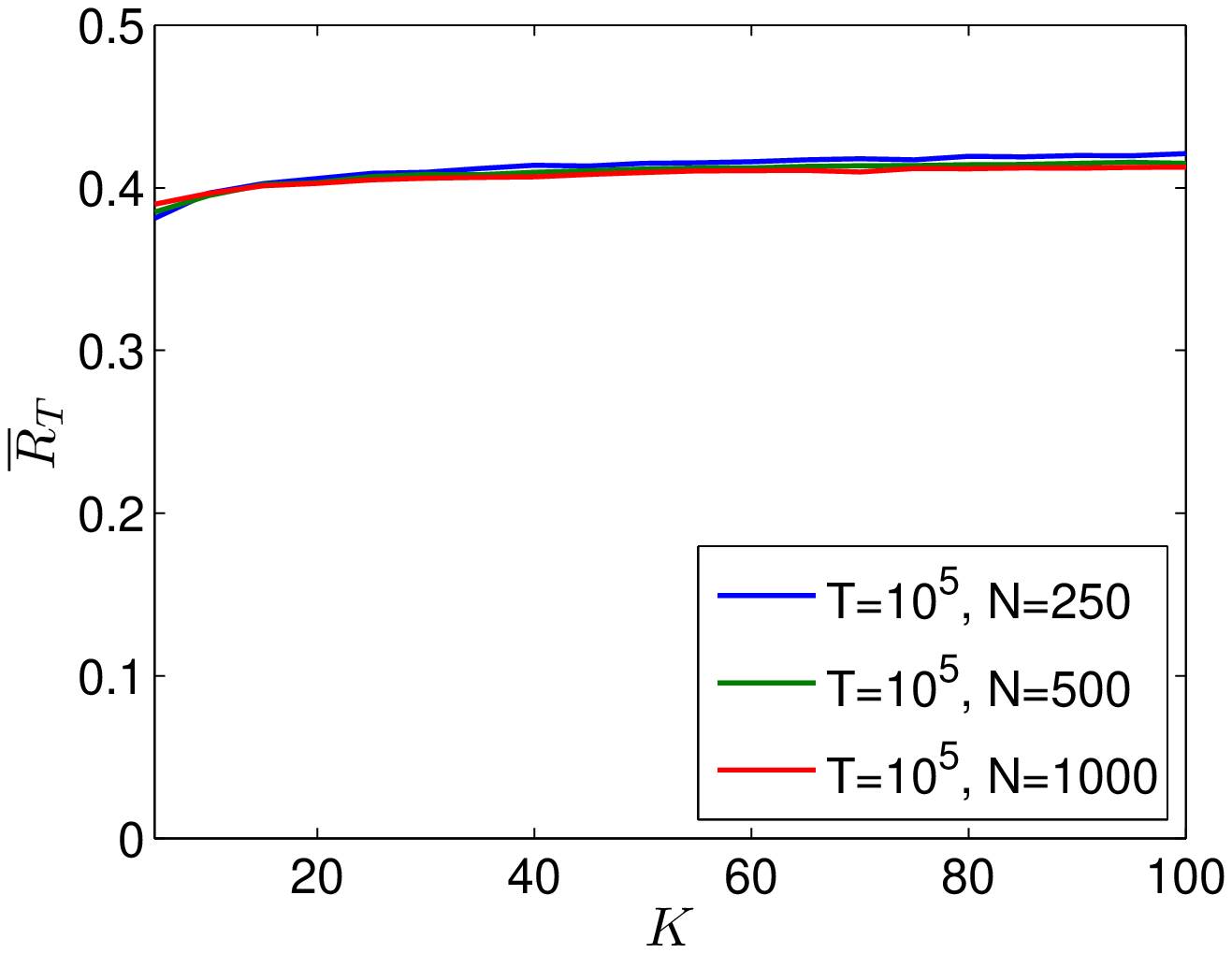}
\end{center}
\vspace{-0.4cm}
\caption{Position--dependent externalities with unknown $\{q_i\}_{i \in \N}$. Dependency of the relative regret on $K$ for two different choice of the the qualities $q$.}\label{f:const2}
\vspace{-0.2cm}
\end{figure*}

First of all we analyze the accuracy of the bound provided in Section~\ref{ssec:uq}, where the model presents only position--dependent externalities and the qualities of the ads are unknown. We design the simulations such that $\lambda_m=\lambda$ for every $m$ with $\Lambda_1 = 1$ and $\Lambda_K = 0.8$ (i.e., $\lambda = \sqrt[K-1]{0.8}$). Thus, $\Lambda_{\min} = 0.8$ in all the experiments. 

In Fig.~\ref{f:const} we analyze the accuracy of the bound w.r.t. the parameters $T$ and $N$. All the three curves in the left plot are completely flat (except for white noise) showing that the value of the relative regret $\bR_T$ for different values of $K$ and $N$ not change as $T$ increases. This suggests that the bound in Theorem~\ref{thm:constant} effectively predicts the dependency of the regret $R_T$ w.r.t. the number of rounds $T$ of the auction as $\tilde O(T^{2/3})$. The right plot represents the dependency of the relative regret $\bR_T$ on the number of ads $N$. In this case we notice that it is relatively accurate as $N$ increases but there is a transitory effect for smaller values of $N$ where the regret grows faster than predicted by the bound (although $B(T,K,N, q_{\min}, \Lambda_{\min})$ is still an upper--bound to $R_T$). Finally, the left plot of Fig.~\ref{f:const2} suggests that the dependency on $K$ in the bound of Theorem~\ref{thm:constant} is over--estimated, since the relative regret $\bR_T$ decreases as $K$ increases. As discussed in the comment to the proof in Section~\ref{s:constant} this might be explained by the over--estimation of the term $\frac{\max_i(\hqp_{i} \hv_i;l)}{\max_i(\hqp_{i} \hv_i;k)}$ in the proof. In fact, this term is likely to decrease as $K$ increases. In order to validate this intuition, we have identified some instances for which the bound seems to accurately predict the dependency on $K$. For these instances $q_1 = 0.1$, $q_2=0.095$, and $q_i=0.09$ for every $2<i\leq K$. As a result, the ratio between the qualities $q_i$ is fixed (on average) and does not change with $K$. The right plot of Fig.~\ref{f:const2} shows that, with these values of $q_i$, the ratio $\bR_T$ is constant for different values of $N$, implying that in this case the bound accurately predicts the behavior of $R_T$. In fact, as commented in Theoerm~\ref{thm:constant}, we derive distribution--independent bounds where the qualities $q_i$ do not appear in the bound. As a result, $R_T$ should be intended as a worst case w.r.t. all the possible configurations of qualities and the externalities.

\subsubsection{Unknown $\{\Lambda_m\}_{m \in \K}$}

\begin{figure*}[th]
\begin{center}
\includegraphics[width=0.45\textwidth]{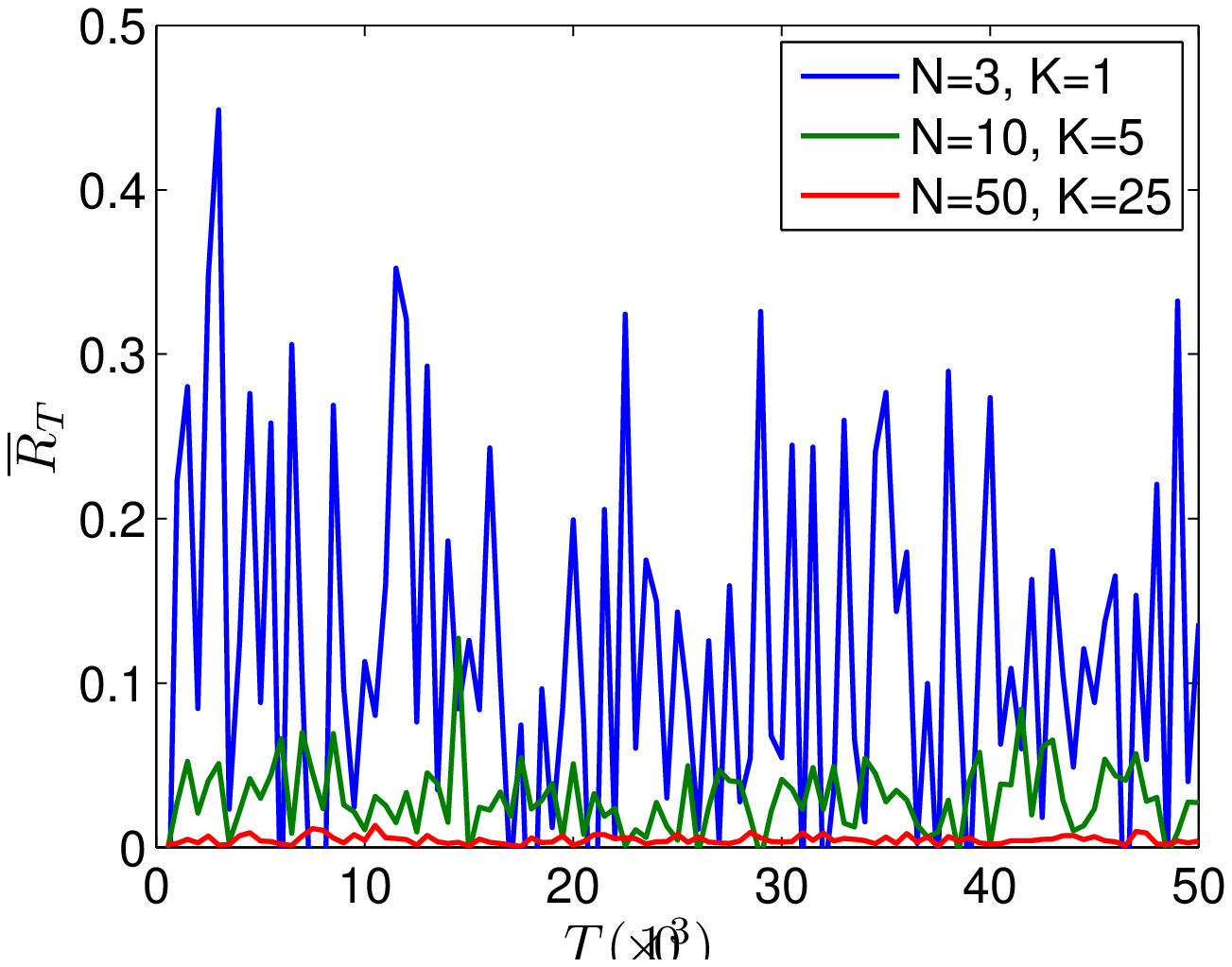}
\includegraphics[width=0.45\textwidth]{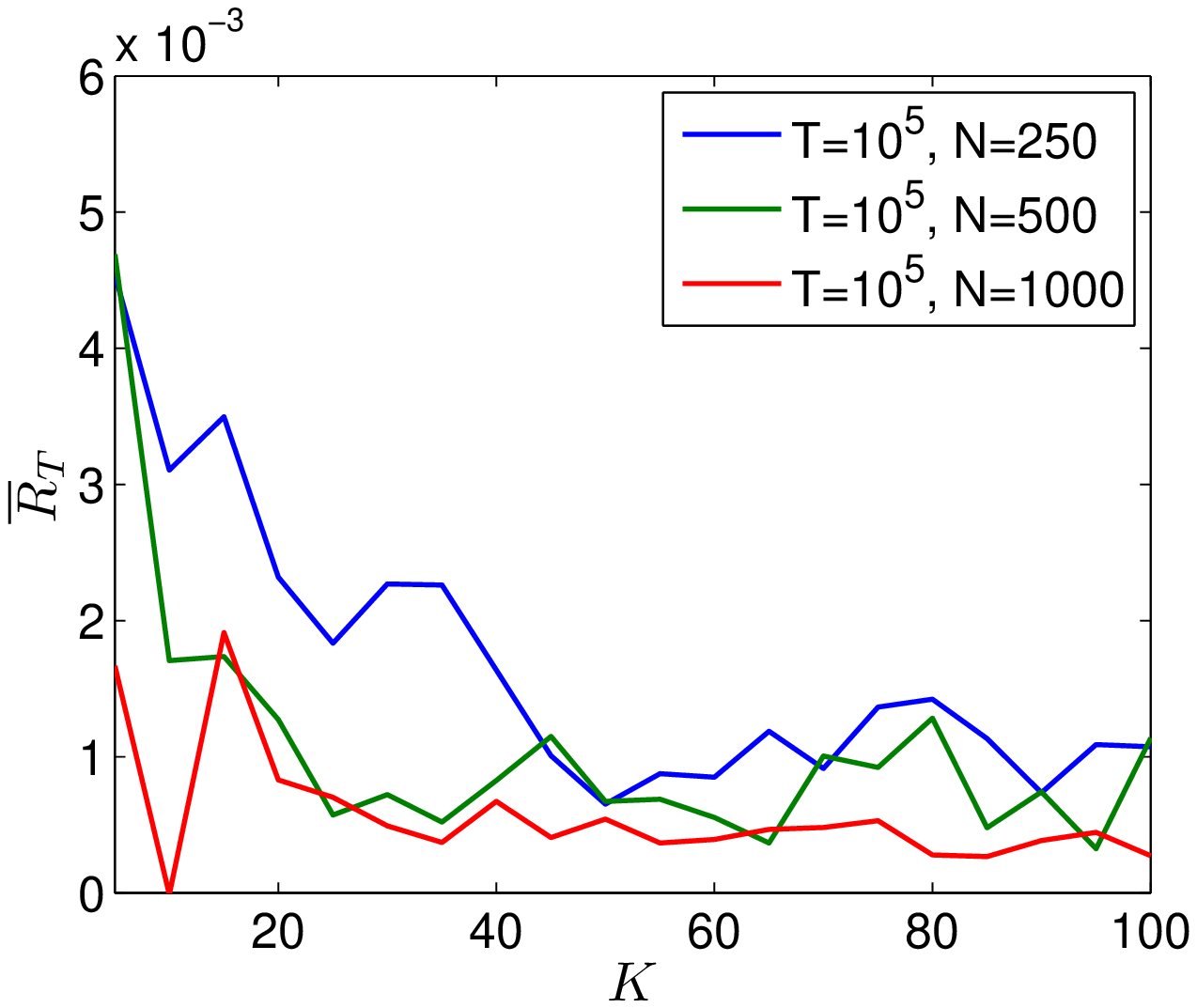}
\end{center}
\vspace{-0.4cm}
\caption{Position--dependent externalities with unknown $\{\Lambda_m\}_{m \in \K}$. Dependency of the relative regret on $T$ and $K$.}\label{f:pd.TK}
\vspace{-0.4cm}
\end{figure*}

\begin{figure*}[t]
\begin{center}
\includegraphics[width=0.45\textwidth]{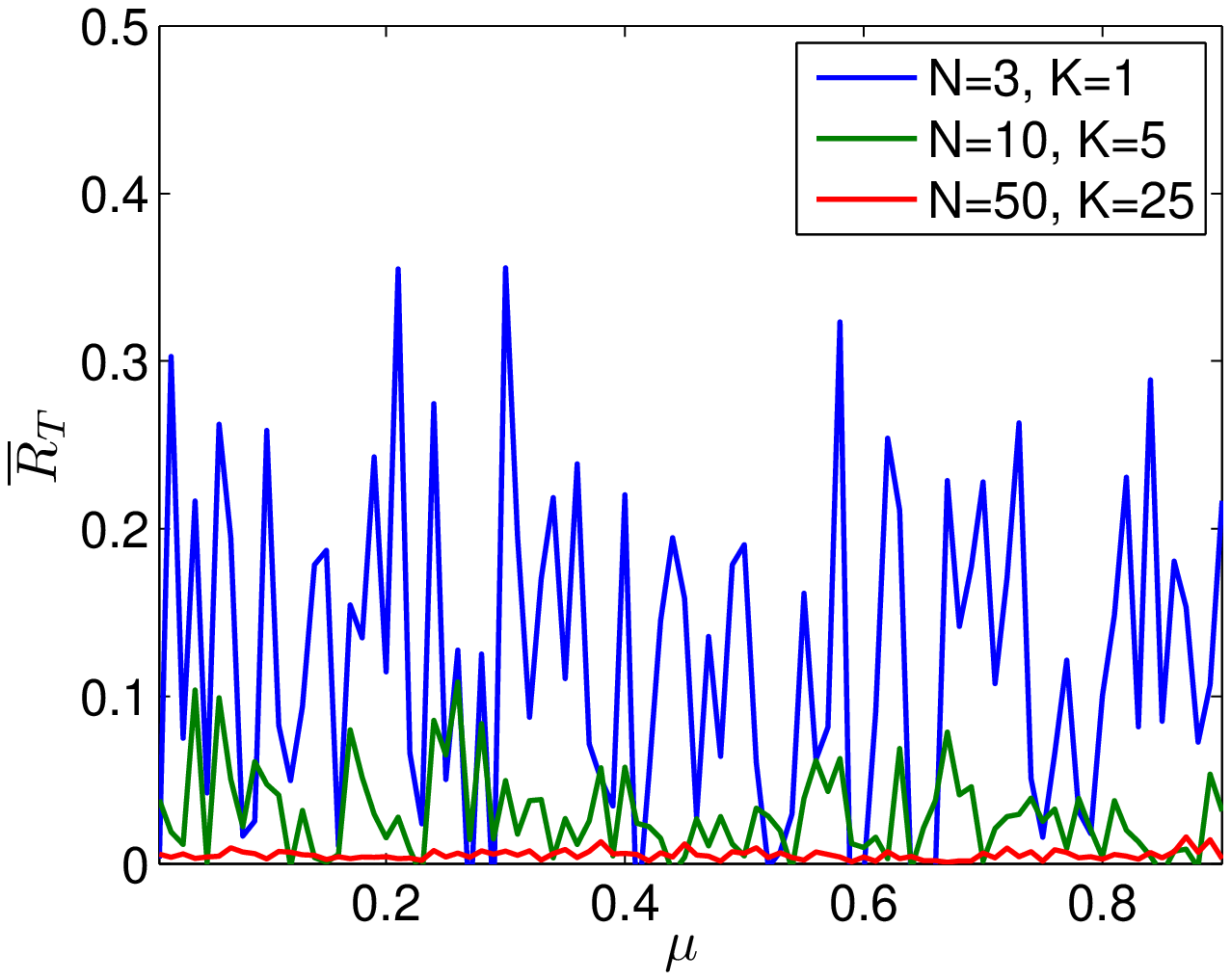}
\includegraphics[width=0.45\textwidth]{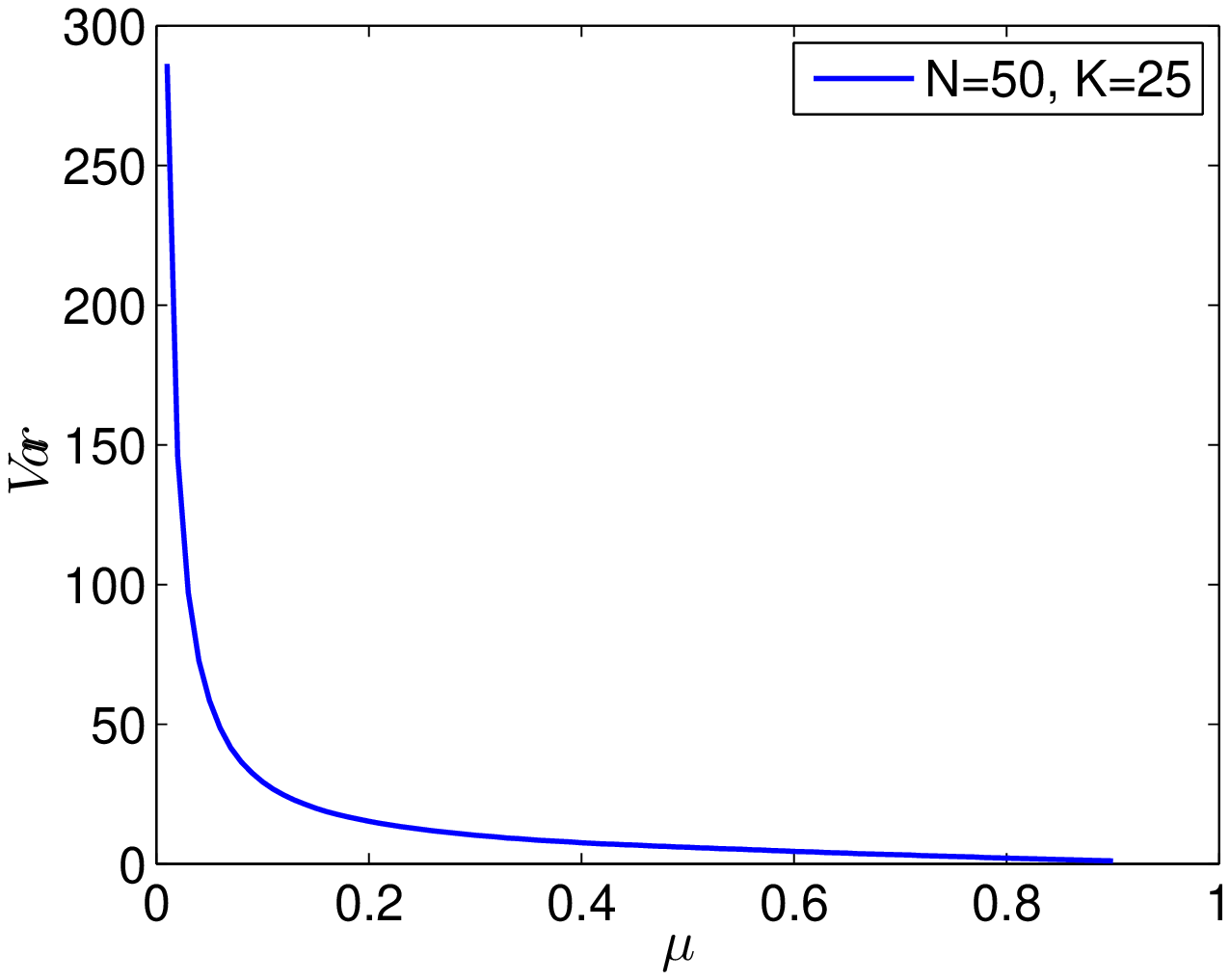}
\end{center}
\vspace{-0.4cm}
\caption{Position--dependent externalities with unknown $\{\Lambda_m\}_{m \in \K}$. Dependency of the relative regret on $\mu$. Variance of the revenue of the auctioneer}\label{f:pd.mu}
\vspace{-0.2cm}
\end{figure*}

We now investigate the accuracy of the bound derived for algorithm A--VCG2$^\prime$ presented in Section~\ref{sssec:l.uc.m}. We used several probability distributions to generate the values of $\{\lambda_m\}_{m \in \K}$. We observed that, when they are drawn uniformly from the interval $[0.98, 1.00]$, the numerical simulations confirm our bound (as we show below), whereas the bound seems to overestimate the dependences over $K$ and $\mu$ when the support of the probability distribution is larger (i.e., $[<0.98, 1.00]$); we do not report any plot for this second case.

The left plot of Figure~\ref{f:pd.TK} shows the dependence of the ratio $\overline{R}_T$ w.r.t. $T$ when $\mu = 0.01$. Despite the noise, the ratio seems not to be affected by the variation of $T$, confirming our bound. In the right plot of Figure~\ref{f:pd.TK}, the ratio follows the same behaviour as $K$ varies when $T = 10^5$ and $\mu = 0.01$ except that the bound seems to overestimate the dependence when $K$ assumes small values (as it happens in practice).
In the left plot of Figure~\ref{f:pd.mu}, the ratio $\overline{R}_T$ seems to be constant as $\mu$ varies when $T = 10^5$. 

We conclude our analysis studying the variance of the payments as $\mu$ varies. The bound over $R_T$, provided in Section~\ref{sssec:l.uc.m}, suggests to choose a $\mu \rightarrow 0$ in order to reduce the regret. Nonetheless, the regret bounds are obtained in expectation w.r.t. all the sources of randomization (including the mechanism) and do not consider the possible deviations. Thus in the right plot of Figure~\ref{f:pd.mu} we investigate the variance of the payments. In fact, The variance is excessively high for small values of $\mu$, making the adoption of these value inappropriate. Thus, the choice of $\mu$ should consider both these two dimensions of the problem: the regret and the variance of the payments.

\subsubsection{Unknown $\{\Lambda_m\}_{m \in \K}$ and $\{q_i\}_{i \in \N}$}

In this section we analyze the bound provided in Section~\ref{ssec:uql} for position--dependent auctions where both the prominences and the qualities are unknown. For these simulations we generate $\{\lambda_m\}_{m \in \K}$ samples from a uniform distribution over $[0.5,1]$. In the simulations we adopted the values of $\tau$, $\delta$ and $\mu$ derived for the bound. In particular, in order to balance the increase of variance of the payments when $\mu$ decreases, the number of rounds is not constant, but it changes as a function of $\mu$, i.e. $\frac{1000}{\mu}$. This means that, in expectation, the bid of a generic ad $a_i$ is modified $1000$ times over the number of the rounds.

\begin{figure*}[t]
\begin{center}
\includegraphics[width=0.45\textwidth]{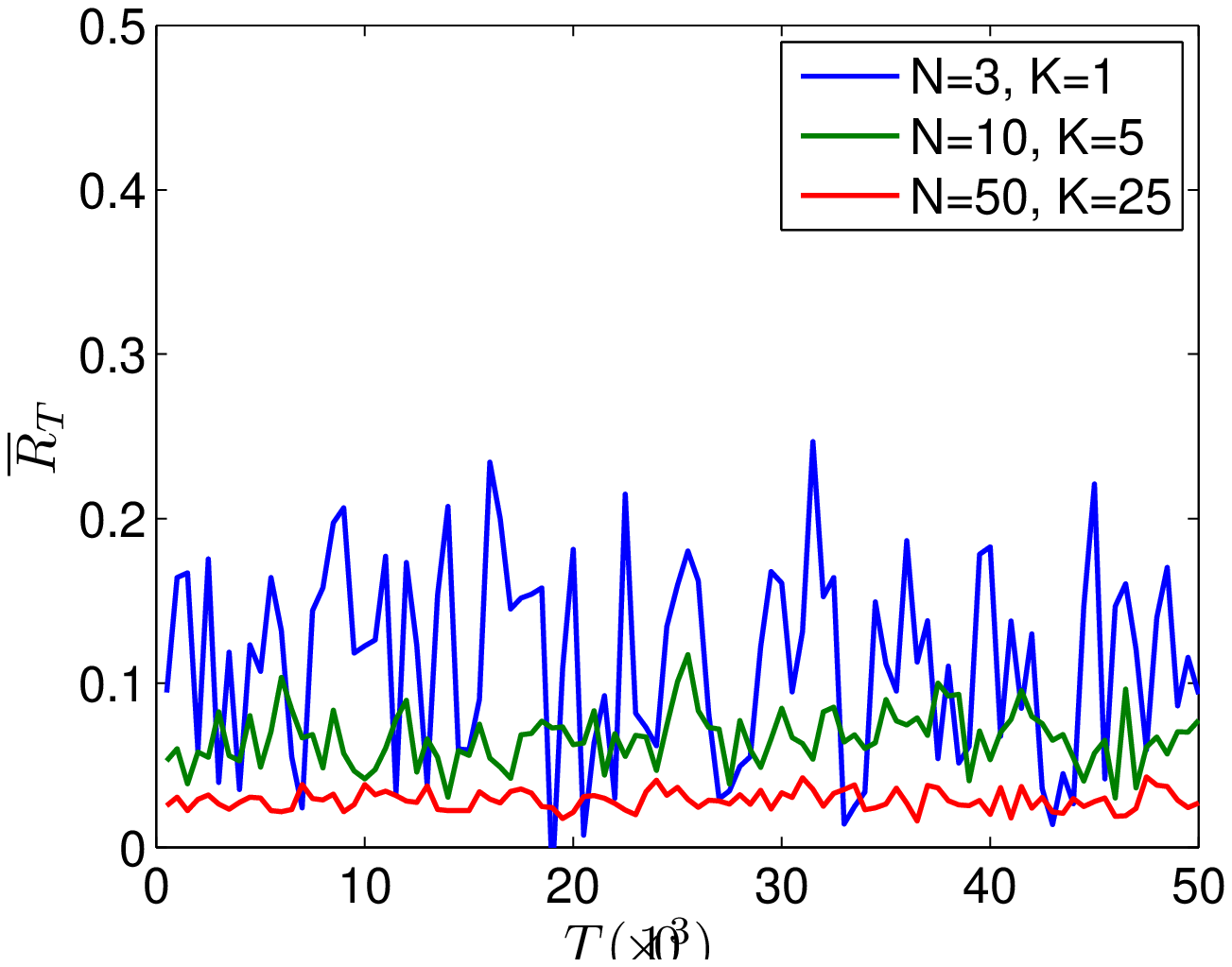}
\includegraphics[width=0.45\textwidth]{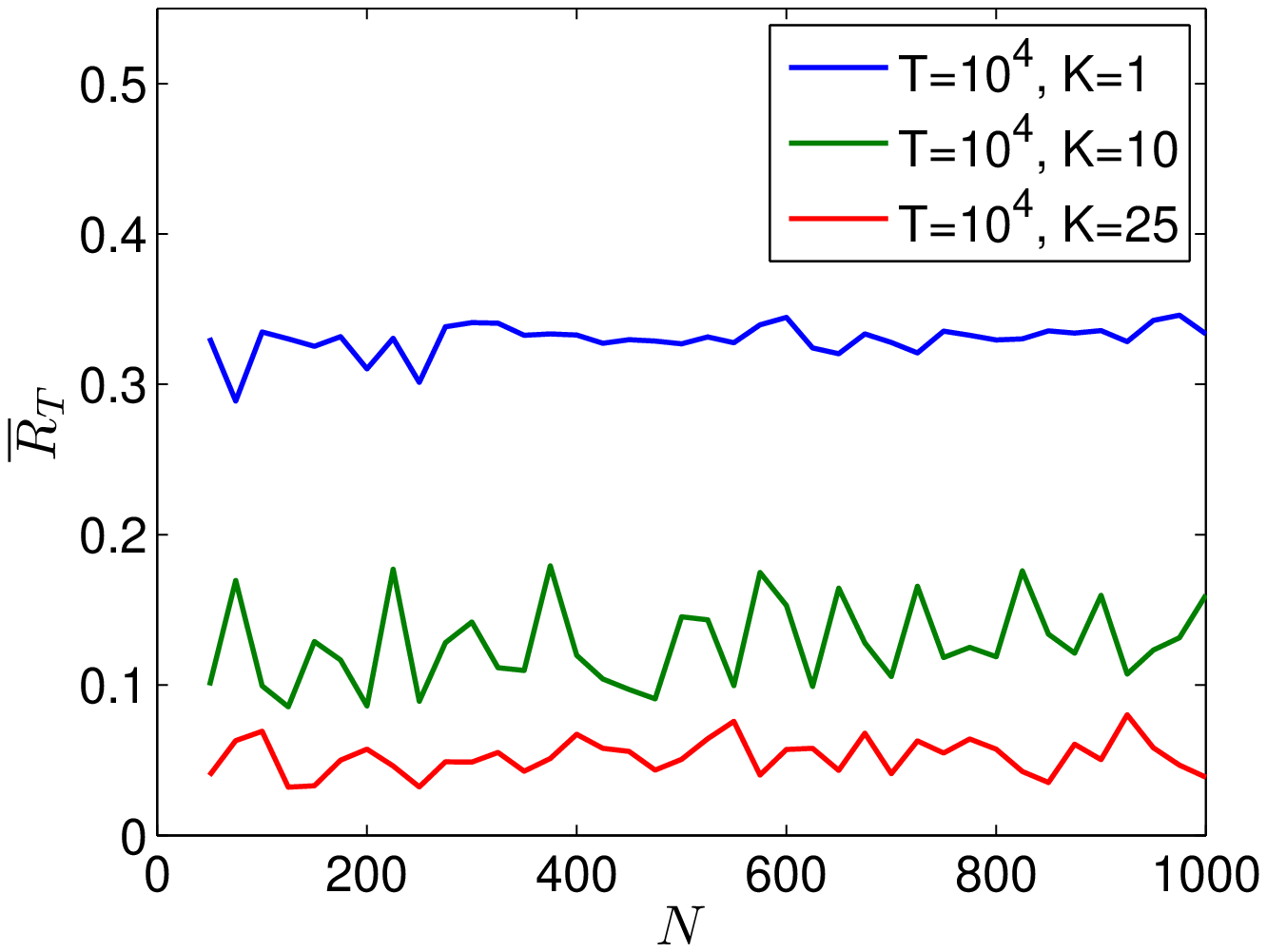}
\end{center}
\vspace{-0.4cm}
\caption{Position--dependent externalities with unknown $\{q_i\}_{i \in \N}$ and $\{\Lambda_m\}_{m \in \K}$. Dependency of the relative regret on $T$, $N$.}\label{f:ql}
\vspace{-0.4cm}
\end{figure*}

In the plots of Fig.~\ref{f:ql}, we show that the bound~(\ref{eq:regret.posdep.qlu}) accurately predicts the dependence of the regret w.r.t. the parameters $T$ and $N$. Indeed, except for the white noise due to the high variance of the payments based on the cSRP, the two plots shows that fixing the other parameters, the ratio $\bR_T$ is constant as $T$ and $N$ increase, respectively.

\begin{figure*}[t]
\begin{center}
\includegraphics[width=0.45\textwidth]{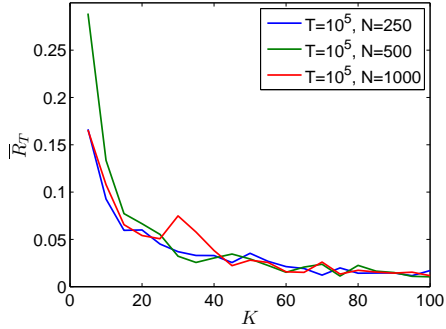}
\end{center}
\vspace{-0.4cm}
\caption{Position--dependent externalities with unknown $\{q_i\}_{i \in \N}$ and $\{\Lambda_m\}_{m \in \K}$. Dependency of the relative regret on $K$.}\label{f:ql_depK}
\vspace{-0.4cm}
\end{figure*}

The plot in Fig.~\ref{f:ql_depK} represents the dependency of the relative regret w.r.t. the parameter $K$. We can deduce that the bound $R_T$ over--estimate the dependency on $K$ for small values of the parameters, while, with larger values, the bound accurately predicts the behavior, the curves being flat.


\subsection{Position/Ad--Dependent Externalities}\label{s:exp.externalities}

\begin{figure*}[t]
\begin{center}
\includegraphics[width=0.45\textwidth]{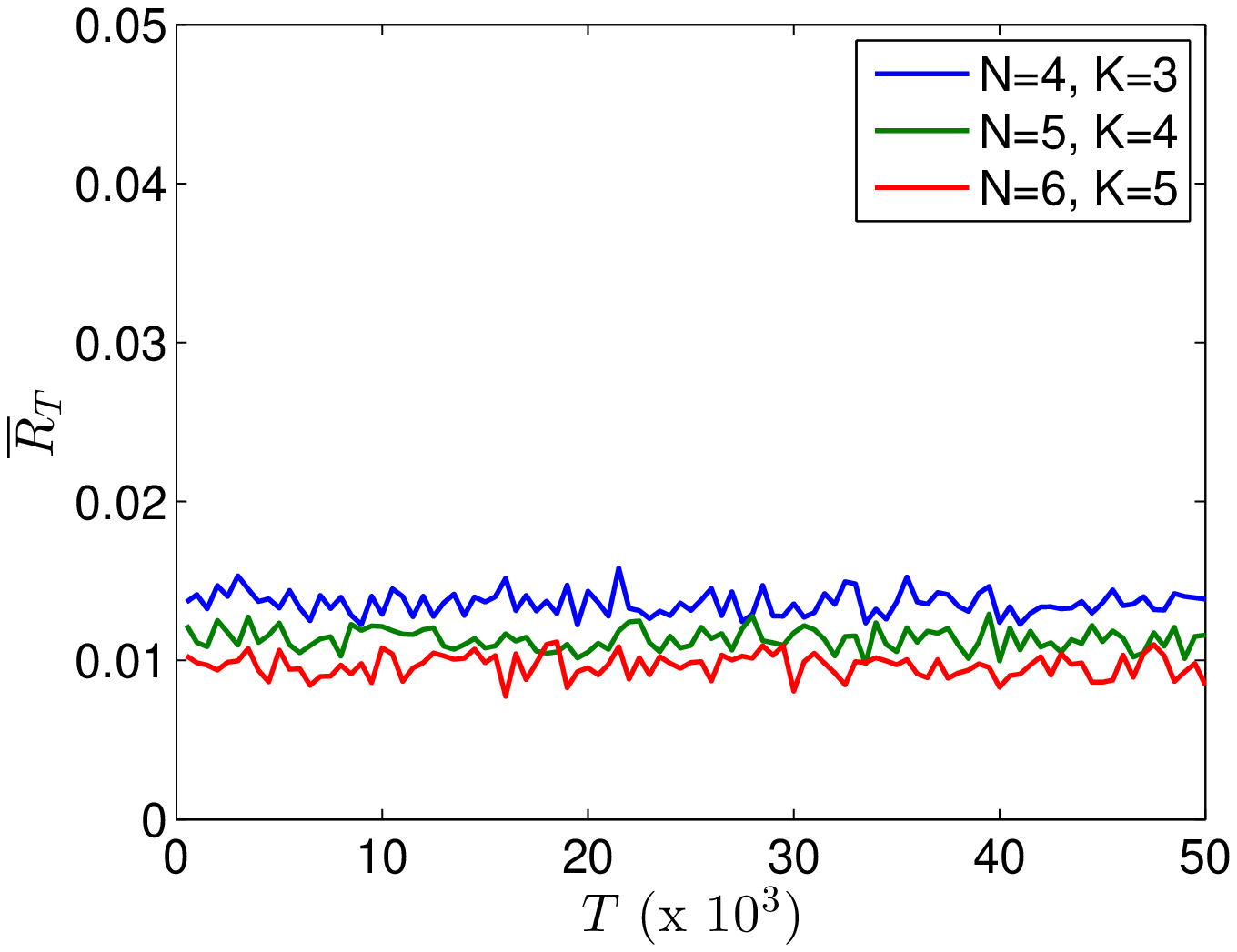}
\includegraphics[width=0.45\textwidth]{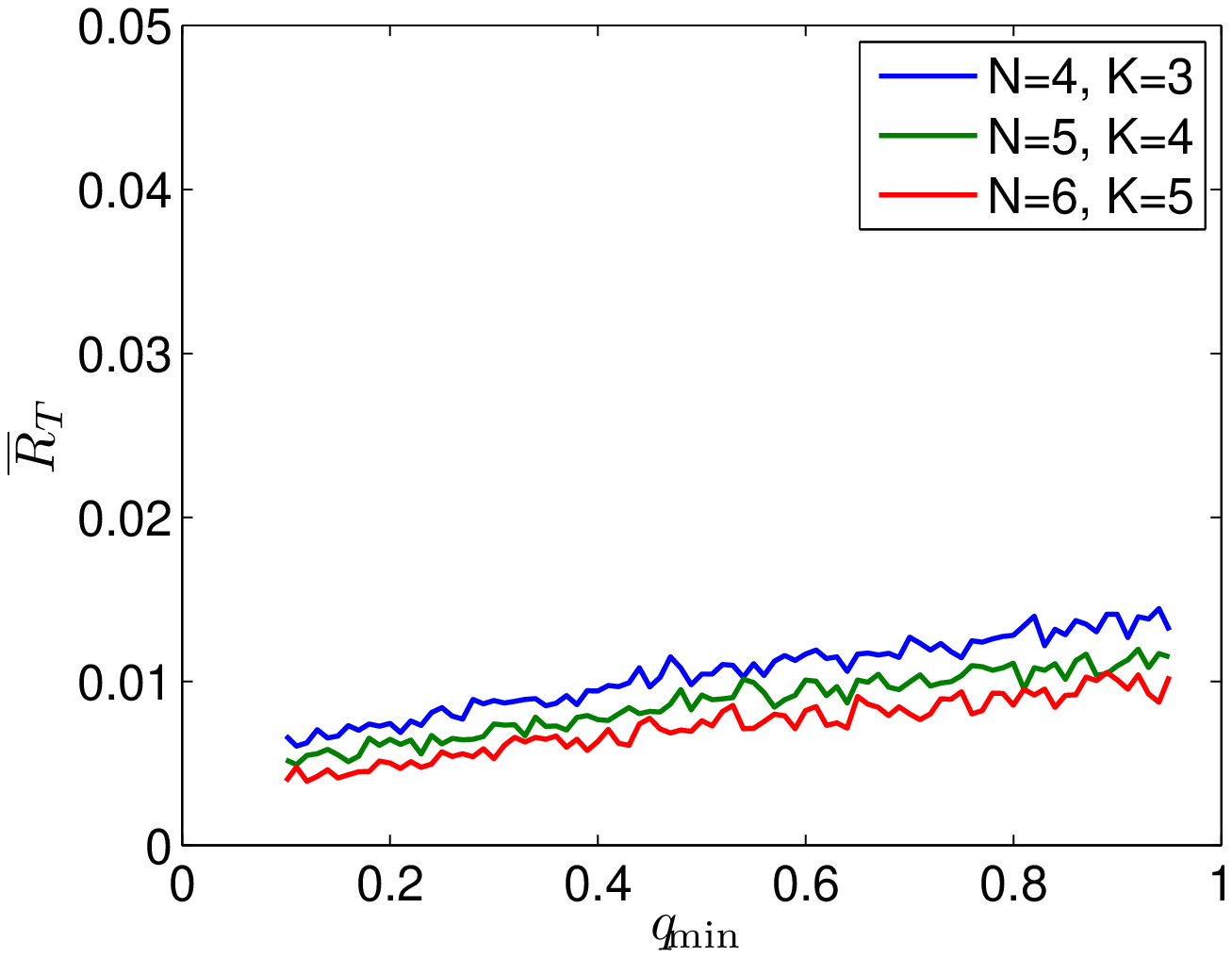}
\end{center}
\vspace{-0.4cm}
\caption{Dependency on $T$ and $q_{\min}$ in auctions with position/ad--dependent externalities.}\label{f:extern}
\vspace{-0.4cm}
\end{figure*}

\begin{figure}[t]
\begin{center}
\includegraphics[width=0.45\textwidth]{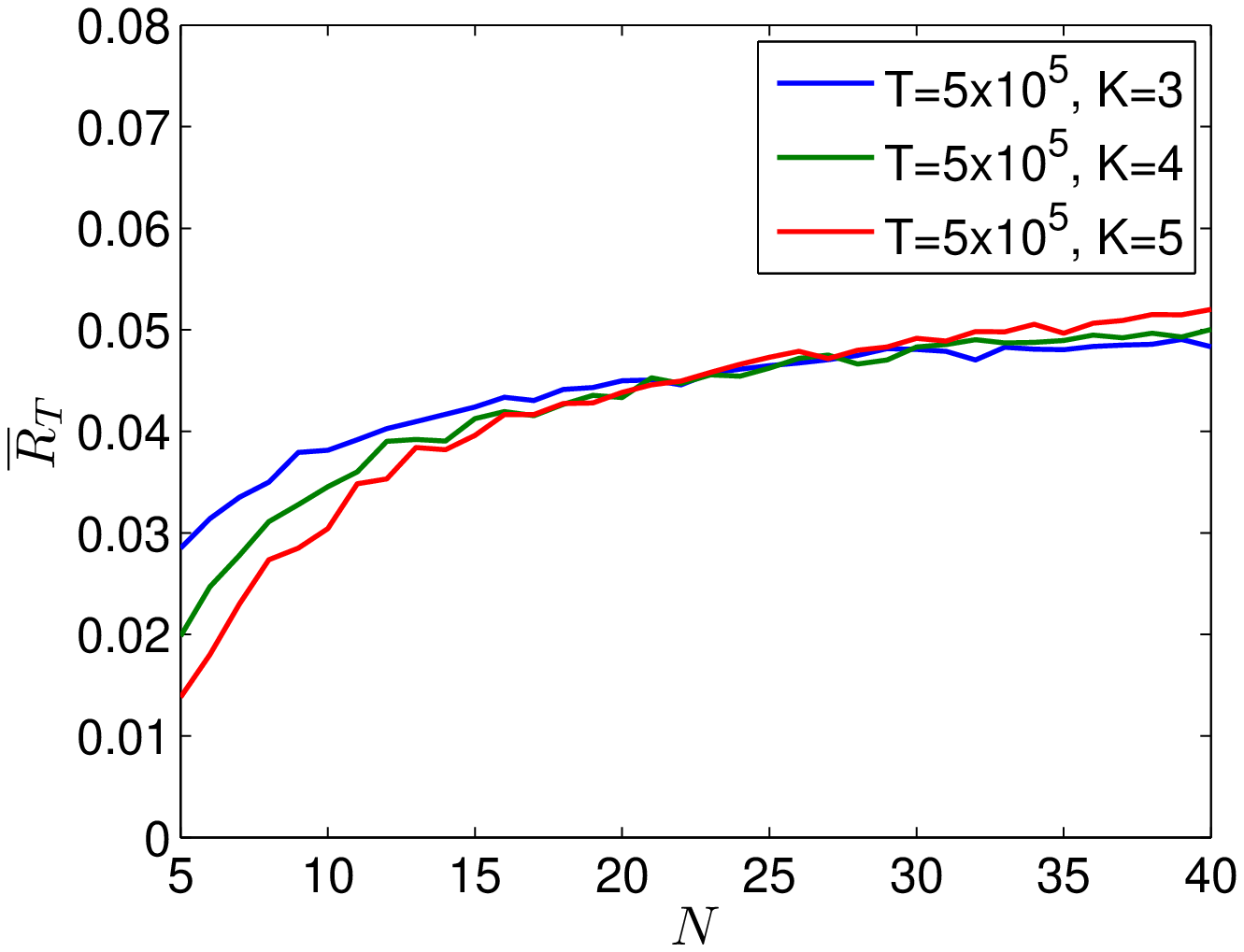}
\end{center}
\vspace{-0.4cm}
\caption{Dependency of the relative regret $\bR_T$ on $N$.}\label{f:add-extern}
\vspace{-0.2cm}
\end{figure}

In this section we analyze the bound provided in Section 5.1 for auctions with position--dependent and ad--dependent externalities where both only the qualities are unknown.

In the bound provided in Theorem~\ref{thm:extern} the regret $R_T$ presents a linear dependency on $N$ and an inverse dependency on the smallest quality $q_{\min}$. The relative regret $\bR_T$ is now defined as $R_T/B$ where $B$ is  bound (\ref{eq:regret.extern}). In the left plot of Fig.~\ref{f:extern} we report $\bR_T$ as $T$ increases. As it can be noticed, the bound accurately predicts the behavior of the regret w.r.t. $T$ as in the case of position--dependent externalities. In the right plot of Fig.~\ref{f:extern} we report $\bR_T$ as we change $q_{\min}$. According to the bound in (\ref{eq:regret.extern}) the regret should decrease as $q_{\min}$ increases (i.e., $R_T \leq \tilde O(q_{\min}^{-1}$)) but it is clear from the plot that $R_T$ has a much smaller dependency on $q_{\min}$, if any\footnote{From this experiment is not clear whether $\bR_T = \tilde O(q_{\min}$), thus implying that $R_T$ does not depend on $q_{\min}$ at all, or $\bR_T$ is sublinear in $q_{\min}$, which would correspond to a dependency $R_T = \tilde O(q_{\min}^{-f})$ with $f<1$.}. Finally, we study the dependency on $N$ (Figure~\ref{f:add-extern}). In this case $\bR_T$ slightly increases and then it tends to flat as $N$ increases.  This result suggests that the, theoretically derived, $N^{1/3}$ dependency of $R_T$ w.r.t. the number of ads might be correct.
We do not report results on $K$ since the complexity of finding the optimal allocation $f^*$ becomes intractable for values of $K$ larger than~8, as shown in~\cite{aamas2013}, making the empirical evaluation of the bound impossible.

%% file: sec/07conclusion.tex

\section{Conclusions and Future Work}\label{s:conclusions}

In this paper, we studied the problem of learning the click through rates of ads in sponsored search auctions with truthful mechanisms. This problem is highly challenging, combining online learning tools (i.e., regret minimization algorithms) together with economic tools (i.e., truthful mechanisms). While almost all the literature focused on single--slot scenarios, here we focused on multi--slot scenarios. With multiple slots it is necessary to adopt a user model to characterize the valuations of the users over the different slots. Here, we adopted the cascade model, that is the most common model used in the literature. In the paper, we studied a number of scenarios, each with a specific information setting of unknown parameters. For each scenario, we designed a truthful learning mechanism, studied its economic properties, derived an upper bound over the regret, and, for some mechanisms, also a lower bound. We considered both the regret over the auctioneer's revenue and the social welfare. 

We showed that for the cascade model with only position--dependent externalities it is possible to design a truthful no--regret  learning mechanism for the general case in which all the parameters are unknown. Our mechanism presents a regret $O(T^{2/3})$ and it is incentive compatible in expectation over the random component of the mechanism. However, it remains open whether or not it is possible to obtain a regret $O(T^{1/2})$. For specific sub cases, in which some parameters are known to the auctioneer, we obtained better results in terms of either incentive compatibility, obtaining dominant strategy truthfulness, or regret, obtaining a regret of zero. We showed that for the cascade model with the position-- and ad--dependent externalities it is possible to design a dominant strategy truthful mechanism with a regret $O(T^{2/3})$ when only the quality is unknown. Instead, even when the cascade model is only with ad--dependent externalities and no parameter is known it is not possible to obtain a no--regret dominant strategy truthful mechanism. The proof of this result would seem to suggest that the same result holds also when truthfulness is in expectation. However, we did not produce any proof for that, leaving it for future works. Finally, we empirically evaluated the bounds we provided, showing that the dependency of each bound from the parameters is empirically confirmed.

Two main questions deserve future investigation. The first question concerns the study of a lower bound for the case in which there are only position--dependent externalities for different notions of truthfulness in expectation, e.g., both in expectation over the click realizations and in expectation over the random component of the mechanism. Furthermore, it is open whether the separation of exploration and exploitation phases is necessary and, in the negative case, whether it is possible to obtain a regret $O(T^{1/2})$. The second question concerns a similar study related to the case with only ad--dependent externalities. 

%% file: sec/AppA-monotonicity.tex

\section{Monotonicity and Myerson's payments} \label{ap:monotonicity}

Consider a generic direct--revelation mechanism $M = (\N, \Theta, V, f, \{p_i\}_{i \in \N})$ as defined in Section~\ref{ssec:md}. A single--parameter linear environment is such that
\begin{itemize}
	\item the type of each agent $i$ is a scalar $v_i$ (single--parameter assumption),
	\item the utility function of agent~$i$ is $u_i(\hbv) = z_i(f(\hbv)) v_i - p_i(\hbv)$ where $z_i: \Theta \rightarrow \Re$ is a function of the allocation (linear assumption).
\end{itemize} 
An allocation function $f$ is \textit{monotone} in a single--parameter linear environment if 
\begin{align*}
z_i(f(\hbv_{-i}, v_i'')) \geq z_i(f(\hbv_{-i}, v_i'))
\end{align*} 
for any $v_i''\geq v_i'$. Essentially, $z_i$ is monotonically increasing in $v_i$ once $\hbv_{-i}$ has been fixed. In such environments, it is always possible to design a DSIC mechanism imposing the following payments~\cite{tardos_sp}:
\begin{equation} \label{eq:p_tardos}
p_i(\hbv) = h_i(\hbv_{-i}) + z_i(f(\hbv)) \hv_i - \int_{0}^{\hv_i} z_i(f(\hbv_{-i}, u)) du
\end{equation}
\noindent where $h_i(\hbv_{-i})$ is a generic function not depending on the type of agent~$i$. 

%% file: sec/App-proofs.tex



\section{Proof of Revenue Regret in Theorem~\ref{thm:constant}}

\noindent We start by reporting the  proof of Proposition~\ref{p:hoeffding}.

\begin{pf}\textit{(Proposition~\ref{p:hoeffding})}
The derivation is a simple application of the Hoeffding's bound. We first notice that each of the terms in the empirical average $\hq_i$ (\ref{eq:est.q}) is bounded in $[0; 1/\Lambda_{\pi(i;\theta_t)}]$. Thus we obtain
\begin{equation*}
\mathbb{P}\left( |q_i - \hq_i | \geq \epsilon \right) \leq 2 \exp\bigg(-\frac{2|B_i|^2\epsilon^2}{\sum_{t \in B_i}\big(\frac{1}{\Lambda_{\pi(i;\theta_t)}}-0\big)^2}\bigg) = \frac{\delta}{N}.
\end{equation*}
By reordering the terms in the previous expression we have
\begin{align*}
\epsilon &= \sqrt{\left(\sum_{t \in B_i}\frac{1}{\Lambda^2_{\pi(i;\theta_t)}}\right) \frac{1}{2|B_i|^2} \log{\frac{2N}{\delta}}},
\end{align*}
which guarantees that all the empirical estimates $\hq_i$ are within $\epsilon$ of $q_i$ for all the ads with probability, at least, $1 - \delta$.
\qed
\end{pf}


Before stating the main result of this section, we need the following technical lemma.

\begin{lemma}\label{lem:ratio}
For any slot $s_m$ with $m \in \K$, with probability $1-\delta$,
\begin{equation}\label{eq:ratio}
		\frac{\max\limits_{i\in \mathcal{N}} (q_i \hv_i; m)}{\max\limits_{i\in \mathcal{N}} (\hqp_i \hv_i; m)} \leq 1,
\end{equation}
where the operator $\max(\cdot;\cdot)$ is defined as in Section~\ref{s:constant}.
\end{lemma}

\begin{pf}
The proof is a straightforward application of Proposition~\ref{p:hoeffding}.
We consider the optimal allocation $\theta^*$ defined in (\ref{eq:efficient-alloc}) and the estimated allocation $\tilde{\theta}$ defined in (\ref{eq:optimalallocationestimatedq}). We denote $h = \alpha(m;\theta^*) = \arg \max\limits_{i \in \N}(q_i \hv_i; m)$, i.e., the index of the ad allocated in a generic slot in position $m$. There are two possible scenarios:
\begin{itemize}
		\item If $\pi(h;\tilde{\theta}) < m$ (the ad is displayed into a higher slot in the approximated allocation $\tilde{\theta}$), then
		$\exists j \in \N$ s.t. $\pi(j;\theta^*) < m \wedge \pi(j;\tilde{\theta}) \geq m$. Thus
		$$ \max\limits_{i \in \N}(\hqp_i \hv_i; m) \geq \hqp_j \hv_j \geq q_j \hv_j \geq q_h \hv_h = \max\limits_{i \in \N}(q_i \hv_i; m)$$
		where the second inequality holds with probability $1 - \delta$;
		\item If $\pi(h;\tilde{\theta}) \geq m$ (the ad is displayed into a lower or equal slot in the approximated allocation $\tilde{\theta}$), then
		$$ \max\limits_{i \in \N}(\hqp_i \hv_i; m) \geq \hqp_h \hv_h \geq q_h v_h = \max\limits_{i \in \N}(q_i \hv_i; m)$$
		where the second inequality holds with probability $1 - \delta$.
\end{itemize}
In both cases, the statement follows.\qed
\end{pf}


\begin{pf}\textit{(Theorem~\ref{thm:constant})}

\noindent\textbf{Step 1: expected payments.} 
The proof follows steps similar to those in~\cite{devanur2009price}. We first recall that for any ad $a_i$ such that $\pi(i; \theta^*)\leq K$, the expected payments of the VCG mechanism in this case reduce to (\ref{eq:pay.opt.click.vcg.posdep.ad}):
\begin{align*}
p^*_i(\hat{\mathbf{v}}) = \sum_{l=\pi(i; \theta^*)+1}^{K+1} \left[(\Lambda_{l-1} - \Lambda_l) \max\limits_{j\in \mathcal{N}}(q_j \hv_j; l)\right],
\end{align*}
while, given the definition of \avcg1 \ reported in Section~\ref{s:constant}.1, the expected payments for at $t$--th iteration of the auction are
\begin{align*}
\tp_i(\hbv) = \left\{
  \begin{array}{ll}
    0 & \text{if } t \leq \tau \text{ (\textit{exploration})}\\
    \tilde{p}_{i}(\hbv) & \text{if } t > \tau \text{ (\textit{exploitation})}
  \end{array} \right.
\end{align*}
where the payment for any ad $a_i$ such that $\pi(i; \tilde\theta)\leq K$ is defined in (\ref{eq:hpay.const}) as
\begin{align*}
\tilde p_i (\hbv) &=\frac{q_i}{\hqp_i} \sum_{l=\pi(i;\tilde{\theta})+1}^{K+1} (\Lambda_{l-1} - \Lambda_l) \max\limits_{j \in \N}(\hqp_j \hv_j; l).
\end{align*}

\noindent\textbf{Step 2: exploration regret.} Since for any $t\leq \tau$ \avcg\ sets all the payments to 0 the per--round regret is
\begin{align}\label{eq:exploration.regret}
r_t = \sum_{m = 1}^K (p^*_{\alpha(m;\theta^*)}(\hbv) - 0) = \sum_{m = 1}^K \sum_{l = m}^{K} \Delta_l \max_{i\in\N}(q_i\hv_i; l+1) \leq \vmax\sum_{m = 1}^K \Lambda_m,
\end{align}
where $\Delta_l = \Lambda_l - \Lambda_{l+1}$. The exploration regret is obtained by summing up $r_t$ over $\tau$ steps.

\noindent\textbf{Step 3: exploitation regret.} Now we focus on the expected (w.r.t. clicks) per--round regret during the exploitation phase. According to the definition of payments, at each round $t \in \{\tau + 1, \ldots,  T\}$ of the exploitation phase we bound the per--round regret $r_t$ as
\begin{align*}
		r_t &= \sum_{m = 1}^K (p^*_{\alpha(m;\theta^*)}(\hbv) - \tilde{p}_{\alpha(m;\tilde{\theta})}(\hbv)) \\
		&= \sum_{m = 1}^K \sum_{l = m}^{K} \Delta_l \left( \max\limits_{i\in \mathcal{N}}(q_i \hv_i; l+1) - \frac{ \max\limits_{i\in \mathcal{N}}(\hqp_i \hv_i; l+1) }{\hqp_{\alpha(m;\tilde{\theta})}} q_{\alpha(m;\tilde{\theta})} \right) \\
		&= \sum_{m = 1}^K \sum_{l = m}^{K} \Delta_l \frac{ \max\limits_{i\in \mathcal{N}}(\hqp_i \hv_i; l+1)}{\hqp_{\alpha(m;\tilde{\theta})}} \left( \frac{ \max\limits_{i\in \mathcal{N}}(q_i \hv_i;l+1)}{\max\limits_{i\in \mathcal{N}}(\hqp_{i} \hv_i;l+1)} \hqp_{\alpha(m;\tilde{\theta})} - q_{\alpha(m;\tilde{\theta})} \right) \\
		&= \sum_{m = 1}^K \sum_{l = m}^{K} \Delta_l\frac{ \max\limits_{i\in \N}(\hqp_{i} \hv_i;l+1)}{\max\limits_{i\in \N}(\hqp_{i} \hv_i;m)} \hv_{\alpha(m; \tilde{\theta})} \left( \frac{\max\limits_{i \in \N}(q_i \hv_i;l+1)}{\max\limits_{i\in \N}(\hqp_{i} \hv_i;l+1)} \hqp_{\alpha(m;\tilde{\theta})} - q_{\alpha(m;\tilde{\theta})} \right).
\end{align*}
By definition of the max operator, since $l+1 > m$, it follows that 
\begin{align}\label{eq:step.loose}
\frac{\max\limits_{i \in \N} (\hqp_{i} \hv_i;l+1)}{\max\limits_{i \in \N} (\hqp_{i} \hv_i;m)} \leq 1.
\end{align}
Finally, from Lemma~\ref{lem:ratio} and $\hv_{\alpha(m; \tilde{\theta})} \leq \vmax$, it follows that
\begin{align} \label{eq:boundVCG.exactvsest}
r_t \leq \sum_{m = 1}^K \sum_{l = m}^{K} \vmax \Delta_l(\hqp_{\alpha(m;\tilde{\theta})} - q_{\alpha(m;\tilde{\theta})}) \leq \vmax \sum_{m = 1}^K \Big[ (\hqp_{\alpha(m;\tilde{\theta})} - q_{\alpha(m;\tilde{\theta})})\sum_{l = m}^{K} \Delta_l\Big],
\end{align}
with probability at least $1-\delta$.
Notice that, by definition of $\Delta_l$, $\sum_{l=m}^{K} \Delta_l = \Lambda_{m} - \Lambda_{K+1} = \Lambda_{m}$. Furthermore, from the definition of $\hqp_i$ and using (\ref{eq:eta}) we have that for any ad $a_i$, $\hqp_{i} - q_i = \hq_i - q_i + \eta \leq 2\eta$, with probability at least $1 - \delta$. Thus, the difference between the payments becomes\footnote{Notice that in the logarithmic term the factor of 2 we have in Proposition~\ref{p:hoeffding} disappears since in this proof we only need the one-sided version of it.}
\begin{align}\label{eq:brs4}
r_t & \leq 2 \vmax\eta\sum_{m = 1}^K \Lambda_m \leq 2\vmax \left(\sum_{m = 1}^K \Lambda_m\right)\sqrt{\Bigg(\sum_{m=1}^{K} \frac{1}{\Lambda_{m}^2}\Bigg) \frac{2N}{K^2 \tau} \log \frac{N}{\delta}}.
\end{align}
with probability $1-\delta$.

\noindent\textbf{Step 4: global regret.} By summing up the regrets reported in (\ref{eq:exploration.regret}) and (\ref{eq:brs4}), we obtain
\begin{align*}
R_T &\leq \vmax \left(\sum_{m = 1}^K \Lambda_m \right)\Bigg( 2(T - \tau) \sqrt{\left(\sum_{m=1}^{K} \frac{1}{\Lambda_{m}^2}\right) \frac{2N}{K^2 \tau} \log \frac{N}{\delta}} + \tau + \delta T \Bigg),
\end{align*}
that can be further simplified give that $\sum_{m = 1}^K \Lambda_m\leq K$ as
\begin{equation}\label{eq:bound.thm.constant}
R_T \leq \vmax K\Bigg( 2(T - \tau) \sqrt{\left(\sum_{m=1}^{K} \frac{1}{\Lambda_{m}^2}\right) \frac{2N}{K^2 \tau} \log \frac{N}{\delta}} + \tau + \delta T \Bigg).
\end{equation}

\noindent\textbf{Step 5: parameters optimization.} 
Beside describing the performance of \avcg1, the previous bound also provides guidance for the optimization of the parameters $\tau$ and $\delta$. We first simplify the bound in (\ref{eq:bound.thm.constant}) as
\begin{align}\label{eq:bound.thm.constant.simple}
R_T &\leq \vmax K\bigg( 2T \sqrt{\left(\sum_{m=1}^{K} \frac{1}{\Lambda_{m}^2}\right) \frac{2N}{K^2 \tau} \log \frac{2N}{\delta}} + \tau + \delta T \bigg) \notag\\
&\leq \vmax K\bigg( \frac{2T}{\Lambda_{\min}} \sqrt{\frac{2N}{K \tau} \log \frac{N}{\delta}} + \tau + \delta T \bigg),
\end{align}
where we used $\tau \leq T$ and $\sum\limits_{m=1}^{K} 1/\Lambda_{m}^2\leq K/\Lambda_{\min}^2$, with $\Lambda_{\min} = \min_{m \in \K} \Lambda_m$. In order to find the optimal value of $\tau$, we take the derivative of the previous bound w.r.t. $\tau$ and set it to zero and obtain
\begin{align*}
\vmax K\Big(-\tau^{-\frac{3}{2}} \frac{T}{\Lambda_{\min}} \sqrt{\frac{2N}{K} \log \frac{N}{\delta}} + 1\Big) = 0,
\end{align*}
which leads to
\begin{equation*}
\tau = 2^\frac{1}{3} K^{-\frac{1}{3}} T^{\frac{2}{3}} N^{\frac{1}{3}} \Lambda_{\min}^{-\frac{2}{3}} \Big( \log{\frac{N}{\delta}}\Big)^{\frac{1}{3}}.
\end{equation*}
Substituting this value of $\tau$ into (\ref{eq:bound.thm.constant.simple}) leads to the optimized bound
\begin{align*}
R_T &\leq v_{\max}  K \bigg(3 \cdot 2^\frac{1}{3} K^{-\frac{1}{3}} T^{\frac{2}{3}} N^{\frac{1}{3}} \Lambda_{\min}^{-\frac{2}{3}} \Big( \log{\frac{N}{\delta}}\Big)^{\frac{1}{3}} + \delta T\bigg).\end{align*}
We are now left with the choice of the confidence parameter $\delta\in (0,1)$, which can be easily set to optimize the asymptotic rate (i.e., ignoring constants and logarithmic factors) as
\[
		\delta = K^{-\frac{1}{3}} T^{-\frac{1}{3}} N^{\frac{1}{3}}\\
\]
\noindent with the trivial constraint that $T > \frac{N}{K}$ (given by $\delta < 1$). We thus obtain the final bound
\[
		R_T \leq 4 \cdot 2^\frac{1}{3} \vmax \Lambda_{\min}^{-\frac{2}{3}} K^\frac{2}{3} T^\frac{2}{3} N^\frac{1}{3} \left[ \log \big( K^\frac{1}{3} T^\frac{1}{3} N^\frac{2}{3} \big)\right]^{\frac{1}{3}},
\]
\noindent which concludes the proof.\qed
\end{pf}



\section{Proof of Revenue Regret in Theorem~\ref{thm:constant.l.baba}}

Unlike the setting considered in Theorem~\ref{thm:constant}, here the regret is only due to the use of a randomized mechanism, since no parameter estimation is actually needed.

\begin{pf}\textit{(Theorem~\ref{thm:constant.l.baba})}

\noindent\textbf{Step 1: payments and additional notation.}
We recall that according to~\cite{tardos_sp} and~\cite{greenLaffont} the expected VCG payments can be written as in (\ref{eq:pay.vcg.emp.tardos}) in the form
\begin{align*}
p^*_i(\hat{\mathbf{v}}) = \Lambda_{\pi(i;f^*(\hbv))} q_i \hv_i - \int_{0}^{\hv_i} \Lambda_{\pi(i;f^*(\hbv_{-i},u))} q_i du,
\end{align*}
while the A--VCG2$^\prime$ mechanism prescribes contingent payments as in (\ref{eq:pay.opt.babaioff.click}), which lead to expected payments
\begin{align}\label{eq:pay.babaioff.exp}
p_i^{B,*}(\hbv) &= \E_{\bx}\big[\Lambda_{\pi(i;f^*(\bx))}|\hbv\big]q_i \hv_i - \int_{0}^{\hv_i} \E_{\bx}\big[\Lambda_{\pi(i;f^*(\bx))}|\hbv_{-i},u\big]q_i du.
\end{align}
Given the randomness of the allocation function of A--VCG2$^\prime$, we need to introduce the following additional notation:
\begin{itemize}
\item $\bs \in \{0,1\}^N$ is a vector where each element $s_i$ denotes whether  the $i$--th bid has been preserved or it has been modified by the self--resampling procedure, i.e., if $x_i=\hv_i$ then $s_i=1$, otherwise if $x_i < \hv_i$ then $s_i=0$. Notice that  $\bs$ does not provide information about the actual modified values $\bx$;
\item $\mathbb{E}_{\bx|\bs}[\Lambda_{\pi(i; f(\bx))}|\hbv]$ is the expected value of prominence associated with the slots allocated to ad $a_i$ conditioned on the declared bids $\hbv$ being perturbed as in $\bs$. 
\end{itemize}
Let $S = \{\bs | \pi(i;f^*(\hbv))\leq K + 1 \Rightarrow s_i = 1\ \forall i \in \N  \}$ be all the realizations where the self--resampling procedure does not modify the bids of the first $K+1$ ads, i.e., the $K$ ads displayed applying $f^*$ to the true bids $\hbv$ and the first non-allocated ad.

\noindent\textbf{Step 2: the regret.}
We proceed by studying the per--ad regret $r_i(\hbv) = p_i^*(\hbv) - p_i^{B,*}(\hbv)$. Given the previous definitions, we rewrite the expected payments $p_i^{B,*}(\hbv)$ as
\begin{align*}
p_i^{B,*}(\hbv) &= \bigg(\Prob[\bs \in S] \Lambda_{\pi(i;f^*(\hbv))} + \Prob[\bs \not \in S] \E_{\bx|\bs \not \in S}[\Lambda_{\pi(i;f^*(\bx))}|\hbv]\bigg)q_i \hv_i \\
&\quad - \int_{0}^{\hv_i} \bigg(\Prob[\bs \in S] \Lambda_{\pi(i; f^*(\hbv_{-i},u))} + \Prob[\bs \not \in S] \E_{\bx|\bs\neq\bone}[\Lambda_{\pi(i;f^*(\bx))}|\hbv_{-i},u]\bigg)q_i du\\
&=\Prob[\bs \in S] \bigg(\Lambda_{\pi(i;f^*(\hbv))}q_i \hv_i - \int_{0}^{\hv_i}\Lambda_{\pi(i; f^*(\hbv_{-i},u))} q_i du\bigg)\\
&\quad+\Prob[\bs \not \in S] \bigg(\E_{\bx|\bs \not \in S}[\Lambda_{\pi(i;f^*(\bx))}|\hbv]q_i \hv_i - \int_{0}^{\hv_i}\E_{\bx|\bs \not \in S}[\Lambda_{\pi(i;f^*(\bx))}|\hbv_{-i},u] q_i du\bigg)\\
&=\Prob[\bs \in S] p_i^*(\hbv) \\
&\quad+ \Prob[\bs \not \in S] \bigg(\E_{\bx|\bs \not \in S}[\Lambda_{\pi(i;f^*(\bx))}|\hbv]q_i \hv_i - \int_{0}^{\hv_i}\E_{\bx|\bs \not \in S}[\Lambda_{\pi(i;f^*(\bx))}|\hbv_{-i},u] q_i du \bigg),
\end{align*}
where in the last expression we used the expression of the VCG payments in (\ref{eq:pay.vcg.emp.tardos}) according to~\cite{tardos_sp} and~\cite{greenLaffont}. 
The per--ad regret is
\begin{align*}
r_i&(\hbv) = p_i^*(\hbv) - p_i^{B,*}(\hbv) \\
&= p_i^*(\hbv) - \Prob[\bs \in S] p_i^*(\hbv) \\
&\quad- \Prob[\bs \not \in S] \bigg(\E_{\bx|\bs \not \in S}[\Lambda_{\pi(i;f^*(\bx))}|\hbv]q_i \hv_i - \int_{0}^{\hv_i}\E_{\bx|\bs \not \in S}[\Lambda_{\pi(i;f^*(\bx))}|\hbv_{-i},u] q_i du\bigg)\\
&= \Prob[\bs \not \in S] p_i^*(\hbv) \\
&\quad- \Prob[\bs \not \in S] \underbrace{\bigg(\E_{\bx|\bs \not \in S}[\Lambda_{\pi(i;f^*(\bx))}|\hbv]q_i \hv_i - \int_{0}^{\hv_i}\E_{\bx|\bs \not \in S}[\Lambda_{\pi(i;f^*(\bx))}|\hbv_{-i},u] q_i du\bigg)}_{r_{i,1}^B}.
\end{align*}
Since we have that $u \leq \hv_i$ in the integral and since the allocation function defined in~\cite{babaioff_impl_pay} is monotone, we have that
\begin{align*}
\E_{\bx | \bs \not \in S}[\Lambda_{\pi(i;f^*(\bx))}|\hbv_{-i},u] \leq \E_{\bx | \bs \not \in S}[\Lambda_{\pi(i;f^*(\bx))}|\hbv],
\end{align*}
which implies that $r_{i,1}^B$ is non--negative. Thus the regret $r_i^B$ can be bounded as
\begin{align}\label{bnd:rSRP}
r_i^B(\hbv) &= \Prob[\bs \not \in S] p_i^*(\hbv) \underbrace{ - \Prob[\bs \not \in S] r_{i,1}^B}_{\leq 0} \nonumber \\
&\leq \Prob[\bs \not \in S] p_i^*(\hbv) \leq \Prob\big[\exists j: s_j = 0 \land \pi(j;f^*(\hbv)) \leq K+1\big] \vmax \nonumber\\
&\leq \sum_{j\in \N: \pi(j;f^*(\hbv)) \leq K+1 }\Prob[s_j = 0] \vmax = \left( K + 1 \right) \mu \vmax \leq 2 K \mu \vmax.  
\end{align}
We can now compute the bound on the global regret $R_T$. Since this mechanism does not require any estimation phase, the regret is simply
\begin{align*}
R_T & \leq 2 K^2 \mu \vmax T.
\end{align*}

\noindent\textbf{Step 3: parameters optimization.} In this case, the bound would suggest to choose a $\mu \rightarrow 0$, but it is necessary to consider that with $\mu \rightarrow 0$ the variance of the payment goes to infinity.

\end{pf}


\section{Proof of Revenue Regret in Theorem~\ref{thm:constant.ql}}

The proof of Theorem~\ref{thm:constant.ql} needs to combine the result of Theorem~\ref{thm:constant.l.baba} and the regret due to the estimation of the parameters similarly to what is done in Theorem~\ref{thm:constant}.

\begin{pf}\textit{(Theorem~\ref{thm:constant.ql})}

\noindent\textbf{Step 1: payments and the regret.}
Similar to the proof of Theorem~\ref{thm:constant.l.baba}, we use the form of the VCG payments as in (\ref{eq:pay.vcg.emp.tardos}):
\begin{align*}
p^*_i(\hat{\mathbf{v}}) = \Lambda_{\pi(i;f^*(\hbv))} q_i \hv_i - \int_{0}^{\hv_i} \Lambda_{\pi(i;f^*(\hbv_{-i},u))} q_i du,
\end{align*}
while A--VCG3 uses the contingent payments in (\ref{eq:pay.babaioff.ppc}), which in expectation become
\begin{align}\label{eq:pay.babaioff.exp.tilde}
\tilde p_i^{B}(\hbv) &= \E_{\bx}\big[\Lambda_{\pi(i;\tilde f(\bx))}|\hbv\big]q_i \hv_i - \int_{0}^{\hv_i} \E_{\bx}\big[\Lambda_{\pi(i;\tilde f(\bx))}|\hbv_{-i},u\big]q_i du.
\end{align}
We also need to introduce the expected payments
\begin{align*}
\tilde p_i(\hat{\mathbf{v}}) = \Lambda_{\pi(i;\tilde f(\hbv))} q_i \hv_i - \int_{0}^{\hv_i} \Lambda_{\pi(i;\tilde f(\hbv_{-i},u))} q_i du,
\end{align*}
which correspond to the VCG payments except from the use of the estimated allocation function $\tilde f$ instead of $f^*$.

Initially, we compute an upper bound over the per--ad regret $r_i = p_i^* - p_i$ for each round of the exploitation phase and we later use this result to compute the upper bound for the regret over the whole time interval ($R_T$). We divide the per--ad regret in two different components:
\begin{align}
r_i(\hbv) &= p_i^*(\hbv) - \tp^B_i(\hbv) \\
&= \underbrace{p_i^*(\hbv) - p_i^{B,*}(\hbv)}_{\text{cSRP regret}} + \underbrace{p_i^{B,*}(\hbv) - \tp^B_i(\hbv)}_{\text{learning regret}} = r_i^B(\hbv) + r_i^L(\hbv) \nonumber,
\end{align}
where
\begin{itemize}
		\item $r_i^B(\hbv)$ is the regret due to the use of the approach proposed in~\cite{babaioff_impl_pay} instead of the VCG payments, when all the parameters are known;
		\item $r_i^L(\hbv)$ is the regret due to the uncertainty on the parameters when the payments defined in~\cite{babaioff_impl_pay} are considered.
\end{itemize}
For the definitions of $\bs$ and $\mathbb{E}_{\bx|\bs}[\Lambda_{\pi(i; f(\bx))}|\hbv]$  refer to the proof of Theorem~\ref{thm:constant.l.baba}.

\noindent\textbf{Step 2: the cSRP regret.}
We can reuse the result obtained in the proof of Theorem~\ref{thm:constant.l.baba}. In particular, we can use the bound in (\ref{bnd:rSRP}), i.e. $r_i^B(\hbv) \leq \left(K+1\right) \mu \vmax$. Given that we have assumed $N > K$, in the remaining parts of this proof we will use the following upper bound: $r_i^B(\hbv) \leq \left(K+1\right) \mu \vmax \leq N \mu \vmax $.

\noindent\textbf{Step 3: the learning regret.}
Similar to the previous step, we write the learning expected payments based on the cSRP in (\ref{eq:pay.babaioff.exp.tilde}) as
\begin{align*}
\tilde p^B_i(\hbv) =\Prob[\bs=\bone] \tp_i(\hbv) + \Prob[\bs\neq\bone] \bigg(\E_{\bx|\bs\neq\bone}[\Lambda_{\pi(i; \tf(\bx))}|\hbv] q_i \hv_i - \int_{0}^{\hv_i}\E_{\bx|\bs\neq\bone}[\Lambda_{\pi(i; \tf(\bx))}|\hbv_{-i},u] q_i du\bigg).
\end{align*}
Then the per-ad regret is
\begin{align*}
r_i^L(\hbv) &= p_i^{B,*}(\hbv) - \tilde p_i^B(\hbv)\\
&= \Prob[\bs=\bone] (p_i^*(\hbv)- \tp_i(\hbv)) +\\
&\quad + \Prob[\bs\neq\bone] \bigg(\underbrace{\E_{\bx|\bs\neq\bone}[\Lambda_{\pi(i;f^*(\bx))}|\hbv]q_i \hv_i - \int_{0}^{\hv_i}\E_{\bx|\bs\neq\bone}[\Lambda_{\pi(i;f^*(\bx))}|\hbv_{-i},u] q_i du}_{\leq \vmax} +\\
&\quad\quad\quad\quad\quad\quad\quad \underbrace{-\E_{\bx|\bs\neq\bone}[\Lambda_{\pi(i;\tf(\bx))}|\hbv]q_i \hv_i + \int_{0}^{\hv_i}\E_{\bx|\bs\neq\bone}[\Lambda_{\pi(i;\tf(\bx))}|\hbv_{-i},u] q_i du}_{= - r_{i,1}^B \leq 0}\bigg)\\
&\leq p_i^*(\hbv)- \tp_i(\hbv) +  N\mu \vmax.
\end{align*}
We now simply notice that payments $\tp_i$ are WVCG payments corresponding to the estimated allocation function $\tf$ and can be written as
\begin{align*}
\tp_i(\hbv) = \frac{q_i}{\hqp_i} \Big[\widetilde{SW}\big(\tf_{-i}\left(\hbv\right),\hbv\big) - \widetilde{SW}_{-i}\big(\tf\left(\hbv\right), \hbv\big)\Big],
\end{align*}
which allows us to use the results stated in proof of Theorem~\ref{thm:constant} and from (\ref{eq:boundVCG.exactvsest}) we can conclude that
\begin{align*}
\sum_{i: \pi(i; f^*(\hbv)\leq K)} \left(p_i^*\left(\hbv\right)- \tp_i\left(\hbv\right)\right) \leq 2\vmax\eta \left( \sum_{m=1}^K \Lambda_m \right) \leq 2K\vmax\eta.
\end{align*}

\noindent\textbf{Step 4: the global regret.} We now bring together the two instantaneous regrets and we have that at each round of the the exploitation phase we have the regret $r = \sum_{i=1}^N r_i$.
We first notice that the expected instantaneous regret $r_i$ for each ad $a_i$ is defined as the difference between the VCG payment $p_i^*(\hbv)$ and the (expected) payments computed by the estimated randomized mechanism $p_i(\hbv)$. We notice that $p_i^*(\hbv)$ can be strictly positive only for the $K$ displayed ads, while $p_i(\hbv) \geq 0 \ \forall i \in \N$, due to the mechanism randomization. Thus, $p_i^*(\hbv) - p_i(\hbv) > 0$ only for at most $K$ ads. Thus we obtain the per--round regret
\begin{align*}
		r &\leq \sum_{i: \pi(i;f^*(\hbv)) \leq K} r_i = \sum_{i: \pi(i;f^*(\hbv)) \leq K} \left(r_i^B + r_i^L\right)\\
		& \leq K N \mu \vmax + \sum_{i: \pi(i;f^*(\hbv)) \leq K} \left(p_i^*\left(\hbv\right)- \tp_i\left(\hbv\right) + N\mu \vmax \right)\\
		&\leq  K N \mu \vmax + 2 K \vmax \eta + K N \mu \vmax =  2 K \vmax \eta + 2 K N \mu \vmax.
\end{align*}
Finally, the global regret becomes
\begin{align*}
R_T & \leq \vmax K \left[\left( T-\tau \right) \left(2 \sqrt{\frac{N}{\tau} \log \frac{2N}{\delta}} + 2 \mu N \right) + \tau + \delta T \right].
\end{align*}

\noindent\textbf{Step 5: parameters optimization.}
We first simplify further the previous bound as
\begin{align}\label{eq:global.regret.simplified}
R_T & \leq \vmax K \left[ T\left(2 \sqrt{\frac{N}{\tau} \log \frac{2N}{\delta}} + 2 \mu N \right) + \tau + \delta T \right].
\end{align}
We first optimize the value of $\tau$, take the derivative of the previous bound w.r.t. $\tau$ and set it to zero and obtain
\begin{align*}
\vmax K\Big(-\tau^{-\frac{3}{2}} T \sqrt{N \log \frac{2N}{\delta}} + 1\Big) = 0,
\end{align*}
which leads to
\begin{equation*}
\tau = T^{\frac{2}{3}} N^{\frac{1}{3}} \left( \log{\frac{2N}{\delta}}\right)^{\frac{1}{3}}.
\end{equation*}
Once replaced into (\ref{eq:global.regret.simplified}) we obtain
\begin{align*}
R_T & \leq \vmax K \left[ 3T^\frac{2}{3}N^\frac{1}{3}\Big( \log{\frac{2N}{\delta}}\Big)^{\frac{1}{3}} + 2 T\mu N + \delta T \right].
\end{align*}
The optimization of the asymptotic order of the bound can then be obtained by setting $\mu$ and $\delta$ so as to equalize the second and third term in the bound. In particular by setting
\begin{align*}
\mu = T^{-\frac{1}{3}} N^{-\frac{2}{3}} \quad \text{ and }\quad \delta=T^{-\frac{1}{3}}N^\frac{1}{3},
\end{align*}
we obtain the final bound
\begin{align*}
R_T & \leq 6\vmax K T^\frac{2}{3}N^\frac{1}{3}\Big( \log \big(2N^\frac{2}{3}T^\frac{1}{3}\big)\Big)^{\frac{1}{3}}.
\end{align*}
\end{pf}


\section{Proof of Revenue Regret in Theorem~\ref{thm:extern}}

Before deriving the proof of Theorem~\ref{thm:extern}, we prove two  lemmas that we use in the following proofs.

\begin{lemma}\label{lem:welfare.mod}
Let $\calG$ be an arbitrary space of allocation functions, then for any $g\in\calG$, when $| q_i - \hqp_i |\leq \eta$ with probability $1 - \delta$, we have
\begin{align*}
-2K \vmax \eta \leq  \SW(g(\hbv), \hbv) - \tSW(g(\hbv), \hbv) \frac{q_i}{\hqp_{i}}  \leq \frac{2K \vmax}{q_{\min}} \eta, 
\end{align*}
with probability $1-\delta$.
\end{lemma}

\begin{pf}
By using the definition of $\SW$ and $\tSW$ we have the following sequence of inequalities
\begin{align*}
\SW(g(\hbv), \hbv) - &\tSW(g(\hbv), \hbv) \frac{q_i}{\hqp_{i}}  \\
&\leq \sum_{j: \pi(j; g(\hbv))\leq K}  \eps_{\pi(j; g(\hbv))} \hv_j \left( q_j - \hqp_{j} \frac{q_i}{\hqp_{i}}\right) \\
& \leq \vmax \sum_{j: \pi(j; g(\hbv))\leq K} \left( q_j - q_j \frac{q_i}{\hqp_{i}} + q_j \frac{q_i}{\hqp_{i}} - \hqp_{j} \frac{q_i}{\hqp_{i}} \right) \\
&= \vmax \sum_{j: \pi(j; g(\hbv))\leq K} \Bigg[ q_j \left( \frac{\hqp_{i} - q_i}{\hqp_{i}} \right) + \underbrace{(q_j - \hqp_{j})}_{\leq 0} \frac{q_i}{\hqp_{i}} \Bigg] \\
& \leq \frac{\vmax}{q_{\min}} \sum_{j: \pi(j; g(\hbv))\leq K} \left( \hq_i - q_i  + \eta \right)  \leq \frac{2K \vmax}{q_{\min}} \eta.
\end{align*}
The second statement follows from
\begin{align*}
\tSW(g(\hbv), \hbv) \frac{q_i}{\hqp_{i}} - &\SW(g(\hbv), \hbv)  \\
&= \sum_{j: \pi(j; g(\hbv))\leq K} \eps_{\pi(j; g(\hbv))}(g(\hbv)) \hv_j \left( \hqp_{j} \frac{q_i}{\hqp_{i}} - q_j \right) \\
& \leq \vmax  \sum_{j: \pi(j; g(\hbv))\leq K} \left( \hqp_{j} \frac{q_i}{\hqp_{i}} - q_j \right) \\
& \leq \vmax \sum_{j: \pi(j; g(\hbv))\leq K} (\hqp_j - q_j) \leq 2K \vmax \eta.
\end{align*}
\qed
\end{pf}

\begin{lemma}\label{lem:welfare}
Let $\calG$ be an arbitrary space of allocation functions, then for any $g\in\calG$, when $| q_i - \hqp_i |\leq \eta$ with probability $1 - \delta$, we have
\begin{align*}
0 \leq \left( \tSW(g(\hbv), \hbv) - \SW(g(\hbv), \hbv) \right) \leq 2K \vmax \eta,
\end{align*}
with probability $1-\delta$.
\end{lemma}

\begin{pf}
The first inequality follows from
\begin{align*}
\SW(g(\hbv), \hbv) &- \tSW(g(\hbv), \hbv) \\
& = \sum_{j: \pi(j; g(\hbv))\leq K} \eps_{\pi(j; g(\hbv))}(g(\hbv)) \hv_j \left( q_j  - \hqp_j \right) \\
&\leq \vmax \sum_{j: \pi(j; g(\hbv))\leq K} ( q_j - \hqp_j ) \leq 0,
\end{align*}
while the second inequality follows from
\begin{align*}
\tSW(g(\hbv), \hbv) - &\SW(g(\hbv), \hbv) \\
& = \sum_{j: \pi(j;g(\hbv))\leq K} \eps_{\pi(j;g(\hbv))}(g(\hbv)) \hv_j \left( \hqp_{j} - q_j \right) \\
&\leq \vmax  \sum_{j: \pi(j; g(\hbv))\leq K} \big( \hqp_j - q_j \big) \\
&= \vmax \sum_{j: \pi(j; g(\hbv))\leq K} \left(\hq_j + \eta - q_j \right) \leq 2K \vmax \eta.
\end{align*}
\qed
\end{pf}

We are now ready to proceed with the proof of Theorem~\ref{thm:extern}.

\begin{pf}\textit{(Theorem~\ref{thm:extern})}

\noindent \textbf{Step 1: per--ad regret.}
We first compute the instantaneous per--ad regret $r_i = p^*_i(\hbv) - \tp_i(\hbv)$ at each round of the exploitation phase for each ad $a_i$. According to the definition of payments we have
\begin{align*}
r_i = \underbrace{\SW(f^*_{-i}(\hbv), \hbv) - \tSW(\tf_{-i}(\hbv), \hbv) \frac{q_i}{\hqp_{i}}}_{r^1_i}  + \underbrace{\tSW_{-i}(\tf(\hbv),\hbv) \frac{q_i}{\hqp_{i}} - \SW_{-i}(f^*(\hbv), \hbv)}_{r^2_i}.
\end{align*}
We bound the first term through Lemma~\ref{lem:welfare.mod} and the following inequalities
\begin{align*}
r^1_i &= \SW(f^*_{-i}(\hbv), \hbv) - \tSW(f^*_{-i}(\hbv), \hbv) \frac{q_i}{\hqp_{i}} + \tSW(f^*_{-i}(\hbv), \hbv) \frac{q_i}{\hqp_{i}} - \tSW(\tf_{-i}(\hbv), \hbv) \frac{q_i}{\hqp_{i}} \\
&\leq \max_{f \in \F_{-i}} \left( \SW(f(\hbv), \hbv) - \tSW(f(\hbv), \hbv) \frac{q_i}{\hqp_{i}} \right) + \underbrace{\left( \tSW(f^*_{-i}(\hbv), \hbv) - \max_{f \in \F_{-i}} \tSW(\tf(\hbv), \hbv) \right)}_{\leq 0} \frac{q_i}{\hqp_{i}} \\
&  \leq \frac{2K \vmax}{q_{\min}} \eta,
\end{align*}
\noindent with probability $1-\delta$. We rewrite $r^2_i$ as
\begin{align*}
r^2_i &= \left( \tSW(\tf(\hbv), \hbv) - \eps_{\pi(i; \tf(\hbv))}(\tf(\hbv)) \hqp_{i} \hv_i \right)\frac{q_i}{\hqp_{i}} - \SW(f^*(\hbv), \hbv) + \eps_{\pi(i; f^*(\hbv))}(f^*(\hbv)) q_i \hv_i \\
&= \underbrace{\tSW(\tf(\hbv), \hbv) \frac{q_i}{\hqp_{i}} - \SW(f^*(\hbv), \hbv)}_{r^{3}_i} + \left(\eps_{\pi(i; f^*(\hbv))}\left(f^*(\hbv)\right) - \eps_{\pi(i; \tf(\hbv))}(\hf(\hbv))\right) q_i \hv_i.
\end{align*}
\noindent We now focus on the term $r^3_i$ and use Lemma~\ref{lem:welfare.mod} to bound it as
\begin{align*}
r^3_i &= \tSW(\tf(\hbv), \hbv) \frac{q_i}{\hqp_{i}} - \SW(\tf(\hbv), \hbv) + \underbrace{\SW(\tf(\hbv), \hbv) - \max_{f \in \F} \SW(f(\hbv), \hbv)}_{\leq 0} \\
&\leq \max_{f \in \F} \left(\tSW(f(\hbv), \hbv) \frac{q_i}{\hqp_{i}} - \SW(f(\hbv), \hbv) \right) \\
&\leq 2K \vmax \eta.
\end{align*}

\noindent\textbf{Step 2: exploitation and global regret.} 
We define $I = \{i |\pi(i; f^∗(\hbv)) \leq K \lor \pi(i; \tf(\hbv)) \leq K, i \in \N\}$, $|I| \leq 2K$. It is clear that only the ads $a_i$ s.t. $i \in I$ have a regret $r_i \not = 0$. The other ads, $i \not \in I$, have both $p^*_i(\hbv) = 0$ and $\tp_i(\hbv)=0$. Thus, we can bound the regret $r$, at each exploitative round, in the following way
\begin{align*}
r &= \sum_{i \in I} (r^1_i + r^2_i) \\
& \leq \sum_{i \in I} \Big( \frac{2K \vmax}{q_{\min}} \eta + 2K \vmax \eta \Big) + \sum_{i \in I}  \left(\eps_{\pi(i; f^*(\hbv))}(f^*(\hbv))  - \eps_{\pi(i;\tf(\hbv))}(\tf(\hbv)) \right)q_i \hv_i \\
& = \sum_{i \in I} \Big( \frac{2K \vmax}{q_{\min}} \eta + 2K \vmax \eta \Big) + \sum_{i=1}^N  \left(\eps_{\pi(i; f^*(\hbv))}(f^*(\hbv))  - \eps_{\pi(i;\tf(\hbv))}(\tf(\hbv)) \right)q_i \hv_i \\
&\leq \frac{8K^2 \vmax}{q_{min}} \eta + \SW(f^*(\hbv), \hbv) - \SW(\tf(\hbv), \hbv) \\
& = \frac{8K^2 \vmax}{q_{min}} \eta + \SW(f^*(\hbv), \hbv) - \tSW(f^*(\hbv), \hbv)+\\
& + \underbrace{\tSW(f^*(\hbv), \hbv) - \max_{f \in \F} \tSW(f)}_{\leq 0} + \tSW(\tf(\hbv), \hbv) - \SW(\tf(\hbv), \hbv) \\
& \leq \frac{8K^2 \vmax}{q_{min}} \eta + \underbrace{\SW(f^*(\hbv), \hbv) - \tSW(f^*(\hbv), \hbv)}_{r^1} + \underbrace{\tSW(\tf(\hbv), \hbv) - \SW(\tf(\hbv), \hbv)}_{r^2}
\end{align*}
The remaining terms $r^1$ and $r^2$ can be easily bounded using Lemma~\ref{lem:welfare} as
\begin{align*}
r^1 \leq 0 \quad \text{ and }\quad r^2 \leq 2K \vmax \eta.
\end{align*}
Summing up all the terms we finally obtain $$r \leq \frac{10K^2 \vmax}{q_{\min}} \eta$$ with probability $1-\delta$. Now, considering the  instantaneous  regret of the exploration and exploitation phases, we obtain the final bound on the cumulative regret $R_T$ as follows
\begin{align*}
R_T \leq \vmax K \left[ (T - \tau) \left( \frac{10 K}{\eps_{\min}\qmin} \sqrt{\frac{N}{2K\tau} \log \frac{N}{\delta}} \right) + \tau + \delta T \right].
\end{align*}

\noindent\textbf{Step 3: parameter optimization.} Let $c := \frac{5}{\sqrt{2} \Gmin \qmin}$, then we first simplify the previous bound as
\begin{align*}
R_T \leq \vmax K \left[ 2c T\sqrt{\frac{NK}{\tau} \log \frac{N}{\delta}} + \tau + \delta T \right].
\end{align*}
Taking the derivative with respect to $\tau$ leads to
\begin{align*}
\vmax K\Big(-\tau^{-\frac{3}{2}} cT \sqrt{NK \log \frac{N}{\delta}} + 1\Big) = 0,
\end{align*}
which leads to
\begin{equation*}
\tau = c^\frac{2}{3}T^{\frac{2}{3}} K^{\frac{1}{3}} N^{\frac{1}{3}} \Big( \log{\frac{N}{\delta}}\Big)^{\frac{1}{3}}.
\end{equation*}
Once replaced in the bound, we obtain
\begin{align*}
R_T & \leq \vmax K \left[ 3T^\frac{2}{3}c^\frac{2}{3}N^\frac{1}{3}K^\frac{1}{3}\Big( \log{\frac{N}{\delta}}\Big)^{\frac{1}{3}} + \delta T \right].
\end{align*}
Finally, we choose $\delta$ to optimize the asymptotic order by setting
\begin{align*}
\delta = K^\frac{1}{3} N^\frac{1}{3} c^\frac{2}{3} T^{-\frac{1}{3}},
\end{align*}
which leads to the final bound
\[
	R_T \leq 4\vmax K^\frac{4}{3} c^\frac{2}{3} T^\frac{2}{3} N^\frac{1}{3} \left(\log{\frac{N^\frac{2}{3} T^\frac{1}{3}}{K^\frac{1}{3} c^\frac{2}{3}}}\right)^\frac{1}{3}
\]
Notice that this bound imposes constraints on the value of $T$, indeed, $T>\tau$, thus $T > c^\frac{2}{3} K^\frac{1}{3} T^\frac{2}{3} N^\frac{1}{3} \left( \log{\frac{N}{\delta}} \right)^\frac{1}{3}$ and $\delta < 1$, thus $T > c^2 K N$,  leading to:
$$
		T >  c^2 K  N  \max\left\lbrace\log{\frac{N}{\delta}}, 1\right\rbrace.
$$

The problem of the previous bound is that $\tau$ and $\delta$ depends on $q_{\min}$, which is an unknown quantity. Thus actually choosing this values to optimize the bound may be unfeasible. An alternative choice of $\tau$ and $\delta$ is obtained by optimizing the bound removing the dependency on $q_{\min}$. Let $d = \frac{5}{\sqrt{2} \Gmin}$, then we choose
\[
		\tau = d^{\frac{2}{3}} K^{\frac{1}{3}} T^{\frac{2}{3}} N^{\frac{1}{3}} \left( \log{\frac{N}{\delta}} \right)^{\frac{1}{3}},
\]
and 
\[
\delta = K^\frac{1}{3} N^\frac{1}{3} d^\frac{2}{3} T^{-\frac{1}{3}},
\]
which leads to the final bound
\[
R_T \leq 4 \vmax K^\frac{4}{3} T^\frac{2}{3} N^\frac{1}{3} \frac{d^\frac{2}{3}}{\qmin} \left(\log{\frac{N^\frac{2}{3} T^\frac{1}{3}}{K^\frac{1}{3} d^\frac{2}{3}}}\right)^\frac{1}{3}
\]
under the constraint that $T \geq K N d^2$.
\qed	
\end{pf}

\section{Deviation Regret}\label{app:deviation.regret}

The definition of regret in (\ref{eq:regret}) measures the cumulative difference between the revenue of a VCG compared to the one obtained by \avcg1\ over $T$ rounds. Upper--bounds over this quantity guarantees that the loss in terms of revenue does not linearly increase with $T$. As illustrated in the previous sections, the key passage in the proofs is the upper--bounding of the regret at each round of the exploitation phase (i.e., $r = \sum_{i=1}^N (p_i^* - \tp_i)$). Nonetheless, we notice that this quantity could be negative. In this section we introduce a different notion of regret ($\tR_T$) that we study only for \avcg1, leaving for the future a more detailed analysis. Let us consider the following simple example. Let $N=3$, $K=1$, $\hv_i=1$ for all the ads, and $q_1=0.1$, $q_2=0.2$, and $q_3=0.3$. Let assume that after the exploration phase we have $\hqp_1 = 0.1$, $\hqp_2 = 0.29$, $\hqp_3 = 0.3$. A standard VCG mechanism allocates ad $a_3$ and asks for a payment $p_3^*(\hbv)=0.2$. During the exploitation phase \avcg1\ also allocates $a_3$ but asks for an (expected) payment $\tp_3(\hbv) = (\hqp_2 / \hqp_3) q_3 = 0.29$. Thus, the regret in each exploitation round is $r = p^*_3(\hbv) - \tp_3(\hbv) = -0.09$. Although this result might seem surprising, it is due to the fact that while both \avcg1\ and VCG are truthful, in general \avcg1\ is not efficient. We recall that a mechanism is efficient if for any set of advertisers it always maximizes the social welfare. In the example, if for instance the true quality of ad $a_3$ is $q_3 = 0.28$, then the allocation induced by $\hqp$s is not efficient anymore. By dropping the efficiency constraint, it is possible to design mechanisms with larger revenues than the VCG. For this reason, we believe that a more complete characterization of the behavior of \avcg1\ compared to the VCG should consider the \textit{deviation} between their payments and not only the loss in the revenue.
In particular, let us define the regret as the deviation between the VCG and the approximated VCG:
\begin{align}\label{eq:aregret}
\tR_T(\mathfrak{A}) = \sum_{t=1}^T \Big| \sum_{i=1}^N (p^*_i - \tp_{it}) \Big|,
\end{align}

We prove an upper--bound for the single--slot case (the extension of the multi--slot results is straightforward).

\begin{theorem}\label{thm:a-constant}
Let us consider a sequential auction with $N$ advertisers, $K$ slots, and $T$ rounds with position--dependent cascade model with parameters $\{\Lambda_m\}_{m=1}^K$ and accuracy $\eta$ as defined in~(\ref{eq:eta}). For any parameter $\tau \in \{0, \ldots, T\}$ and $\delta \in [0,1]$, the A--VCG1  achieves a regret:
\begin{align}\label{eq:a-regret.const.exact}
\tR_T \leq K \vmax \left( \tau + \left( T-\tau \right) \frac{2\eta}{q_{\min}} + \delta T \right)
\end{align}
where $q_{\min} = \min_{i \in \N} q_i$. By setting the parameters to
\begin{align*}
\delta &=  N^\frac{1}{3} K^{-\frac{1}{3}} T^{-\frac{1}{3}} \\
\tau &= 2^\frac{1}{3} \frac{ K^{-\frac{1}{3}} N^\frac{1}{3}  T^\frac{2}{3}}{\Lambda_{\min}^\frac{2}{3}} \left(\log{\frac{N}{\delta}}\right)^\frac{1}{3},
\end{align*}
the regret is
\begin{align}\label{eq:a-regret.const}
\tR_T \leq 4 \cdot 2^\frac{1}{3} \frac{ K^{-\frac{1}{3}} N^\frac{1}{3}  T^\frac{2}{3}}{\qmin \Lambda_{\min}^\frac{2}{3}} \left(\log{N^\frac{2}{3} K^\frac{1}{3} T^\frac{1}{3}}\right)^\frac{1}{3}.
\end{align}
\end{theorem}

\begin{pf}

We initially provide a  bound over the instantaneous regret during the exploitation phase. We consider the two sides of the bound separately. We have that for the first side of the bound we can use the result provided in Step~3 in the proof of Theorem~\ref{thm:constant}, i.e., 
\begin{align*}
r_1 &= \sum_{m = 1}^K ( p^*_{\alpha(m;\theta^*)}(\hbv) - \tilde{p}_{\alpha(m;\tilde{\theta})}(\hbv) ) \\
& \leq 2 K\vmax \eta,
\end{align*}
with probability $1-\delta$.

Now we bound the other side.
\begin{align*}
		r_2 &= \sum_{m = 1}^K \left(\tilde{p}_{\alpha(m;\tilde{\theta})}(\hbv) - p^*_{\alpha(m;\theta^*)}(\hbv) \right) \\
		&= \sum_{m = 1}^K \sum_{l = m}^{K} \Delta_l \left(  \frac{ \max\limits_{i\in \mathcal{N}}(\hqp_i \hv_i; l+1) }{\hqp_{\alpha(m;\tilde{\theta})}} q_{\alpha(m;\tilde{\theta})}   - \max\limits_{i\in \mathcal{N}}(q_i \hv_i; l+1)  \right) \\
		&\leq \max\limits_{i\in \mathcal{N}}(q_i \hv_i; l+1)  \sum_{m = 1}^K \sum_{l = m}^{K} \Delta_l \left(  \frac{ \max\limits_{i\in \mathcal{N}}(\hqp_i \hv_i; l+1) }{\max\limits_{i\in \mathcal{N}}(q_i \hv_i; l+1) }  - 1  \right)
\end{align*}

In order to proceed  with the bound, notice that, for a generic ad $a_i$ we have that $\hqp_i \hv_i = \left(\hq_i + \eta\right) \hv_i \leq \left(q_i + 2\eta\right) \hv_i \leq q_i \hv_i + \frac{2\eta}{\qmin} q_i \hv_i $.

Now, consider $i'=\arg\max\limits_{j\in \mathcal{N}}(q_j \hv_j; l+1)$, the ad displayed in $s_{l+1}$ when the true qualities are known, we can face two different situation:
\begin{itemize}
	\item $\pi\left(i'; \tf(\hbv)) \geq \pi(i'; f^*(\hbv)\right)$: in this case we can easily conclude that $\hqp_{\alpha(l+1; \tf(\hbv))} \hv_{\alpha(l+1; \tf(\hbv))} \leq \hqp_{i'} \hv_{i'} \leq q_{i'} \hv_{i'} + \frac{2\eta}{\qmin} q_{i'} \hv_{i'} $;
	\item $\pi\left(i'; \tf(\hbv)\right) < \pi\left(i'; f^*(\hbv)\right)$: in this case we can observe that $ q_{i'} \hv_{i'} + \frac{2\eta}{\qmin} q_{i'} \hv_{i'} \geq  q_j \hv_j + \frac{2\eta}{\qmin} q_j \hv_j$ $\forall j \in \N$ s.t. $\pi(j; f^*(\hbv)) < \pi(i'; f^*(\hbv))$. Thus, considering that $\exists j \in \N$ s.t. $\pi(j; f^*(\hbv)) < \pi(i'; f^*(\hbv))$ and $\pi(j; \tf(\hbv)) \geq l+1$, we can conclude $ \hqp_{\alpha(l+1; \tf(\hbv))} \hv_{\alpha(l+1; \tf(\hbv))} \leq  \hqp_j \hv_j \leq q_j \hv_j + \frac{2\eta}{\qmin} q_j \hv_j \leq q_{i'} \hv_{i'} + \frac{2\eta}{\qmin} q_{i'} \hv_{i'} $. 
\end{itemize}

Using these results we obtain
\begin{align*}
&\max\limits_{i\in \mathcal{N}}(\hqp_i \hv_i; l+1) =  \hqp_{\alpha(l+1; \tf(\hbv))} \hv_{\alpha(l+1; \tf(\hbv))} \leq \\
&\leq q_{i'} \hv_{i'} + \frac{2\eta}{\qmin} q_{i'} \hv_{i'} =  \max\limits_{i\in \mathcal{N}}(q_i \hv_i; l+1) +\frac{1}{\qmin} 2\eta \max\limits_{i\in \mathcal{N}}(q_i \hv_i; l+1)
\end{align*}
and thus
\begin{align*}
		r_2&\leq \vmax \sum_{m = 1}^K \sum_{l = m}^{K} \Delta_l \left(  \frac{ \max\limits_{i\in \mathcal{N}}(q_i \hv_i; l+1) +\frac{1}{\qmin} 2\eta \max\limits_{i\in \mathcal{N}}(q_i \hv_i; l+1) }{\max\limits_{i\in \mathcal{N}}(q_i \hv_i; l+1)} -1 \right) \\
		&\leq \vmax \sum_{m = 1}^K \sum_{l = m}^{K} \Delta_l \left(  1 +\frac{1}{\qmin} 2\eta -1 \right) \\
		&\leq \vmax \frac{1}{\qmin} 2 \eta \sum_{m = 1}^K \underbrace{\sum_{l = m}^{K} \Delta_l}_{=\Lambda_m} \leq \vmax \frac{1}{\qmin} 2 \eta K.
\end{align*}
with probability $1-\delta$. As a result we have 
\begin{align*}
\left|\sum_{m = 1}^K ( p^*_{\alpha(m;\theta^*)}(\hbv) - \tilde{p}_{\alpha(m;\tilde{\theta})}(\hbv) ) \right| \leq 2\vmax K \frac{\eta}{q_{\min}},
\end{align*}
with probability $1-\delta$. The final bound on the expected regret is thus
\begin{align}
\tR_T \leq K \vmax \left( \tau + \left( T-\tau \right) \frac{2\eta}{q_{\min}} + \delta T \right)
\end{align}

We first simplify the previous bound as
\begin{align*}
\tR_T  & \leq K \vmax \left( \tau + \frac{2T}{q_{\min}} \sqrt{\left(\sum_{m=1}^{K} \frac{1}{\Lambda_{m}^2}\right) \frac{2N}{K^2 \tau} \log \frac{N}{\delta}} + \delta T \right)\\
	   &  \leq K \vmax \left( \tau + \frac{2T}{q_{\min} \Lambda_{\min}} \sqrt{\frac{2N}{K \tau} \log \frac{N}{\delta}} + \delta T \right)
\end{align*}

and choosing the parameters

$$\tau = 2^\frac{1}{3} \frac{ K^{-\frac{1}{3}} N^\frac{1}{3}  T^\frac{2}{3}}{\Lambda_{\min}^\frac{2}{3}} \left(\log{\frac{N}{\delta}}\right)^\frac{1}{3} $$

$$ \delta =  N^\frac{1}{3} K^{-\frac{1}{3}} T^{-\frac{1}{3}}$$

the final bound is

$$ \tR_T  \leq  4 \cdot 2^\frac{1}{3} \frac{ K^{-\frac{1}{3}} N^\frac{1}{3}  T^\frac{2}{3}}{\qmin \Lambda_{\min}^\frac{2}{3}} \left(\log{N^\frac{2}{3} K^\frac{1}{3} T^\frac{1}{3}}\right)^\frac{1}{3}$$
\end{pf}

\myremark{(the bound).} We notice that the bound is very similar to the bound for the regret $R_T$ but now an inverse dependency on $q_{\min}$ appears. This suggests that bounding the deviation between the two mechanisms is more difficult than bounding the revenue loss and that as the qualities become smaller, the \avcg1\ could be less and less efficient and, thus, have a larger and larger revenue. This result has two important implications. \textit{(i)} If social welfare maximization is an important requirement in the design of the learning mechanism, we should analyze the loss of \avcg1\ in terms of social welfare and provide (probabilistic) guarantees about the number of rounds the learning mechanism need in order to be efficient (see~\cite{gonen2007incentive-compatible} for a similar analysis). \textit{(ii)} If social welfare is not a priority, this result implies that a learning mechanism could be preferable w.r.t. to a standard VCG mechanism. We believe that further theoretical analysis and experimental validation are needed to understand better both aspects.

%% file: sec/App-proofsSW.tex
\section{Proofs of Social-Welfare Regret in Theorems~\ref{th:pd_q_sw} and~\ref{th:pad_q_sw}}

Before stating the main result of this section, we need the following technical lemma.

\begin{lemma}\label{lem:SW.rexpl}
Let us consider an auction with N advertisers, K slots, and T rounds, and a mechanism that separates the exploration ($\tau$ rounds) and the exploitation phases ($T-\tau$ rounds).
Consider an arbitrary space of allocation functions $\calG$, $\tilde{g} \in \arg \max_{g' \in \calG} \tSW\left(g'(\hbv), \hbv \right)$ and $| q_i - \hqp_i |\leq \eta$ with probability $1 - \delta$. For any $g \in \calG$, an upper bound of the global regret over the SW ($R_T^{SW}$) of the mechanism adopting $\tilde{g}$ instead of $g$ is:

$$
R_T^{SW} \leq \vmax K \left[ 2 (T - \tau) \eta + \tau + \delta T \right].
$$

\end{lemma}

\begin{pf}
We now prove the bound on the social welfare, starting from the cumulative instantaneous regret during the exploitation phase.
\begin{align*}
r &= \SW(g(\hbv), \hbv) - \SW(\tg(\hbv), \hbv) \\
& = \SW(g(\hbv), \hbv) - \tSW(g(\hbv), \hbv)+\\
& + \underbrace{\tSW(g(\hbv), \hbv) - \max_{g' \in \calG} \tSW(g'(\hbv), \hbv)}_{\leq 0} + \tSW(\tg(\hbv), \hbv) - \SW(\tg(\hbv), \hbv) \\
&\leq \underbrace{\SW(g(\hbv), \hbv) - \tSW(g(\hbv), \hbv)}_{r^1} + \underbrace{\tSW(\tg(\hbv), \hbv) - \SW(\tg(\hbv), \hbv)}_{r^2}
\end{align*}
\noindent The two remaining terms $r^1$ and $r^2$ can be easily bounded by using Lemma~\ref{lem:welfare}
\begin{align*}
r &\leq r_1 + r_2 \leq 0 + 2K\vmax \eta = 2 K \vmax\eta
\end{align*}
\noindent with probability $1 - \delta$.

Thus, we can conclude that:
$$
R^{SW}_T \leq \vmax K \left[ 2 (T - \tau) \eta + \tau + \delta T \right].
$$
\end{pf}

\begin{pf}\textit{(Theorem~\ref{th:pd_q_sw}) }

\noindent \textbf{Step 1: global regret.}
We apply Lemma~\ref{lem:SW.rexpl} to the position--dependent cascade model with $\{q_i\}_{i \in \N}$ unknowns, obtaining
\begin{align*}
R^{SW}_T &\leq \vmax K \left[ 2 (T - \tau) \eta + \tau + \delta T \right]\\
		 &\leq \vmax K \left[2 (T - \tau) \frac{\sqrt{2}}{\Lambda_{\min}} \sqrt{\frac{N}{K \tau} \log \frac{2N}{\delta}} + \tau + \delta T \right]
\end{align*}
\noindent \textbf{Step 2: parameter optimization.}
First we notice that adopting the value of the parameters identified in Theorem~\ref{thm:constant} we obtain an upper bound $\tilde{O}(T^\frac{2}{3})$ for the global regret $R_T^{SW}$.
 
In order to find values that better optimize the bound over $R_T^{SW}$, let $e := \frac{\sqrt{2}}{\Lambda_{\min}}$, then we first simplify the previous bound as
\begin{align*}
R^{SW}_T &\leq \vmax K \left[ 2 e \sqrt{\frac{N}{K \tau} \log \frac{2N}{\delta}} + \tau + \delta T \right]
\end{align*}
Taking the derivative of the previous bound w.r.t. $\tau$ leads to
$$ \vmax K \left( - \tau^{-\frac{3}{2}} e T \sqrt{\frac{N}{K} \log \frac{2N}{\delta}} + 1  \right) = 0,$$
which leads to
$$\tau = e^\frac{2}{3} T^\frac{2}{3} N^\frac{1}{3} K^{-\frac{1}{3}} \left( \log \frac{2N}{\delta} \right)^\frac{1}{3}$$
Once replaced in the bound, we obtain
$$ R_T^{SW} \leq \vmax K \left[ 3  e^\frac{2}{3} T^\frac{2}{3} N^\frac{1}{3} K^{-\frac{1}{3}} \left( \log \frac{2N}{\delta} \right)^\frac{1}{3} + \delta T \right]$$
Finally, we choose $\delta$ to optimize the asymptotic order by setting
$$ \delta = e^\frac{2}{3} K^{-\frac{1}{3}} N^\frac{1}{3} T^{-\frac{1}{3}} $$
given that $\delta < 1$ this imply that $T > e^2 K^{-1} N$. 

The final bound is
$$ R_T^{SW} \leq 4 \vmax e^\frac{2}{3} K^\frac{2}{3} N^\frac{1}{3} T^\frac{2}{3} \left( \log 2 e^{-\frac{2}{3}} N^\frac{2}{3} K^\frac{1}{3} T^\frac{1}{3} \right)^\frac{1}{3}$$

\end{pf}

\begin{pf}\textit{(Theorem~\ref{th:pad_q_sw}) }

\noindent \textbf{Step 1: global regret.}
We apply Lemma~\ref{lem:SW.rexpl} to the model with position-- and ad--dependent externalities with $\{q_i\}_{i \in \N}$ unknowns, obtaining
\begin{align*}
R^{SW}_T &\leq \vmax K \left[ 2 (T - \tau) \eta + \tau + \delta T \right]\\
		 &\leq \vmax K \left[ 2 (T - \tau) \frac{\sqrt{2}}{\eps_{\min}} \sqrt{\frac{N}{K \tau} \log \frac{2N}{\delta}} + \tau + \delta T \right]
\end{align*}
\noindent \textbf{Step 2: parameter optimization.}
First we notice that adopting the value of the parameters identified in Theorem~\ref{thm:extern} we obtain an upper bound $\tilde{O}(T^\frac{2}{3})$ for the global regret $R_T^{SW}$.
 
In order to find values that better optimize the bound over $R_T^{SW}$, it is possible to use the procedure followed in the proof of Theorem~\ref{th:pd_q_sw} with $e := \frac{\sqrt{2}}{\eps_{\min}}$:
$$ R_T^{SW} \leq 4 \vmax e^\frac{2}{3} K^\frac{2}{3} N^\frac{1}{3} T^\frac{2}{3} \left( \log 2 e^{-\frac{2}{3}} N^\frac{2}{3} K^\frac{1}{3} T^\frac{1}{3} \right)^\frac{1}{3}$$

\end{pf}


\section{Proof of Social-Welfare Regret in Theorem~\ref{thm:constant.l.sw.baba}}

\begin{pf}\textit{(Theorem~\ref{thm:constant.l.sw.baba})}

The bound over the global regret on the social welfare ($R_T^{SW}$) can be easily derived considering that each bid is modified by the self--resampling procedure with a probability of $\mu$.
Thus we can define $S' = \{\bs' | \bs'\in\{0,1\}^N,  \pi(i;f^*(\hbv))\leq K  \Rightarrow s'_i = 1    \}$, i.e. all the random realization where the self--resampling procedure does not modify the bids of the ads displayed when the allocation function is $f^*$ is applied to the true bids $\hbv$.
Thus we have:
\begin{align*}
	R_T^{SW} &\leq T \left( \mathbb{P}\left[\bs \in S'\right] \cdot 0 + \underbrace{\mathbb{P}\left[\bs \not \in S'\right]}_{\leq K \mu} K \vmax \right) \leq K^2 \mu \vmax T
\end{align*} 
\end{pf}


\section{Proof of Social-Welfare Regret Theorem~\ref{th:pd_lq_sw}}

\begin{pf}\textit{(Theorem~\ref{th:pd_lq_sw})}

\noindent \textbf{Step 1: instantaneous regret.}
We start computing the instantaneous regret over the SW during the exploitation phase.

First of all we introduce the following definition: $S' = \{\bs' |\bs'\in\{0,1\}^N, \pi(i;f^*(\hbv) )\leq K \Rightarrow s'_i = 1  \}$, i.e. all the random realization where the self--resampling procedure does not modify the bids of the ads displayed when the allocation function is $f^*$ is applied to the true bids $\hbv$.  

We now provide the bound over the regret.

\begin{align*}
r &= \SW(f^*(\hbv), \hbv) - \E_{\bx}\left[\SW(\tf(\bx), \hbv)|\hbv\right] \\
&= \underbrace{\Prob[\bs \in S']}_{\leq 1} \left(\SW(f^*(\hbv), \hbv) - \E_{\bx|\bs \in S'}\left[\SW(\tf(x), \hbv)|\hbv\right]\right) +\\
& +  \underbrace{\Prob[\bs  \not \in S']}_{\leq K\mu} \left(\SW(f^*(\hbv), \hbv) - \E_{\bx | \bs \not \in S'}\left[\SW(\tf(x), \bv)|\hbv\right]\right)\\
& \leq \SW(f^*(\hbv), \hbv) - \E_{\bx|\bs \in S'}\left[\SW(\tf(\bx), \hbv)|\hbv\right] + \\ & + K \mu \underbrace{\left(\SW(f^*(\hbv), \hbv) - \underbrace{\E_{\bx | \bs \not \in S'}\left[\SW(\tf(x), \bv)|\hbv\right]}_{\geq 0}\right)}_{\leq K \vmax}\\
& \leq \underbrace{\SW(f^*(\hbv), \hbv) - \E_{\bx|\bs \in S'}\left[\tSW(f^*(\bx), \hbv)|\hbv\right]}_{r_1 \leq 0}+ \\&+ \underbrace{\E_{\bx|\bs \in S'}\left[\tSW(f^*(\bx), \hbv)|\hbv\right] - \E_{\bx|\bs \in S'}\left[\tSW(\tf(\bx), \hbv)|\hbv\right]}_{r_2 \leq 0}+ \\
& + \E_{\bx|\bs \in S'}\left[\tSW(\tf(\bx), \hbv)|\hbv\right] - \E_{\bx|\bs \in S'}\left[\SW(\tf(\bx), \hbv)|\hbv\right] + \vmax \mu K^2\\
& \leq \max_{f \in \F} \left( \E_{\bx|\bs \in S'}\left[\tSW(f(\bx), \hbv) - \SW(f(\bx), \hbv)|\hbv\right] \right) + \vmax \mu K^2\\
& \leq \max_{f \in \F} \left( \sum_{j: \pi(j; f(x)) \leq K} \Lambda_{\pi(j; f(x))} v_j (\hq_j - q_j) \right) + \vmax \mu K^2\\
& \leq \vmax \max_{f \in \F} \left( \sum_{j: \pi(j; f(x)) \leq K} (\hq_j - q_j) \right) + \vmax \mu K^2\\
& \leq 2 \vmax K \eta + \vmax \mu K^2 = \vmax K \left( 2 \eta + K \mu   \right)
\end{align*}

We provide a brief intuition of bounds $r_1$ and $r_2$. The bound $r_1$ can be explained noticing that when the bids of the ads displayed in $f^*(\hbv)$ are not modified we have that $\alpha(m; f^*(\hbv)) = \alpha(m; f^*(\bx))$ where $m\leq K$ and $\bx$ s.t. $\bs \in S'$. The bound for $r_2$ can be understood noticing that when the bids of the ads s.t. $\pi(j; f^*(\bx)) \leq K$ are not modified and $x_i \leq \hv_i \ \forall i \in \N$, we obtain $\tSW(f^*(\bx), \hbv) = \tSW(f^*(\bx), \bx) \leq \max_{\theta \in \Theta} \tSW(\theta, \bx) = \tSW(\tf(\bx), \bx) \leq \tSW(\tf(\bx), \hbv)$. 

%

\noindent \textbf{Step 2: global regret.}
We can now compute the upper bound for the global regret

\begin{align*}
R_T^{SW} & \leq \vmax K \left[ (T - \tau) ( 2 \eta + K \mu) + \tau + \delta T \right]\\
& \leq \vmax K \left[ (T - \tau) \left( 2 \sqrt{\frac{N}{\tau} \log{\frac{2N}{\delta}}} + K \mu \right) + \tau + \delta T \right]
\end{align*}

\noindent \textbf{Step 3: parameter optimization.}
We first simplify the previous bound as
\begin{align*}
R_T^{SW} & \leq \vmax K \left[ 2 T \sqrt{\frac{N}{\tau} \log{\frac{2N}{\delta}}} + K \mu T  + \tau + \delta T \right]
\end{align*}

Taking the derivative of the previous bound w.r.t. $\tau$ leads to

$$ \vmax K  \left( -\tau^{-\frac{3}{2}} T \sqrt{N\log{\frac{2N}{\delta}}} + 1 \right)  = 0,$$

which leads to
 
$$\tau = N^\frac{1}{3}  T^\frac{2}{3} \left(\log{\frac{2N}{\delta}}\right)^\frac{1}{3}$$

Once replaced in the bound, we obtain

\begin{align*}
 R_T^{SW} & \leq 3 \vmax  K N^\frac{1}{3} T^\frac{2}{3} \left(\log{\frac{2N}{\delta}}\right)^\frac{1}{3}+ \mu K^2 \vmax T + \delta \vmax K  T
\end{align*}

Finally, we choose $\delta$  and $\mu$ to optimize the asymptotic order by setting

\begin{align*}
\delta & =  N^\frac{1}{3} T^{-\frac{1}{3}}\\
\mu & = K^{-1}  T^{-\frac{1}{3}} N^\frac{1}{3}
\end{align*} 

given that $\delta < 1$ this imply that $T > N$ and, given that $\mu < 1$ we have that $T>\frac{N}{K^3}$. 

The final bound is

$$ R_T^{SW} \leq  5 \cdot \vmax  K N^\frac{1}{3} T^\frac{2}{3} \left( \log{2 N^\frac{2}{3}  T^\frac{1}{3}} \right)^\frac{1}{3}$$

\end{pf}